\documentclass[aps,prb,showpacs,twocolumn,floats,superscriptaddress,a4paper]{revtex4}
\usepackage{amssymb}
\usepackage{amsbsy}
\usepackage{amsmath}
\usepackage{graphicx}
\usepackage{textcomp}
\usepackage[body={18cm,24.3cm},top={2.5cm},left={1.5cm}]{geometry}

\let\a=\alpha  \let\g=\gamma \let\d=\partial 
   
\let\l=\lambda     
\let\s=\sigma

   \let\Y=\Psi
     \let\da=\dagger
\let\la=\label  \let\re=\ref
  
\def\dR{\delta{\mathbf{R}}}
\def\DF{\Delta{F}}

\def\nn{\nonumber}

\let\fr=\frac

\def\bpm{\begin{pmatrix}}
\def\epm{\end{pmatrix}}
\def\be{\begin{equation}}
\def\ee{\end{equation}}
\def\bea{\begin{eqnarray}}
\def\eea{\end{eqnarray}}
\def\ba{\begin{array}}
\def\ea{\end{array}}

\def\vp{\varphi}
\def\td{\tilde}

\def\ep{{\epsilon}}

\def\var{\text{var}}
\def\re{\text{Re}}
\def\im{\text{Im}}
\def\var{\sigma^2_\nu}
\def\Var{\chi^2_\nu}

\def\Varo{\chi^2_1}
\def\Vart{\chi^2_2}
\def\hth{\delta\theta}

\def\hph{\delta\phi}
\def\hvp{\delta\varphi}

\begin{document}
\title{Anderson localization of one-dimensional hybrid particles}

\author{Hong-Yi Xie}
\affiliation{Condensed Matter Theory Sector, SISSA-ISAS, Trieste,
Italy}
\author{V. E. Kravtsov}
\affiliation{ The Abdus Salam International Centre for
Theoretical Physics, P.O.B. 586, 34100 Trieste, Italy and\\
Landau Institute for Theoretical Physics, 2 Kosygina st.,117940
Moscow, Russia.}
\author{M. M\"uller}
\affiliation{The Abdus Salam International Centre for Theoretical
Physics, P.O.B. 586, 34100 Trieste, Italy.}
%\author{Hong-Yi Xie$^{1}$, M.Muller$^{2}$, V.E.Kravtsov$^{2,3}$}
%\affiliation{Condensed Matter Theory Sector, SISSA-ISAS, Trieste,
%Italy\\ $^{2}$The Abdus Salam International Centre for
%Theoretical Physics, P.O.B. 586, 34100 Trieste, Italy.\\
%$^{3}$Landau Institute for Theoretical Physics, 2 Kosygina
%st.,117940 Moscow, Russia.}

\date{\today}

\begin{abstract}
We solve the Anderson localization problem on a two-leg ladder by
the Fokker-Planck equation approach. The solution is exact in the
weak disorder limit at a fixed inter-chain coupling.
The study is motivated by progress in investigating the hybrid
particles such as cavity polaritons. This application corresponds to
parametrically different intra-chain hopping integrals (a ``fast'' chain
coupled to a ``slow'' chain). We show that the canonical
Dorokhov-Mello-Pereyra-Kumar (DMPK) equation is insufficient for
this problem. Indeed, the angular variables describing the
eigenvectors of the transmission matrix enter into an extended DMPK
equation in a non-trivial way, being entangled with the two
transmission eigenvalues. This extended DMPK equation is solved
analytically and the two Lyapunov exponents are obtained as
functions of the parameters of the disordered ladder. The main
result of the paper is that near the resonance
energy, where the dispersion curves of the two decoupled and
disorder-free chains intersect, the localization properties of the
ladder are dominated by those of the slow chain. Away from the resonance  they are
dominated by the fast chain: a local excitation on the slow
chain may travel a distance of the order of the localization length of the fast chain.
\end{abstract}

\maketitle

\section{introduction}
Despite more than half a century of history, Anderson
localization\cite{and} is still a very active field whose influence
spreads throughout all of physics, from condensed matter to wave
propagation and imaging. \cite{fify} A special field where most of
rigorous results on Anderson localization have been obtained, are
one-dimensional and quasi-one-dimensional systems with uncorrelated
disorder. Most of the efforts in this direction were made to obtain the
statistics of localized wave functions in strictly one-dimensional
continuous systems~\cite{berez,mel} or tight-binding chains (see the
recent work Ref.~\onlinecite{krav-yud} and references therein).
Alternatively, the limit of thick multi-channel $N \gg 1$ wires has
been studied by the nonlinear super-symmetric
sigma-model.\cite{efet}

A transfer matrix approach which allows one to consider any number of
channels $N$ was suggested by Gertsenshtein and Vasil'ev in the field of random waveguides.\cite{ger-vas}
This approach has been applied to the problem of Anderson localization by Dorokhov\cite{dorok} and later on
by Mello, Pereyra and Kumar (DMPK).\cite{mpk2}
It is similar in
spirit to the derivation of the Fokker-Planck equation (the
diffusion equation) from the Langevin equation of motion for a
Brownian particle. However, in the present case an elementary step
of dynamics in time is replaced by the scattering off an ``elementary
slice'' of the $N$-channel wire. As a result, a kind of
Fokker-Planck equation arises which describes diffusion in the space
of parameters of the scattering matrix $\mathbf{M}$, in which the
role of time is played by the co-ordinate along the
quasi-one-dimensional system. Usually the scattering matrix
$\mathbf{M}$ is decomposed in a multiplicative way by the Bargmann's
parametrization\cite{mpk2} which separates the ``angle variables''
of the $U(N)$-rotation matrices and the $N$ eigenvalues
$T_{\rho=1,...,N}$ of the transmission matrix. If the probability
distribution of the scattering matrix is assumed invariant under
rotation  of the local basis (\emph{isotropy assumption}), the
canonical DMPK equation\cite{dorok,mpk1,mpk2} may be
obtained, which has the form of a Fokker-Planck equation in the
space of $N$ transmission eigenvalues. This equation was solved in
Ref.~\onlinecite{bee} for an arbitrary number $N$ of transmission
channels.

The isotropy condition is not automatically fulfilled. It is believed that the
isotropy condition is valid for a large number $N \gg 1$ of well
coupled chains where the ``elementary slice'' is a macroscopic
object and the ``local maximum entropy ansatz'' applies.\cite{mpk2}
It is valid at weak disorder in a strictly one-dimensional chain in
the continuum limit $a \rightarrow 0$, or for a one-dimensional
chain with finite lattice constant $a$ outside the center-of-band
anomaly. In this case the distribution of the only angular variable
describing a $U(1)$ rotation, the scattering phase, is indeed
flat.\cite{krav-yud}

However, the case of \emph{few} ($N \gtrsim 1$) coupled chains is much
more complicated. 
 As was pointed out originally by
Dorokhov\cite{dorok}, and later on by Tartakovski,\cite{tar} in this case the angular and radial variables,
are entangled in the Fokker-Planck equation. These are the variables determining
the eigenvectors and eigenvalues of the transmission matrix, respectively. We will refer to
this generic Fokker-Planck equation as the \emph{extended} DMPK
equation in order to distinguish it from the \emph{canonical} DMPK
equation which contains only the radial part of the Laplace-Beltrami
operator.
 The minimal model where such an entanglement is unavoidable, is the two-leg model of $N=2$ coupled
 disordered chains.

Yet this case is important not only as a minimal system where
the canonical DMPK equation breaks down.
 It is relevant for the Anderson localization of linearly mixed hybrid particles
 such as
 polaritons.\cite{hh-swk} Polaritons are the result of coherent mixing of the electromagnetic field in a
 medium (photons in a waveguide for example) and excitations of matter (excitons). In the absence of disorder photons
have a much larger group velocity than excitons, and thus one
subsystem is fast while the other one is slow. As a specific example, quasi-one-dimensional resonators
 were recently fabricated by confining electromagnetic fields inside a semiconductor rod~\cite{tri} or to
 a sequence of quantum wells.~\cite{man}
 In such resonators the dispersion of transverse-quantized photons is quadratic in the small momentum, with an
 effective mass as small as $10^{-4}$ of the effective mass of the Wannier-Mott exciton which is of the
 order of the mass of a free electron. 

Disorder is unavoidable in such systems due to the imperfections
 of the resonator boundary and impurities. In many cases one can consider only one mode of transverse
 quantization for both the photon and the exciton. Thus a model of two dispersive modes (particles)
 with parametrically different transport properties arise. Due to the large dipole moment of the exciton
 these particles are mixed, resulting in avoided mode crossing. On top of that, disorder acts
 on both of them, whereby its effect on the two channels can be rather different~\cite{sav}. It is easy to see~\cite{sav} that this system maps one-to-one onto a single particle model of two coupled chains in the presence of disorder.
Ref.~\onlinecite{sav} solved the coupled Dyson equations for the
Green's functions of exciton and cavity phonon numerically,
focussing on  the so-called ``motional narrowing'' in the
reflectivity spectra of normal incidence~\cite{whi}. However, the
issue of  localization of cavity polaritons was not raised. The
latter was addressed in Ref.~\onlinecite{kos}, which analyzed the
scattering of electromagnetic waves in a disordered quantum-well
structure supporting excitons. The random susceptibility of excitons
in each quantum well was shown to induce disorder for the light
propagation, and the Dyson equation for the Green's function of the
electromagnetic wave was then solved by the self-consistent theory
of localization. The author reached the conclusion that the
localization length of light with frequencies within the polariton
spectrum is substantially decreased due to enhanced backscattering
of light near the excitonic resonance. This is in qualitative
agreement with our exact and more general study of the coupled
disordered two-leg problem. The latter also finds natural
applications in nanostructures and electronic propagation in
heterogeneous biological polymers, such as DNA molecules~\cite{wei}.

The main question we are asking in the present paper is: What happens to the localization
properties when a fast chain is coupled to a slow one? Will the
fast chain dominate the localization of the hybrid particle (e.g. a
polariton) or the slow one? In other words: will the smallest
Lyapunov exponent of the two-leg system (the inverse localization
length)
 be similar to the one of the isolated fast chain, or rather to the one of the isolated slow chain? Can the presence of
 the ``more strongly quantum'' component (photon) help the ``more classical'' component (exciton) to get out of
 the swamp of localization? This latter question can be asked in many different physical situations. It has been referred to as the ``M\"{u}nchhausen effect'' in Ref.~\onlinecite{tho},
to describe the following effect predicted for a dc SQUID (superconducting quantum
interference device) with two biased Josephson junctions, one
with small plasma frequency (large mass), the other one with large plasma
frequency (small mass): The junction with small mass can actually drag the
``slower'' junction (larger mass) out of its metastable state.
%{\bf HYX: Baron M\"{u}nchhausen allegedly pulled himself and the horse which he was riding out of a swamp by his own hair. The ``M\"{u}nchhausen effect'' is realized in a dc superconducting quantum interference device with  two biased Josephson junctions in Ref. \onlinecite{tho}.}

 Here we give an answer in the specific situation of a single hybrid particle.
 More interesting situations may arise when interacting and non-equilibrium polaritons are
 approaching Bose-condensation.\cite{tri,man,marcht,wertz,ale}

A further question of more general interest can be addressed by the
same model problem. Namely, consider two or more coupled channels
with similar propagation speed (i.e., inverse effective mass), but
different disorder level: Which channel will dominate the
localization, the cleaner or the more disordered one? This type of
question arises not only in these hybrid single particle problems,
but is an important element in the analysis of many particle
problems, where few and many particle excitations have various
channels of propagation (e.g., all particles moving together, or
moving in subgroups of fewer particles). It is an important, but
scarcely understood question, what determines the character of the
propagation of such excitations when many parallel, but coupled
channels with different transport characteristics exist. Intuitively
one expects the fastest, and least disordered channel to dominate
the delocalization. 

However, our analytical solution of the hybrid
two-leg chain shows that in the one-dimensional case, this intuition
is not always correct. Instead we find that, when the channels are strongly
mixing with each other, it is the largest rate of back scattering,
i.e., the more disordered chain, which dominates the physics. This
may be seen as one of the many manifestations of the fact that in one dimension
the localization length is essentially set by the mean free path.
Our solution of the two chain problem furnishes a useful benchmark for approximate solutions
in more complex and interacting situations. However, we caution
that the phenomenology may be quite different in higher
dimensions. We will discuss this further in the conclusion.

The answer to the above questions will be obtained analytically from
the exact solution of the two-leg (two-chain) Anderson localization
model. This solution represents a major technical advance, because
for the first time a model, which leads to an extended DMPK equation
with non-separable angular and radial variables, is exactly solved.
Without going into details our results are the following:

(i) The answer depends qualitatively on whether or not the system is
close  to the \emph{resonance energy} $E_R$, which is defined as the
energy where the dispersion curves of the two corresponding
decoupled disorder-free chains intersect (see Fig.~\ref{disper}).

(ii) \emph{Near the resonance} the presence of the fast leg does not
help to substantially delocalize the slow component (see
Fig.~\ref{scale}). The localization length of a hybrid particle is
at most by a factor of $\approx 3$ larger than the one of the
slow particle (see Eqs.~(\ref{loc-length-e-0}) and (\ref{C1})), being parametrically smaller than that of the fast
particle. Thus the slow particle dominates the localization
properties of the hybrid particle near the resonance energy $E_R$.

(iii) A particular case where the resonance happens at all energies
is the case of two coupled identical chains subject to different
disorder (see Fig.~\ref{neq-disorder}). In this case the dominance of
the more disordered chain extends to all energies thus pushing the
localization length of the ladder sharply down compared to that of
the less disordered isolated chain.

(iv)  \emph{Away from the resonance} the wavefunctions stay either mostly on the slow leg, being strongly localized. Or they have their main weight on the fast leg, and hybridize here and there with the slow leg
(see Fig.~\ref{disper} and Fig.~\ref{band-edge}). It is this second type of wavefunctions which helps excitations on the slow leg to delocalize due to the presence of the faster leg, 
even though this happens with small probability far from the resonance.

(v) A very peculiar behavior occurs \emph{near the band-edges} of
the slow particle, where the system switches from two to one
propagating channels. Just below the band-edge the localization
length of the hybrid particle decreases dramatically being driven
down by the localization length of the slow chain that vanishes
 at the band-edge (neglecting the Lifshitz tails). Above the
band-edge the localization length of a hybrid particle sharply
recovers, approaching the value typical for the one-chain problem.
Thus near the band-edge the localization length of the two-leg
system has a sharp minimum, which is well reproduced by direct
numerical simulations (see Fig.~\ref{r-energy}).

The paper is organized as follows. In Section II the problem is
formulated and the main definitions are given.  In Section III the
extended DMPK equation is derived. In Section IV the exact solution
for the localization lengths is given and the main limiting cases
are discussed. In Section V numerical results concerning the wave
functions in each leg are presented. In Sec.~\ref{one-channel} a problem of one propagating channel and one evanescent channel is considered. The application of the
theory to hybrid particles such as polaritons, as well as considerations about higher dimensions, are discussed,
in the Conclusion.

\section{Two-leg Anderson model and transfer matrix for ``elementary'' slice}   \label{tran-sing-slic}
\subsection{The model}
The Anderson model on a two-leg ladder is determined by the tight-binding Hamiltonian
\bea      \la{full-ham}
H= &&\sum_{\nu=1,2} \sum_{x}\left( {\ep_{x\nu}c_{x\nu}^{\da}c_{x\nu}}-
{t_{\nu} {\left(c_{x\nu}^{\da}c_{x+1\nu}+h.c.\right)}} \right)  \nn \\
   &&-t\sum_{x}{\left(c_{x1}^{\da}c_{x2}+h.c.\right)}+\delta{e}\sum_{x}{c_{x2}^{\da}c_{x2}},
\eea where $x \in \mathbb{Z}$ is the co-ordinate along the
ladder, and $\nu \in \{1,2\}$ is the index labelling the two legs. In this model
the on-site energies $\ep_{x\nu}$ are independently distributed Gaussian random variables with
zero mean, and $t_{\nu}$ is the
hopping strength between nearest-neighbor sites on the $\nu$-th leg.
In general, the two legs will be subject to different random
potentials, characterized by the two variances:
\begin{equation}
\var = \overline{\ep_{x\nu}^2}.   \label{var1}
\end{equation}
We also consider different hopping strengths, for which we assume
\begin{equation}
t_1 \ge t_2.
\end{equation}
%Below we may thus refer to chain $1$ and $2$ as the ``fast'' and the ``slow'' chain, respectively.
The transverse hopping strength between the legs is $t$. Finally, it is natural to  consider a homogeneous
potential $\delta{e}$ (i.e., a detuning) on leg $2$.
%This may influence the localization properties of the system at a given energy as well.

%The continuous counterpart of Eq.~(\ref{full-ham}) which can be
%derived in the limit of vanishing lattice constant $a\rightarrow 0$
%for energies close to $E_{2}^{(-)}\sim E_{1}^{(-)}$ (see Fig.1) near
%the bottom of the dispersion curves for isolated chains, takes the
%form: \bea H=
%\sum_{\nu=1,2}\left(-\frac{\hbar^2}{m_\nu}\frac{d^2}{dx^2}
%+\epsilon_\nu(x)\right)\, P_{\nu}+\delta e\sigma_z -t\sigma_x \eea
%where $P_{\nu}=\frac{1-(-1)^\nu\sigma_z}{2}$ and the Pauli matrices
%$\sigma_{x,z}$ act in the spinor space of the two legs, and the weak
%Gaussian disorder is defined by the correlation function: \bea
%\overline {\epsilon_\nu(x) \epsilon_{\nu'}(x')} = W_\nu^2
%\delta(x-x')\delta_{\nu\nu'}. \eea The masses $m_{\nu}=t_{\nu}a^{2}$
%are controlled by the hopping integrals $t_{\nu}$  and are supposed
%to be finite in the limit $a\rightarrow 0$.

The Hamiltonian~(\ref{full-ham}) is a generic model describing two
coupled, uniformly disordered chains. Moreover, the model can also be
adopted as an effective model to describe non-interacting excitations with two linearly
mixing channels of propagation in the presence of disorder. An important example is polaritons, the two channels
correspond to the photon mode and the exciton mode, respectively. 
%{\bf In principle, we could also study the effects of correlations in the random onsite energies $\ep_{x\nu}$. 
%It has be shown numerically that combinations of long-range correlations along the longitudinal direction and 
%short-range correlations in the transverse direction may lead to delocalized states in some energy intervals.\cite{guo-xia}
%However, in the present work we only study uncorrelated disorders.}  

The model (\ref{full-ham}) has been studied analytically previously in the literature, focusing on the special case $t_1 =t_2$ and $\s_1^2 = \s_2^2$. 
The continuous limit was solved long ago by Dorokhov.\cite{dorok} The tight-binding model was considered later on by Kasner and Weller.\cite{kas-wel} Their
results will be reference points for our more general study in the present
work. 

The Schr\"{o}dinger equation of the Hamiltonian~(\ref{full-ham}) at a given energy $E$ has the form
\begin{equation}  \label{sch-equ-1}
\Y(x-1)+\Y(x+1)=(\mathbf{h}(E)+\boldsymbol{\ep}_x) \Y(x),
\end{equation}
where $\Y(x)$ is a single particle wavefunction with two components,
representing the amplitudes on the leg 1 and 2,
\be \label{ham1-p}
\mathbf{h}(E) = \begin{pmatrix} -\fr{E}{t_1} & -\fr{t}{t_1} \\
-\fr{t}{t_2} & -\fr{E-\delta{e}}{t_2} \end{pmatrix}, \ee and \be
\label{ham1-ip}
\boldsymbol{\ep}_x=\text{diag}\left(\fr{\ep_{x1}}{t_1},\,\fr{\ep_{x2}}{t_2}\right).
\ee The terms $\mathbf{h}(E)$ and $\boldsymbol{\ep}_x$ can be
considered as the disorder-free and disordered part of the local
Hamiltonian at the co-ordinate $x$. Notice that the disordered part~(\ref{ham1-ip})
is expressed as an \emph{effective disorder} on
the two legs, i.e., it is measured in units of the  hopping
strengths. In the analytical part of the present work, following the
Fokker-Planck approach, we solve  the problem exactly in the case of
small disorder, $||\boldsymbol{\epsilon}_{x}||\ll 1$.

\subsubsection{Disorder free part}
The disorder-free ladder can easily be solved by diagonalizing $\mathbf{h}(E)$
in Eq.~(\ref{ham1-p}). Thereby, the Schr\"{o}dinger equation
transforms into
\begin{equation}   \label{schro-2}
\td{\Y}(x-1)+\td{\Y}(x+1)=(\td{\mathbf{h}}+\td{\boldsymbol{\ep}}_x) \td{\Y}(x),
\end{equation}
where \be \label{ham2-p} \td{\mathbf{h}}=\text{diag} (
\l_{1},\,\l_{2} ), \ee and the ``rotated'' disorder potential is given
by:

\be \label{ham2-ip} \td{\boldsymbol{\ep}}_x = \begin{pmatrix}
                   \ep_{x+}+\ep_{x-}\cos{\g} & \ep_{x-}\sin{\g} \\
                   \ep_{x-}\sin{\g}   &   \ep_{x+}-\ep_{x-}\cos{\g}
                   \end{pmatrix}.
\ee
Both depend implicitly on $E$ via $\lambda_\tau(E)$ and $\gamma(E)$. In Eqs.~(\ref{ham2-p}) and (\ref{ham2-ip}) the following definitions
are used:

(i) In the disorder-free part~(\ref{ham2-p}),
\bea   \label{two-bands}
\l_{\tau}(E)=&-&\fr{1}{2} \left( \fr{E}{t_1}+\fr{E-\delta{e}}{t_2} \right)   \\
          &-&(-1)^{\tau} \sqrt{\fr{1}{4} \left( \fr{E}{t_1}-\fr{E-\delta{e}}{t_2}
          \right)^2+\fr{t^2}{t_1 t_2}},\nn
\eea\\
where $\tau \in \{1,2\}$ is the channel or band index. As we will see in Eq.~(\ref{dispers}), $\tau=1$
labels the conduction band, and $\tau=2$ the valence band of the pure ladder.~\cite{ftn}

(ii) In the disordered part~(\ref{ham2-ip}),
\begin{equation}  \label{epm}
\ep_{x\pm}=\fr{1}{2}\left(\fr{\ep_{x1}}{t_1} \pm \fr{\ep_{x2}}{t_2}\right)
\end{equation}
are the symmetric and anti-symmetric combination of the disorder on the two legs. The
``mixing angle'' $\g=\g(E)$ is defined through
\begin{equation}\label{clean-angle}
\tan{\g}(E)= \fr{\sqrt{2}t\sqrt{t_1^2+t_2^2}} {(t_1-t_2)(E-E_R)},
\end{equation}
with a resonance pole at
\begin{equation}   \label{res-en}
E_R=\delta{e}\frac{t_1}{t_1-t_2}.
\end{equation}
The value of $\g$ is chosen as: $\g \in [0,\pi/2]$ if $E \ge E_R$; $\g \in [\pi/2,\pi]$ if $E \le E_R$.

\begin{figure}[t]
\centering
\includegraphics[height=4.3cm,width=8.5cm]{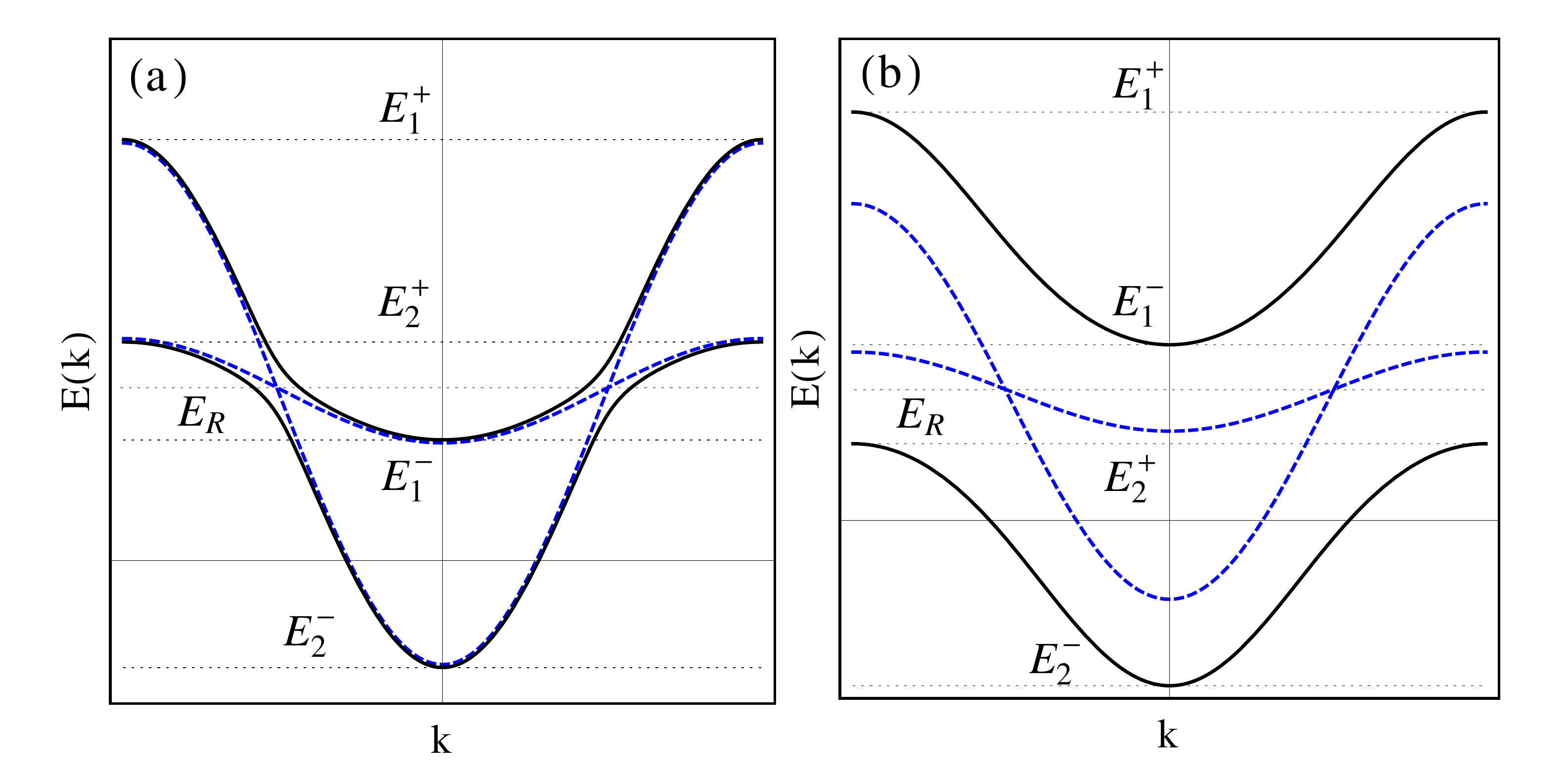}
\caption{Two situations of clean energy dispersions. The dashed and solid curves correspond to decoupled and coupled chains.
The decoupled dispersion curves intersect at the resonance energy $E_R$. (a) {\em no
gap:} $E_{1}^{-} \le E_{2}^{+}$. There are two propagating channels
at a given energy for $E_{1}^{-}\le E \le E_{2}^{+}$. (b) {\em
gapped:} $E_{1}^{-}>E_{2}^{+}$. Apart from a forbidden band, only
one propagating channel exists.} \label{disper}
\end{figure}

The pure system can be solved easily. In the absence of disorder the eigenfunctions
$\td{\Y}(x)$ at energy $E$ are composed of plane waves with momenta $k_\tau$
satisfying \be 2\cos{k_{\tau}}=\lambda_{\tau}.     \label{two-ch}
\ee $\pm k_\tau$ are degenerate solutions of Eq.~(\ref{two-ch}),
which is due to the space-inversion symmetry along the longitudinal
direction of the pure ladder. Eq.~(\ref{two-bands}) and
(\ref{two-ch}) determine the energy dispersions of the conduction
band and the valance band,
\bea  \label{dispers}
E_{\tau}(k)=&-&\left( t_1+t_2 \right) \cos{k}+\frac{\delta{e}}{2}  \\
            &-&(-1)^{\tau}\sqrt{\left[(t_1-t_2)\cos{k}+\frac{\delta{e}}{2}\right]^2+t^2}.\nn
\eea
Generally, if $t_1 \neq t_2$, the two \emph{decoupled} bands
(i.e., $t=0$ in Eq.~(\ref{dispers})) cross at the energy $E_R$ [cf. Fig.~\ref{disper}], if
$|\delta{e}|\le 2(t_1-t_2)$. When the energy $E$ is close to
the resonance energy $E_{R}$, the two legs mix with almost equal weights, even if we turn on
a very small inter-chain coupling $t$.  In the particular case of equal chain hoppings
$t_1=t_2$ and no detuning $\delta{e}=0$, there is a resonance at all energies
since the two decoupled bands coincide.

The top ($+$) and bottom
($-$) edges of the $\tau$-band are
\begin{equation}
\begin{split}
E_{\tau}^{\pm} = &\pm(t_1+t_2)+\frac{\delta{e}}{2}\\
                 &-(-1)^{\tau} \sqrt{\left( t_2-t_1 \pm \frac{\delta{e}}{2} \right)^2+t^2}.
                    \label{band-dedge}
\end{split}
\end{equation}
According to Eq.~(\ref{dispers}), there are two cases of energy
dispersions, which may arise depending on the choice of parameters:

(i) In the case of $E_{1}^{-} \le E_{2}^{+}$ [see Fig.~\ref{disper}(a)], there is \emph{no gap} between the two bands. This
is the case if the detuning $\delta e$ and the interchain coupling $t$ are both not too large. More
precisely, one needs {\bf $|\delta e|< 2(t_1+t_2)$} and $t \le t_c$, where
\begin{equation}   \label{tc}
t_c=\frac{\sqrt{t_1t_2 \left[ 4 \left( t_1+t_2 \right)^2-\delta{e}^2 \right]}}{t_1+t_2}.
\end{equation}
In the energy interval $E_{1}^{-} \le E \le E_{2}^{+}$, we have two
propagating channels; otherwise, at most one propagating channel
exists.

(ii) In the opposite case, $E_{1}^{-} > E_{2}^{+}$ [see Fig.~\ref{disper}(b)], there is a \emph{gap} between the two bands. We
therefore have at most one propagating channel at any energy.

Moreover, if $k_\tau$ is the wavevector of a propagating channel, we call
$k_\tau \in ( -\pi,\pi ]$, and $k_\tau \ge 0$ and $k_\tau < 0$ the \emph{right-} and \emph{left-moving} branch, resp. From Eq.
(\ref{two-ch}) we also define a \emph{rapidity} for each
propagating channel as
\begin{equation}
v_{\tau} \equiv  \left| \frac{\d{\l_{\tau}}}{{\d{k_{\tau}}}} \right| =\sqrt{4-\l_{\tau}^2}.
\label{rapidity}
\end{equation}

\subsubsection{Disordered part}
The impurity matrix~(\ref{ham2-ip}) contains two ingredients which
determine the localization properties of the model. One is
$\ep_{x\pm}$ [see Eq.~(\ref{epm})], which are the equally weighted
(either symmetric or anti-symmetric) combinations of \emph{effective
disorder} on the two legs. The other is the mixing angle $\g$
[see Eq.~(\ref{clean-angle})], which describes the \emph{effective
coupling} between the two legs. We refer to $\gamma$ as the
\emph{bare} mixing angle because it will be renormalized by
disorder. The \emph{renormalized} mixing angle
$\td{\gamma}$ [see Eq.~(\ref{re-angle})] will be discussed in Sec.~\ref{local-lenth}. Being functions of these two quantities, the
diagonal elements of $\td{\boldsymbol{\ep}}_x$ are local random
potentials applied on the two channels $\tau=1,2$, and the off-diagonal
elements describe the random hopping between them.

We  analyze the model qualitatively in terms of effective disorder
and bare mixing angle before carrying out the detailed
calculation. As already discussed above, either one or two
propagating channels are permitted at a given energy. This leads to two distinct mechanisms of
localization in the bulk of the energy band:

(i) \emph{Two-channel regime}. In this case, the physics is
dominated by the mixing angle $\gamma$. If $\gamma \sim 0$ or
$\pi$, the mixing of the two channels is weak: The magnitudes of off-diagonal elements of matrix~
(\ref{ham2-ip}) are much smaller than the magnitudes of the diagonal
elements. This means that the two legs are \emph{weakly entangled},
and the transverse hopping $t$ can be treated as a perturbation. A perturbative study of wavefunctions in
this regime is presented in Sec.~\ref{wavefuntions}. However, if $\gamma \sim \pi/2$, the
magnitudes of the off-diagonal elements are of the same order as the
diagonal elements. This implies that the two legs are \emph{strongly
entangled}. The localization properties are
controlled by the leg with strong disorder, because in Eq.~(\ref{epm}) it always
dominates over the weaker disorder on the other leg.

(ii) \emph{One-channel regime}.
The single-channel case has been
solved by Berezinskii\cite{berez} and Mel'nikov\cite{mel} in the case of a single chain. The
results they obtained can be applied in our problem by substituting
the variance of disorder and rapidity of the propagating channel
with the corresponding quantities. However, we have to emphasize
here that even if only one channel exists, coupling effects
still present, since both the effective disorder and the rapidity in the
remaining channel depend on the transport properties of both legs.
In the one-channel regime the second channel is still present,
but supports only evanescent modes. We show in Sec.~\ref{one-channel} that the effect of the evanescent channel on the propagating one is subleading when disorder is weak.

\subsection{Transfer matrix approach} \label{tra-mat-app}
The Fokker-Planck approach and its related notations, such as
the transfer matrix, the S-matrix etc. are
introduced in detail in Refs.~\onlinecite{mpk1} and \onlinecite{mpk2}. We only
outline the methodology here. The Fokker-Planck approach to one- or quasi-one dimensional systems with static disorder at zero
temperature is based on  studying the statistical distribution of
random transfer matrices for a system of finite length. An
ensemble of such transfer matrices is constructed by imposing
appropriate symmetry constraints. In the present model there are
two underlying symmetries: \emph{time-reversal invariance} and
\emph{current conservation}, which dramatically reduce the
number of free parameters of transfer matrices. After a proper
parametrization, the probability distribution function of these
parameters completely describes the ensemble of transfer matrices,
and therefore totally determines the statistical distributions of many
macroscopic quantities of the system, such as the conductance, etc. In
order to obtain the probability distribution function of the free
parameters, a stochastic evolution-like procedure is introduced by
computing  the variation of the probability distribution function of
these parameters in a ``bulk'' system as an extra impurity ``slice'' is
patched on one of its terminals, under the assumption that the
patched slice is statistically independent of the bulk. Thereby, we
construct a Markovian process for the probability distribution
function. This is described by a kind of Fokker-Planck equation in
the parameter space of the transfer matrix with the length of the
system as the time variable. Essentially, this procedure is
analogous to deriving the diffusion equation from the Langevin
equation for a Brownian particle. In practice, taking the length to
infinity, we can analytically extract asymptotic properties of the
model, such as localization lengths, etc., from the fixed
point solution of the Fokker-Planck equation.

As discussed above, the only microscopic quantity needed in
order to write down the Fokker-Planck equation of our model
Hamiltonian (\ref{full-ham}) is the transfer matrix of an
``elementary slice'' at any co-ordinate $x$. The Schr\"{o}dinger
equation (\ref{schro-2}) can be represented in the following
``transfer-matrix'' form:
\begin{equation}
\td{\Phi}(x+1)= \tilde{\mathbf{m}}_x  \td{\Phi}(x),  \label{sch-tran}
\end{equation}
where the $4$-component wave function $\td{\Phi}(x)$ and the $4\times 4$
transfer matrix $\tilde{\mathbf{m}}_x$ is explicitly shown in the
$2\times 2$ ``site-ancestor site'' form:
\begin{equation}
\td{\Phi}(x) \equiv \begin{pmatrix} \tilde{\Psi}(x) \\
\tilde{\Psi}(x-1) \end{pmatrix}, \quad \tilde{\mathbf{m}}_x \equiv
\begin{pmatrix} \tilde{\mathbf{h}}+\tilde{\boldsymbol{\epsilon}}_x & -\mathbf{1}
\\ \mathbf{1} & \mathbf{0} \end{pmatrix}.   \label{tran-eq}
\end{equation}
with $\td{\Psi}(x)$ and $\td{\bf{h}}$, $\td{\boldsymbol{\epsilon}}_{x}$ being the
$2$-component vector and $2\times 2$ matrices in the space of channels as
defined in Eq.~(7).

The transfer matrix $\tilde{\mathbf{m}}_x$ is manifestly \emph{real} (which reflects
the time-reversal symmetry) and \emph{symplectic} (which reflects the current conservation):
\begin{equation}
\label{sympl-cond} \tilde{\bf m}_{x}^{T}J\tilde{\bf m}_{x}=J,
\end{equation}
where $J$ is the standard skew-symmetric matrix:
\begin{equation}
\label{J} J=\left( \begin{matrix}0 & \mathbf{1} \cr -\mathbf{1} & 0
\end{matrix}\right).
\end{equation}
Note, however, that the transfer matrix $\tilde{\mathbf{m}}_x$ is
\emph{not}  a convenient representation to construct a Fokker-Planck
equation. The reason is simple: because it is not diagonal without
impurities, the perturbative treatment of impurities is hard to
perform. The proper transfer matrix $\mathbf{m}_x$ is a certain rotation,
which does not mix the two channels of the matrix $\tilde{\mathbf{m}}_{x}$, but transforms to a more convenient basis within each 2-dimensional channel subspace
[see Appendix~\ref{app-a}]. The latter corresponds to the basis of solutions
to the disorder-free Schr\"odinger equation $\psi_{\tau}(x)$
($\tau=1,2$ labeling the channels) which conserves the current along
the ladder:
\begin{equation}
\label{curr}
%\Psi_{x}=\left(\begin{matrix}\psi_{1}(x)\cr \psi_{2}(x)
%\end{matrix} \right)_{{\rm ch}},\;\;\;\;\;
j_{x}=-i\,[\psi_{\tau}^{*}(x)\psi_{\tau}(x+1)-h.c.]={\rm const}=\pm 1.
\end{equation}
For propagating modes with real wave vectors $k_{\tau}$ these are the
right- and left- moving states
\begin{equation}
\label{j-states prop}  \psi_{\tau}^{\pm}(x)= e^{\pm
ik_{\tau}\,x}/\sqrt{2 \sin k_{\tau}},
\end{equation}
which obey the conditions:
\begin{equation}
\label{PT} \psi_{\tau}^{\pm}(x)=(\psi_{\tau}^{\pm}(-x))^{*}.
\end{equation}
For the {\it evanescent} modes with {\it imaginary} $k=i\kappa$
the corresponding current-conserving states obeying Eq.~(\ref{PT})
can be defined, too:
\begin{equation}
\label{j-states-ev} \psi_{\tau}^{\pm}(x)=\frac{{\rm exp}[\mp i \pi/4-
\kappa_{\tau} x] + {\rm exp}[\pm i \pi/4 + \kappa_{\tau} x]}{\sqrt{4
\sinh \kappa_{\tau}}}
\end{equation}
In this new basis of current-conserving states, the transfer matrix
takes the form (see Appendix~\ref{app-a}):
\be
 \mathbf{m}_x=\mathbf{1}+\delta\mathbf{m}_x , \quad
 \delta\mathbf{m}_x=\begin{pmatrix}
                  -i \boldsymbol{\a}_{x}^{\ast} \td{\boldsymbol{\ep}}_{x} \boldsymbol{\a}_x &
                  -i \boldsymbol{\a}_{x}^{\ast} \td{\boldsymbol{\ep}}_x \boldsymbol{\a}_x^{\ast} \\
                  \,\,i \boldsymbol{\a}_{x} \td{\boldsymbol{\ep}}_x \boldsymbol{\a}_x & \,\,i
                  \boldsymbol{\a}_{x} \td{\boldsymbol{\ep}}_x \boldsymbol{\a}_x^{\ast}
                   \end{pmatrix},      \la{slice}
\ee where the matrix $\boldsymbol{\a}_x$ is diagonal in channel space,
\begin{equation}
\boldsymbol{\a}_x=\text{diag}\left(\psi_{1}^{+}(x),\psi_{2}^{+}(x)\right)_
{{\rm ch }}. \label{alp}
\end{equation}
Note that it is expressed in terms of the two components of the current
conserving states Eqs.~(\ref{j-states prop}, \ref{j-states-ev})
corresponding to the first and the second channel.

In Eq.~(\ref{slice}), the unit matrix $\mathbf{1}$ is the \emph{pure
part} of $\mathbf{m}_x$, which keeps the two incident plane waves
invariant, and $\delta\mathbf{m}_x$ describes the \emph{impurities},
which break the momentum conservation and induce intra- and
inter-channel scattering. The physical meaning of $\mathbf{m}_x$ can
be understood from the scattering processes described below. If
there is only one right-moving plane wave in the 1-channel on the
left-hand side (l.h.s.) of the slice, which is represented by a four
dimensional column vector with the first component one and the
others zero, we can detect four components on the right-hand side
(r.h.s.) of the slice, including the evanescent modes. In the case of
two propagating channels these four components are right- and
left-moving plane waves in the 1- and 2-channel, whose magnitudes
and phase-shifts form the first row of $\mathbf{m}_x$. The other
rows can be understood in the same manner. In short, the $11$-,
$12$-, $21$- and $22$- block of $\delta\mathbf{m}_x$ represent
respectively the right-moving forward-scattering, right-moving
backward-scattering, left-moving backward-scattering and left-moving
forward-scattering on the slice. In each block, the
diagonal elements represent intra-channel scattering and the
off-diagonal elements represent inter-channel scattering.

It is important that  $\mathbf{m}_x$, Eq.~(\ref{slice}), fulfills the
same constraints regardless of the propagating or evanescent
character of the modes [see Appendix~\ref{app-a}]:  \be
\mathbf{m}_x^{\ast}=\boldsymbol{\varSigma}_1 \mathbf{m}_x
\boldsymbol{\varSigma}_1, \quad
\mathbf{m}_x^{\dagger}\boldsymbol{\varSigma}_3\mathbf{m}_x=\boldsymbol{\varSigma}_3,
\label{requ-symp} \ee where $\boldsymbol{\varSigma}_1$ and
$\boldsymbol{\varSigma}_3$ are the four dimensional generalization
of the first and third Pauli matrix with zero and unit entries
replaced by $2 \times 2$ zero and unit matrices in the channels
space. The first condition follows from $\tilde{\bf
m}_x^{\ast}=\tilde{\bf m}_x$, while the second condition is a consequence of
the symplecticity Eq.~(\ref{sympl-cond}). Thus these conditions are a
direct consequence of the fact that $\tilde{\bf m}_x$ belongs to the
\emph{symplectic group} $\text{Sp}(4,\mathbb{R})$. As is obvious from the choice of the basis (\ref{curr}-\ref{j-states-ev}), their physical
meaning is the time-reversal symmetry and the current
conservation.

The representation Eq.~(\ref{slice}) of the transfer matrix of an
``elementary slice'' renders both the physical interpretation and
the symmetry constraints very transparent, and it will be seen to be
a convenient starting point to construct the Fokker-Planck equation.
On the other hand, since $\tilde{\mathbf{m}}_x$ [see Eq.
(\ref{tran-eq})] is real and has a relatively simple form, it is
more suitable for numerical calculations.
\section{fokker-planck equation for the distribution function of parameters}
\label{fpe-probability distribution function}
\subsection{Parametrization of transfer matrices}
Once the ``building block'' (\ref{slice}) is worked out, we can
construct the Fokker-Planck equation by the blueprint of the
Fokker-Planck approach.\cite{mel,dorok,mpk1,mpk2} The transfer
matrix of a disordered sample with length $L$ is
\begin{equation}
\mathbf{M}(L)=\prod_{x=1}^{L} \mathbf{m}_x= \mathbf{m}_L \cdot \mathbf{m}_{L-1} \cdots \mathbf{m}_1 ,   \label{tm-l}
\end{equation}
which is a $4\times4$ complex random matrix. It is easy to verify
that $\mathbf{M}(L)$ also satisfies the time reversal invariance and
current conservation conditions~(\ref{requ-symp}). It has been
proved in Ref.~\onlinecite{mpk1} that all the $4 \times 4$ matrices
satisfying Eq.~(\ref{requ-symp}) form a group which is identified
with the symplectic group $\text{Sp}(4,\mathbb{R})$. By the
Bargmann's parametrization of $\text{Sp}(4,\mathbb{R})$,\cite{mpk2}
one can represent $\mathbf{M}(L)$ as
\begin{equation}
\mathbf{M}=\begin{pmatrix}
            \mathbf{u} & 0  \\  0  & \mathbf{u}^{\ast}
           \end{pmatrix} \begin{pmatrix}
            \sqrt{\fr{\mathbf{F}+\mathbf{1}}{2}}  &   \sqrt{\fr{\mathbf{F}-\mathbf{1}}{2}} \\
            \sqrt{\fr{\mathbf{F}-\mathbf{1}}{2}}  &   \sqrt{\fr{\mathbf{F}+\mathbf{1}}{2}}
           \end{pmatrix} \begin{pmatrix}
            \tilde{\mathbf{u}} & 0  \\  0  & \tilde{\mathbf{u}}^{\ast}
           \end{pmatrix} ,  \label{param}
\end{equation}
where $\mathbf{u}$ and $\tilde{\mathbf{u}}$ are elements of the unitary group $\text{U}(2)$, and
statistically independent from each other, and
\begin{equation}
\mathbf{F}=\text{diag}(F_1,F_2),    \label{F-varb}
\end{equation}
with $F_{\varrho} \in [1,\infty)$ and $\varrho \in \{1,2\}$. Because
$\text{U}(2)$~\cite{hamsh} has four real parameters, the group
$\text{Sp}(4,\mathbb{R})$ has ten real parameters.
Furthermore, it is convenient to parametrize a $\text{U}(2)$ matrix
by three Euler angles and a total phase angle, i.e.,
\begin{equation}
\mathbf{u}(\phi,\vp,\theta,\psi)= e^{-i \frac{\phi}{2}}
e^{-i \frac{\varphi}{2}\hat{\sigma}_3} e^{-i \frac{\theta}{2}
\hat{\sigma}_2} e^{-i \frac{\psi}{2} \hat{\sigma}_3},
\end{equation}
in which $\hat{\sigma}_{2}$ and $\hat{\sigma}_{3}$ are the second and
third Pauli matrix, and the four angles take their values in the
range $\phi,\varphi \in [0,2\pi)$, $\theta \in [0,\pi)$ and $\psi
\in [0,4\pi)$. In matrix form in the channels space, $\mathbf{u}$
can be written as
\begin{equation}
 \mathbf{u}= e^{-i \frac{\phi}{2}} \begin{pmatrix} \cos{\fr{\theta}{2}}
 e^{-\frac{i}{2}(\varphi+\psi)}  &  -\sin{\fr{\theta}{2}}e^{-\frac{i}{2}(\varphi-\psi)}  \\
 \sin{\fr{\theta}{2}}e^{\frac{i}{2}(\varphi-\psi)}  &
 \cos{\fr{\theta}{2}}e^{\frac{i}{2}(\varphi+\psi)} \end{pmatrix}_{{\rm ch}},    \label{u-matrix}
\end{equation}
which is convenient for the perturbative calculation
below. The $\text{U}(2)$  matrix $\tilde{\mathbf{u}}$ can be
parametrized independently in the same form as Eq.~(\ref{u-matrix}).

The probability distribution function of these
ten real parameters determines completely the transfer matrix
ensemble of the ladder described by the Hamiltonian
~(\ref{full-ham}). The goal of the Fokker-Planck approach is to
obtain the Fokker-Planck equation satisfied by this probability
distribution function, in which the role of time is played by the
length $L$.

From Eq.~(\ref{param}) we obtain the transmission matrix
\begin{equation}  \label{transmiss-mat}
\mathbf{t}:= (M_{++}^{\dagger})^{-1} =\mathbf{u} \left(\frac{\mathbf{F}+\mathbf{1}}{2}\right)^{-1/2}
\td{\mathbf{u}},
\end{equation}
by a simple relation between the transfer matrix and its
corresponding S-matrix.\cite{mpk1,mpk2,imp-trassm-mat} Due to the
unitarity of $\tilde{\mathbf{u}}$, the transmission co-efficients of
the two channels are the two eigenvalues of the Hermitian matrix
\begin{equation}
\mathbf{T}=\mathbf{t}\mathbf{t}^{\dagger}=\mathbf{u}
\left( \frac{\mathbf{F}+\mathbf{1}}{2} \right)^{-1}\mathbf{u}^{\dagger},  \label{Tmatx}
\end{equation}
which are
\begin{equation}
T_{\varrho}=\fr{2}{F_{\varrho}+1},    \la{tran-prob}
\end{equation}
where $\varrho \in \{1,2\}$ is the index of the two-dimensional
eigenspace of the matrix $\mathbf{T}$. Now  the physical meaning of
the parametrization~(\ref{param}) becomes clear. The $F_\varrho$'s
are related to the \emph{two transmission co-efficients} by the
simple form Eq.~(\ref{tran-prob}). The matrix  $\mathbf{u}$
diagonalizing the matrix $\mathbf{T}$, contains the two eigenvectors
of $\mathbf{T}$, describing the polarization of the plane
wave eigenmodes incident from the l.h.s. of the sample. For
instance, if $\theta=0$ ($\mathbf{u}$ is a diagonal matrix of
redundant phases), the two channels do not mix, and the incident
waves are fully polarized in the basis of channels. On the other
hand, if $\theta=\pi$, the two channels are equally mixed, and the
incident waves are unpolarized. In analogy to spherical
co-ordinates, we will refer to the $F_\varrho$'s as the
\emph{radial} variables, while the angles in $\mathbf{u}$ or
$\td{\mathbf{u}}$ are called \emph{angular} variables.

In principle, using the ``building block'' (\ref{slice}) and the
parametrization (\ref{param}), we can solve the full problem by
writing down a Fokker-Planck equation for the joint probability
distribution function of all the ten parameters of $\mathbf{M}$. However, since we
are merely interested in the transmission co-efficients which are
determined by the probability distribution function of $\mathbf{T}$,
instead of manipulating $\mathbf{M}$, we study 
\begin{equation}
\begin{split}
\mathbf{R} &=\mathbf{M}\mathbf{M}^{\dagger}\\
&\\
&=\begin{pmatrix}
            \mathbf{u} & 0  \\  0  & \mathbf{u}^{\ast}
           \end{pmatrix} \begin{pmatrix}
            \mathbf{F}  &   \sqrt{\mathbf{F}^2-\mathbf{1}} \\
            \sqrt{\mathbf{F}^2-\mathbf{1}}  &   \mathbf{F}
           \end{pmatrix} \begin{pmatrix}
            \mathbf{u}^{\dagger} & 0  \\  0  & \mathbf{u}^{\text{T}}
           \end{pmatrix}.
\end{split}     \label{R-matrix}
\end{equation}
$\mathbf{R}$ is a Hermitian matrix and contains only \emph{six} parameters:
\begin{equation}
\vec{\lambda}(\mathbf{R})=(F_1,F_2,\theta,\psi,\phi,\varphi).    \label{variables}
\end{equation}
The probability distribution function of $\vec{\lambda}$, denoted by
$P_L(\vec{\lambda})$, determines the transmission properties of the
sample with length $L$. 
 $P_L(\vec{\lambda})$ is defined by 
\be   \label{pdf-def}
P_L(\vec{\lambda}) = \overline{ \delta{\left(  \vec{\lambda} - \vec{\lambda}{\left( \mathbf{R}(L) \right)} \right)} },
\ee
where the overline denotes the average over realizations of the random potentials in the sample. It is convenient to introduce the characteristic function of $P_L(\vec{\lambda})$ 
\be  \label{cha-fun}
\td{P}_L(\vec{p}) %= \mathcal{F}\left[ P_L(\vec{\lambda}) \right] 
= \int{d\vec{\lambda} e^{i \vec{p} \cdot \vec{\lambda}} P_L(\vec{\lambda})} = \overline{ e^{ i \vec{p} \cdot \vec{\lambda}{\left( \mathbf{R}(L) \right)} }}.
\ee

%In order to calculate expectation values of certain
%quantities with respect to $P_L(\vec{\lambda})$, we have to know the
%Haar measure on $\vec{\lambda}(\mathbf{R})$-space under
%the group $\text{Sp}(4,\mathbb{R})$,\cite{mpk1} which is
%\begin{equation}
%\mu(\vec{\lambda}) \propto |\DF|\sin\theta,  \quad \DF=F_1-F_2,
%\label{haar}
%\end{equation}
%where the factor $\sin{\theta}$ comes from the Haar measure of the group $\text{U}(2)$ in the parametrization (\ref{u-matrix}).\cite{hamsh}

Our main goal in this paper is to calculate the two localization lengths,
defined as the inverse Lyapunov exponents of the transfer
matrix (\ref{param}):
\begin{equation}
\xi_{1(2)}^{-1} \equiv -\lim_{L\rightarrow\infty}{\frac{1}{2}\frac{d}{dL}
\langle \ln{T_{\max(\min)}} \rangle_L},    \label{def-loc}
\end{equation}
in which the subscripts ``max'' and ``min'' denote the larger and
smaller of the two real values $T_{1,2}$, and the averaging $\langle
\cdot \rangle_L$ is earned out with the probability distribution
$P_L(\vec{\lambda})$. Therefore, by definition
\begin{equation}   \label{xi1gxi2}
\xi_{1} > \xi_2.
\end{equation}

\subsection{Physical interpretation of $\xi_1$ and $\xi_2$}
\begin{figure}
\begin{center}
\includegraphics[height=5.9cm,width=8.2cm]{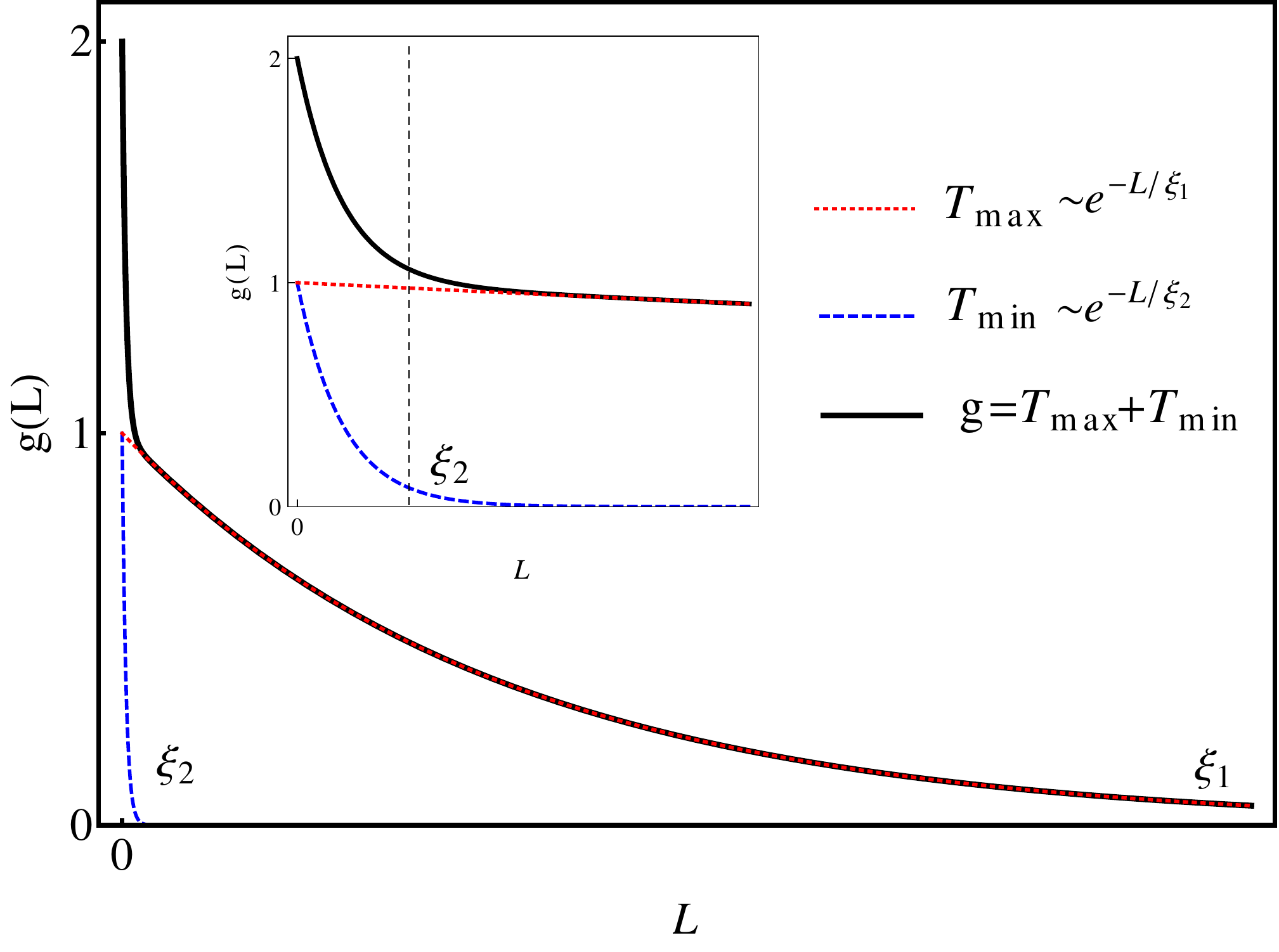}
\caption{Schematic diagram for typical values of dimensionless conductance g as a function of the length $L$ in the case $\xi_1 \gg \xi_2$.
A crossover happens at $L \sim \xi_2$. In the region $L < \xi_2$, g decays as fast as $T_\text{min}$ (see the insert).
Once $L > \xi_2$, g decays as slowly as $T_\text{max}$. The system is well conducting if $L < \xi_1$ and crosses to an insulating regime for $L > \xi_1$.}
\label{tran-eigenv}\
\end{center}
\end{figure}

It is worthwhile to visualize how two \emph{parametrically different} localization lengths ($\xi_1 \gg \xi_2$) manifest themselves in transport properties. For instance, let us discuss the dimensionless conductance $\text{g}=T_1+T_2$, a typical behavior of which is shown as a function of the sample length $L$ in Fig.~\ref{tran-eigenv}. If $L \ll \xi_2$, $T_\varrho \approx 1$, and $\text{g} \approx 2$ corresponds to a nearly
perfect transmission. As $L$ increases, $T_\varrho$ and $\text{g}$ decay exponentially. $T_2$ decays much faster than $T_1$ since $\xi_1 \gg \xi_2$. As long as $L < \xi_1$ the system still conducts well since $\text{g}$ is still appreciable. For $L \sim \xi_1$ it crosses over
to an insulating regime. On the other hand, $\xi_2$ marks the crossover length scale below which $\text{g}(L)$ (black curve) decreases as fast as $T_2$ (blue dashed curve) until the conductance saturates to a plateau $\text{g} \approx 1$.
For $L > \xi_2$, $\text{g}$ decays with the slow rate $\xi_1^{-1}$, like $T_1$ (red dotted curve).
Therefore, the two parametrically different localization lengths can be identified by two distinct decay rates of $\text{g}$ at small and large length scales.
 
The statistics of transmission eigenvalues and localization lengths of disordered multi-channel micro-waveguides 
have been visualized in experiments .\cite{shi-gen} However, only more or less isotropically disordered cases (identical hopping and disorder strength in each channel) were realized, while a  situation where $\xi_1 \gg \xi_2$ is hard to achieve in such systems (see Ref.~\onlinecite{shi-gen} and references therein). In contrast such anisotropic situations are rather natural in exciton polariton systems.    
%Nevertheless, one can expect that the indication of localization lengths we proposed above could be observed in random waveguides but with certain anisotropic setup in order to fulfil the condition $\xi_1 \gg \xi_2$.

We will see in Sec.~\ref{wavefuntions}  that the two localization lengths $\xi_1$ and $\xi_2$ also characterize the spatial variations of the eigenfunctions $\Psi(x)$ on the two legs.

%{\bf MM added: In general $\xi_2$ governs the initial decay of amplitude of a
%generic incident wave, while $\xi_1$ determines the decrease of the exponential
%tail of the transmission through a long chain.}
%In Eq.~(\ref{def-loc}) we actually define the localization lengths
%as the inverses of the \emph{Lyapunov exponents} of the transfer
%matrix (\ref{param}). The from of Eq.~(\ref{def-loc}) is convenient
%for calculating localization lengths from the Fokker-Planck equation
%of $P_L(\vec{\lambda})$. It should be noticed that even if
%$T_{\max}$ and $T_{\min}$ are functions of radial variable
%$F_\varrho$s, they are correlated with the angular variables, since
%the radial variables are \emph{entangled} with angular variables in
%the Fokker-Planck equation (\ref{fokker-planck-3}).

\subsection{Fokker-Planck equation for the distribution function of
$\vec{\lambda}(\mathbf{R})$} \label{Fokker-Planck-eq}
Having a disordered sample of length $L$, whose transfer matrix is
$\mathbf{M}(L)$ and adding one more slice, we obtain the transfer
matrix of the sample with length $L+1$:
\begin{equation}
\mathbf{M}(L+1)=\mathbf{m}_{L+1}\mathbf{M}(L).
\end{equation}
 Simultaneously, according to Eqs.~(\ref{R-matrix}) and (\ref{slice}), $\mathbf{R}(L)$ is updated to
\begin{subequations}  \label{delta-R} 
\be
 \mathbf{R}(L+1)=\mathbf{m}_{L+1} \mathbf{R}(L) \mathbf{m}_{L+1}^{\dagger} = \mathbf{R}(L)+\dR,
\ee
\be
 \dR =(\mathbf{R} \delta{\mathbf{m}^\dagger_{L+1}}+
h.c.)+\delta{\mathbf{m}_{L+1}} \mathbf{R}
\delta{\mathbf{m}^{\dagger}_{L+1}}.
\ee
\end{subequations}
Accordingly, $\vec{\lambda}{\left( \mathbf{R}(L) \right)}$ is incremented by
\be   \label{perturb}
\vec{\lambda}{\left(\mathbf{R}(L+1) \right)}=\vec{\lambda}{\left( \mathbf{R}(L) \right)}+\delta{\vec{\lambda}}.
\ee

According to Eqs.~(\ref{cha-fun}) and (\ref{perturb}), we obtain the characteristic function of $P_{L+1}(\vec{\lambda})$: 
\be  \label{Markovian} 
\td{P}_{L+1}(\vec{p}) = \overline{ e^{ i \vec{p} \cdot \vec{\lambda}{\left( \mathbf{R}(L+1) \right)} }}  
= \overline{ e^{ i \vec{p} \cdot \vec{\lambda}{\left( \mathbf{R}(L) \right)} } e^{ i \vec{p} \cdot \delta{\vec{\lambda}} } }. 
\ee
We can expand $e^{ i \vec{p} \cdot \delta{\vec{\lambda}} }$ on the r.h.s. of Eq.~(\ref{Markovian}) into a Taylor series 
$e^{ i \vec{p} \cdot \delta{\vec{\lambda}} }=\sum_{n}^{\infty}{(i \vec{p} \cdot \delta{\vec{\lambda}})^n}$.
Using Eqs.~(\ref{delta-R},\ref{perturb}) 
%defines the variation of $\mathbf{R}$ described by $\delta{\vec{\lambda}}$ under
%the ``perturbation'' $\dR$, 
standard perturbation theory
yields an expansion of $\delta{\vec{\lambda}}$ in powers of the disorder
potential as
$\delta{\vec{\lambda}}=\sum_{n\ge1}\delta{\vec{\lambda}}^{(n)}$,
where $\delta{\vec{\lambda}}^{(n)}$ is of $n$-th order in
$\tilde{\boldsymbol{\epsilon}}$. 
With this, the r.h.s. of Eq.~(\ref{Markovian}) can be expanded in
powers of the disorder potential. In principle, we can proceed with
this expansion to arbitrarily high orders.
Thereafter, the average
over disorder on the slice $L+1$ can be performed. 
Eqs.~(\ref{delta-R})-(\ref{Markovian}) fully define
our problem. However, it is impossible to solve it analytically
without further simplification.

Progress can be made by considering the
\emph{weak disorder} limit. In the two-channel regime, the weak
disorder limit implies that \emph{both} of the mean free paths are
much larger than the lattice constant.   As a first estimation,
applying the Born approximation to an ``elementary slice'', the inverse
mean free paths of the two propagating channels can be expressed as
certain linear combinations of the variances of the effective
disorders on the two chains, defined as
\begin{equation}
\Var = \frac{\var}{t_{\nu}^2}.   \label{eff-var}
\end{equation}
for chain $\nu$. In the weak disorder limit where the smaller of the two localization
lengths is much larger than the lattice constant,
\begin{equation}  \label{wdl}
l\gg 1,
\end{equation}
only the terms proportional to $\Var$ on the r.h.s of Eq.
(\ref{Markovian}) have to be taken into account. Hence,
we calculate $\delta{\vec{\lambda}}$ perturbatively up to the second
order [see Appendix~\ref{per-cal}]. 
If $L \gg 1$, as we always assume, $\td{P}_{L+1}-\td{P}_{L} \simeq
\partial_L{\td{P}_L}$.
 Under these conditions, Eq.~(\ref{Markovian}) leads to
\be  \label{fp-eq-f}
\partial_L{\td{P}_L} = \overline{i \vec{p} \cdot \delta{\vec{\lambda}}^{(2)} 
e^{ i \vec{p} \cdot \vec{\lambda}{\left( \mathbf{R}(L) \right)} } } - \frac{1}{2}  
\overline{ \left( \vec{p} \cdot \delta{\vec{\lambda}}^{(1)} \right)^2 e^{ i \vec{p} 
\cdot \vec{\lambda}{\left( \mathbf{R}(L) \right)} } }.
\ee 
Note that because the random potentials in different slices are \emph{uncorrelated}, the $\delta{\vec{\lambda}}^{(n)}$ 
terms can be averaged independently of $e^{ i \vec{p} \cdot \vec{\lambda}{\left( \mathbf{R}(L) \right)} }$. By the inverse of the Fourier transform defined in Eq.~(\ref{cha-fun}) we obtain 
 the Fokker-Planck equation for $P_L(\vec{\lambda})$:
\be
\partial_L{P}=-\sum_{i=1}^{6}{
\partial_{\lambda_i} \left[ {\overline{\delta{\lambda_i^{(2)}}} P}-\frac{1}{2}\sum_{j=1}^{6}{
\partial_{\lambda_j}{\left( \overline{\delta{\lambda_i^{(1)}}\delta{\lambda_j^{(1)}}} P \right)}} \right]}.
 \label{fokker-planck-2}
\ee
%where we have used
%\begin{subequations}
%\be
%\mathcal{F}\left[ \sum_{i}{ \partial_{\lambda_i} \left( \overline{\delta{\lambda}_{i}^{(2)}} P \right)} \right] = 
%- \overline{i \vec{p} \cdot \delta{\vec{\lambda}} e^{ i \vec{p} \cdot \vec{\lambda}{\left( \mathbf{R}(L) \right)} } },
%\ee
%\be
%\begin{split}
%& \mathcal{F}\left[ \sum_{i,j}{\partial_{\lambda_i} \partial_{\lambda_j}\left( \overline{\delta{\lambda}_{i}^{(1)} 
%\delta{\lambda}_{j}^{(1)}} P \right)} \right] \\
% & \hspace{2.5cm} = - \overline{ \left( \vec{p} \cdot \delta{\vec{\lambda}}^{(1)} \right)^2 
%e^{ i \vec{p} \cdot \vec{\lambda}{\left( \mathbf{R}(L) \right)} } }.
%\end{split} 
%\ee
%\end{subequations}
In Eq.~(\ref{fokker-planck-2}) the averages are taken over the realizations of random potentials in the slice at $L+1$. 

The Fokker-Planck equation (\ref{fokker-planck-2}) can be rewritten in the form of a continuity equation:
\begin{equation}
\partial_L{P}=-\sum_{i=1}^{6}{
\partial_{\lambda_i} J_i},   \label{fokker-planck-2-s}
\end{equation}
where the generalized current density $J_i$ takes the form:
\be   \label{pro-cur}
J_i = {\text{v}_i(\vec{\lambda}) P}-\sum_{j=1}^{6} D_{ij}(\vec{\lambda}) \partial_{\lambda_j}{P},
\ee
with
\begin{subequations}  \label{coe-vel-dif}
 \be
\text{v}_i(\vec{\lambda}) = \overline{\delta{\lambda_i^{(2)}}} + \partial_{\lambda_j} D_{ij}(\vec{\lambda}) ,
\ee
\be
D_{ij}(\vec{\lambda}) = \frac{1}{2}\overline{\delta{\lambda_i^{(1)}}\delta{\lambda_j^{(1)}}}.
\ee
\end{subequations} 
$\text{v}_i(\vec{\lambda})$ and $D_{ij}(\vec{\lambda})$ are a generalized stream velocity and a generalized diffusion tensor, 
respectively.
%{\bf Here note that $P_L(\vec{\lambda})$ is defined 
%in the Haar measure $\mu{(\vec{\lambda})}$ in Eq.~(\ref{haar}). 
%The probability distribution function $\mu{(\vec{\lambda})}P_L(\vec{\lambda})$, which includes the Haar mesure, should 
%satisfy the Fokker-Planck equation in a different form where the disorder averages are inside the derivatives.\cite{bee2}}
 
%
In order to solve Eq.~(\ref{fokker-planck-2}), we have
to add the initial condition, namely, $P_0(\vec{\lambda})$. Usually
$P_0(\vec{\lambda})$ is chosen as the probability distribution
function in the ballistic limit,\cite{bee}
\begin{equation}P_0(\vec{\lambda})=\delta{(F_1-1)}
\delta{(F_2-1)}\delta{(\theta)}\delta{(\phi)}\delta{(\psi)}\delta{(\varphi)},
\end{equation}
where $\delta(x)$ is the Dirac delta function. However, as we will
see later, a unique fixed point of $P_{L}(\vec{\lambda})$ exists in
the limit $L \rightarrow \infty$, which does not depend upon the
initial condition. Essentially, the existence of a fixed-point
solution of Eq.~(\ref{fokker-planck-2}) is protected by  Anderson
localization which prevents the system from chaos.~\cite{fgp}

\subsection{Coarsegraining}  \label{coa-gra}
Let us analyze the r.h.s. of Eq.~(\ref{fokker-planck-2})
qualitatively. From Eqs.~(\ref{1st}) and (\ref{2nd}) in the Appendix, it is clear that the co-efficients
$\overline{\delta{\lambda_i^{(1)}}\delta{\lambda_j^{(1)}}}$ and
$\overline{\delta{\lambda_i^{(2)}}}$ are sums of terms
carrying phase factors $1,e^{\pm i(k_1-k_2)L},...$ and so on. These phase factors come from the disorder average of products of
two elements of the matrices (\ref{slice}). Their phases correspond to the possible wave vector transfers of 
two scatterings from a slice, similarly as found in
the Berezinskii technique~\cite{berez}.
They are thus linear combinations of two or four values of $\pm k_{1,2}$:
\begin{eqnarray}
{\bf K}_{\rm osc} &=&\left\{ \,\pm\Delta{k}, \quad  \pm 2\Delta{k}, \right.\\
&& \quad\pm 2k_{1(2)}, \quad  \pm(k_1+k_2), \quad \pm\left[ 3k_{1(2)}-k_{2(1)}\right], \nonumber\\
&& \quad \left. \pm 4k_{1(2)}, \quad \pm 2(k_1+k_2), \quad \pm\left[ 3k_{1(2)}+k_{2(1)}\right] \right\},\nonumber
\label{wav-com}
\end{eqnarray}
where
\be
\Delta{k} = k_1-k_2.
\ee
Terms with phase ``$0$'' do not oscillate. The largest spatial period of the
oscillating terms is 
\be
L_{\rm osc}= \max_{\delta k\in {\bf K}_{\rm osc}} \delta k^{-1}
\ee
Under the condition that 
\begin{equation}
L_{\rm osc}\ll l,   \label{scale-3}
\end{equation}
a \emph{coarsegrained} probability distribution function
can be defined as the average of $P_L(\vec{\lambda})$ over $L_{\rm osc}$. From now on, we use the same
symbol $P_L(\vec{\lambda})$ to denote its coarsegrained
counterpart, which satisfies Eq.~(\ref{fokker-planck-2}), but
neglecting the oscillating terms.

Additionally, at special energies it may happen that an oscillation period becomes commensurate with the lattice spacing , $\delta k= \pi/n$. An important example of this \emph{commensurability} is the situation where
$\delta{e}=0$, $2(k_1+k_2)=2\pi$ at $E=0$. In this case the terms
with the phase factor $e^{\pm 2i(k_1+k_2)L}$ do not average and give anomalous
contributions to the non-oscillating co-efficients. This effect leads to the so-called center-of-band anomaly in the
eigenfunction statistics of the one-chain Anderson model (see Ref.~\onlinecite{krav-yud} and references
therein). While they are not included in our analytical study, the
commensurability-induced anomalies can be seen clearly in the numerical results
for localization lengths
(cf. Figs.~\ref{interm-coupling}, \ref{neq-disorder}, \ref{band-edge} and \ref{sig-chanl}).

The coarsegraining procedure leads to a significant simplification:
the co-efficients on the r.h.s. of Eq.~(\ref{fokker-planck-2}) do
not depend on $L$, $\phi$ and $\varphi$ any longer, which renders
the solution of Eq.~(\ref{fokker-planck-2}) much easier. Its
non-oscillating co-efficients are evaluated in Appendix~\ref{per-cal}. We do not reproduce them explicitly here, since we
further transform the Fokker Planck equation below. However it is
worthwhile pointing out a formal property of its co-efficients. From
Eqs.~(\ref{per-term}), (\ref{1st}) and (\ref{2nd}), it is easy to see
that the ingredients for evaluating
$\overline{\delta{\lambda_i^{(1)}}\delta{\lambda_j^{(1)}}}$ and
$\overline{\delta{\lambda_i^{(2)}}}$ are the disorder-averaged
correlators between any two elements of matrices (\ref{slice}).
During the calculation, three Born cross sections appear naturally,
being covariances of the effective disorder variables,
\begin{subequations}   \label{born-sec}
\begin{equation}  \label{born-sec1}
V_1 = \fr{1}{4 v_{1}^{2}} \left( \Varo \cos^4{\fr{\g}{2}}+\Vart \sin^4{\frac{\g}{2}} \right),
\end{equation}
\begin{equation}   \label{born-sec2}
V_2 = \fr{1}{4 v_{2}^{2}} \left( \Varo\sin^4{\fr{\g}{2}}+\Vart\cos^4{\fr{\g}{2}} \right),
\end{equation}
\begin{equation}   \label{born-sec3}
V_3 = \fr{1}{4v_1v_2} \left(\Varo+\Vart\right) \sin^2{\fr{\g}{2}}\cos^2{\fr{\g}{2}},
\end{equation}
\end{subequations}\\
in which $V_{1(2)}$ corresponds to intra-channel scattering
processes $k_{1(2)} \leftrightarrow -k_{1(2)}$, and $V_3$
corresponds to inter-channel scattering processes $k_{1(2)}
\leftrightarrow \pm k_{2(1)}$. Note that the effective disorder
variances (\ref{eff-var}) enter into the three Born cross-sections,
instead of the bare variances (\ref{var1}). We will see that
the above three Born cross-sections completely define the
localization lengths and most phenomena can be
understood based on them. 

We note that the coarsegraining, through
Eq.~(\ref{scale-3}), imposes a crucial restriction on the
applicability of the simplified Fokker-Planck equation. According to
Eqs. (\ref{two-bands}) and (\ref{two-ch}), if $t$ is small enough, at
$E=E_R$,
\begin{equation}  \label{restr}
|\Delta{k}| \propto t.
\end{equation}
In this case, Eqs. (\ref{scale-3}) and (\ref{restr}) require that
\begin{equation}   \label{re-str}
t \gg \delta{E},
\end{equation}
where
\begin{equation}   \label{deltaE}
\delta{E} \propto l^{-1},
\end{equation}
is the characteristic disorder energy scale (essentially the level
spacing in the localization volume). In other words,
Eq.~(\ref{re-str}) imposes a ``strong coupling'' between the two
legs, as compared with the disorder scale.
However, from the point of view of the strength of disorder,
Eq.~(\ref{re-str}) is a more restrictive condition than $\xi_{2}\gg 1$ on
the smallness of disorder. However, it is automatically
fulfilled in the limit $\sigma_{\nu}\rightarrow 0$ at  fixed values
of coupling constants $t$ and $t_{\nu}$.

Eq.~(\ref{re-str}) restricts the region of applicability of the
simplified equation (\ref{fokker-planck-3}) which we will derive
below. Indeed, we will see that by simply taking the limit
$t\rightarrow 0$ in the solution of that equation one does not
recover the trivial result for the uncoupled chains. This is because
the equation is derived under the condition that $t$ is limited from
below by Eq.~(\ref{re-str}). The ``weak coupling'' regime is studied
numerically in Sec. \ref{band-centre} and the cross-over to the
limit of uncoupled chains is observed at a scale of $t\sim
l^{-1}$ as expected.

%In order to calculate the localization lengths from Eq.~(\ref{def-loc}),
%the Fokker-Planck equation for the probability distribution function
%$\mu(\vec{\lambda})P_L(\vec{\lambda})$, which includes the
%$L$-evolution of the Haar measure, is more convenient to work with than $P_L(\vec{\lambda})$ itself. In addition,
Since the definition of localization lengths (\ref{def-loc}) only
involves the $F_\varrho$'s, and since the co-efficients of Eq.
(\ref{fokker-planck-2}) do not contain $\phi$ and $\varphi$, we
define the marginal probability distribution function
\begin{equation}  \label{marg-def}
W_L (F_1,F_2,\theta,\psi)=\int{{d\phi}{d\varphi}P_L(\vec{\lambda})}.
\end{equation}\\
Further we change variables to the set
\begin{equation}
\vec{\eta}=(F_1,F_2,u,\psi),
\end{equation}
where
\begin{equation}           \label{u-variab}
u=\cos{\theta}, \quad u\in(-1,1].
\end{equation}
%As compared to Eq.~(\ref{haar}) the Haar measure on $\vec{\eta}$-space has the simpler form
%\begin{equation}
%\tilde{\mu}(\vec{\eta}) \propto |\Delta F|, \label{haar-2}
%\end{equation}
We thus have
\begin{equation}
W_L(\vec{\eta})=\int{d\phi d\varphi P_L(F_1,F_2,\theta(u),\psi,\phi,\varphi)}.
\label{marg-pdf}
\end{equation}
Substituting Eq.~(\ref{marg-pdf}) into (\ref{fokker-planck-2}), and
replacing the differential operators $\partial_\theta \rightarrow
-\sqrt{1-u^2}\partial_u$ and $\partial_\theta^2 \rightarrow -u
\partial_u + (1-u^2)\partial_u^2$, we obtain the Fokker-Planck
equation for $W_L(\vec{\eta})$:
\begin{widetext}
\be              \label{fokker-planck-3}
\partial_{L}{W} = \sum_{i=1}^{4}{\left[ \partial_{\eta_i} \left( c_{ii}
\partial_{\eta_i} W \right) + \partial_{\eta_i} \left( c_{i} W \right) \right] } +\sum_{j>i=1}^{4}{\partial_{\eta_i}\partial_{\eta_j} \left( c_{ij} W \right)}.
\ee
\end{widetext}
The co-efficients $c_i,c_{ij}$ are relatively simple functions of $\vec{\eta}$. They can be obtained from the averages of the matrix elements computed in App.~\ref{per-cal}, and are given in App.~\ref{coe}. However, only a small number of them will turn out to be relevant for the quantities of interest to us.

One can see that in Eq.~(\ref{fokker-planck-3}) the radial variables, $F_\varrho$, are entangled  with the angular variables $u$ and
$\psi$. Thus  Eq.~(\ref{fokker-planck-3}) is more general than the
\emph{canonical} DMPK equation\cite{dorok,mpk1,mpk2} where only
radial variables appear. To emphasize the difference we refer to
Eq.~(\ref{fokker-planck-3}) as the \emph{extended} DMPK equation.
The derivation of Eq.~(\ref{fokker-planck-3}) for the two-leg problem is
our main technical achievement  in the present paper. It allows us to obtain the
evolution (as a function of $L$) of the expectation value of any quantity
defined in $\vec{\eta}$-space.

\section{Calculating the localization length}                     \label{local-lenth}
It is well-known that in quasi-one dimensional settings single
particles are always localized at any energy in arbitrarily
weak (uncorrelated) disorder.~\cite{abra-and-Licc-rama} The
localization length quantifies the localization tendency in
real space. In this section we calculate the localization lengths
for the present model.

The analytic expression of $\ln{T_{\max(\min)}}$ in Eq.~(\ref{def-loc}) can be written as \be
%\begin{split}
\ln{T_{\max(\min)}}=\Theta(\DF)\ln{T_{2(1)}}+\Theta(-\DF)\ln{T_{1(2)}},
\label{def-loc-2}
%\end{split}
\ee
where $\Theta(z)$ is the Heaviside step function and
\be  \label{haar}
\Delta{F} = F_1 -F_2.
\ee
 Multiplying
both sides of Eq.~(\ref{fokker-planck-3}) by the r.h.s. of Eq.~(\ref{def-loc-2}) and integrating over all the variables, we obtain from Eq.~(\ref{def-loc})
\begin{equation}      \label{loc-len-1}
\xi_{\rho}^{-1} = \lim_{L \rightarrow \infty} \left\langle (D_1+D_2)
-(-1)^\rho (D_1-D_2)\text{sgn}{(\Delta{F})}\right\rangle_L,
\end{equation}
with
\begin{equation}
D_i = \frac{1}{2} \left[ \frac{c_{ii}}{(F_i+1)^2} + \frac{c_{i}}{F_i+1}-
\frac{\partial_{F_i} c_{ii}}{F_i+1} \right],  \quad i \in \{1,2\},
\end{equation}
in which $\rho \in \{1,2\}$, $\text{sgn}(z)$ is the sign
function and the co-efficients $c_i,c_{ii}$ [see App.~\ref{coe}] are
\begin{equation*}
 \begin{split}
  c_i & = (-1)^i 2 \frac{F_i^2-1}{\Delta{F}}\Gamma_6,\\
  c_{ii} & =(F_i^2-1) \Gamma_i,
 \end{split}
\end{equation*}
with
\begin{equation*}
\begin{split}
\Gamma_{i}(u) = & V_1+V_2+4V_3+(-1)^i 2 \left(V_2-V_1\right) u\\
                &+\left(V_1+V_2-4 V_3\right)u^2,\\
\Gamma_{6}(u) = & V_1+V_2-(V_1+V_2- 4 V_3)u^2.
\end{split}
\end{equation*}
The formula (\ref{loc-len-1}) for the localization lengths can be
further simplified in the limit $L \gg 1$. When $L$ is large, the
typical value of $F_{\min(\max)}$ is of the order of
$e^{L/\xi_{1(2)}}$, which is exponentially large. Therefore,
\begin{equation}
F_{\max} \gg F_{\min} \gg 1,  \label{hierar}
\end{equation}
as we assume $\xi_1>\xi_2$  (see Eq.~(\ref{xi1gxi2})). The hierarchy (\ref{hierar}) largely
simplifies the co-efficients of Eq.~(\ref{fokker-planck-3}), which leads to
\begin{subequations}   \label{c1-c11-c2-c22}
\begin{equation}
\lim_{L\rightarrow \infty}\frac{c_{1}}{F_1+1} = -2 \Gamma_6 \Theta\left(\Delta{F}\right),
\end{equation}
\begin{equation}
\lim_{L\rightarrow \infty}\frac{c_{2}}{F_2+1} = -2 \Gamma_6 \Theta\left(-\Delta{F}\right),
\end{equation}
\begin{equation}
\lim_{L\rightarrow \infty} \frac{c_{ii}}{(F_i+1)^2} = \Gamma_i,
\end{equation}
\begin{equation}
\lim_{L\rightarrow \infty} \frac{\partial_{F_i} c_{ii}}{F_i+1} =2 \Gamma_i.
\end{equation}
\end{subequations}
As a result, Eq.~(\ref{loc-len-1}) reduces to
\begin{equation}     \label{locl-2}
\begin{split}
\xi_{\rho}^{-1} = &  V_1+V_2+2V_3 \\
                  & +(-1)^\rho \left( \frac{1}{2} \langle \Gamma_6
                  \rangle + |V_1-V_2| \langle u \rangle \right),\\
\end{split}
\end{equation}
where $\langle \cdot \rangle \equiv
\lim_{L\rightarrow\infty}{\langle \cdot \rangle_L}$, $V_1$, $V_2$
and $V_3$ are the Born cross sections defined in Eq.~(\ref{born-sec}).
The main simplification is that  $\Gamma_6$
depends only on $u$, but not on the other parameters of the scattering matrix. Therefore, the localization lengths are fully determined by the marginal probability distribution function of $u$ defined by
\begin{equation}  \label{def-w}
\text{w}_{L}(u) \equiv \int{dF_{1}dF_{2}d\psi \,W_L(\vec{\eta})}.
\end{equation}

Integrating over $F_1$, $F_2$ and $\psi$ on both sides of Eq.~(\ref{fokker-planck-3}),
we obtain the Fokker-Planck equation for $\text{w}_{L}(u)$:
\begin{equation} \label{ext-dmpk}
\partial_L{\text{w}} = \partial_{u} (c_{33} \partial_{u}\text{w}) +\partial_{u}(c_3 \text{w}),
\end{equation}
where $c_3$ and $c_{33}$ are derived in App.~\ref{coe}. It has a fixed-point solution satisfying
\begin{equation}         \label{u-fix}
\partial_{u} (c_{33} \partial_{u}\text{w}) +\partial_{u}(c_3 \text{w})=0.
\end{equation}
In the large $L$ limit the co-efficients  are given by
\begin{subequations}   \label{c3-c33}
\begin{equation}
\lim_{L\rightarrow \infty}c_{3} = \left( \left|V_1-V_2\right|-
\partial_u \Gamma_6 \right) (1-u^2),
\end{equation}
\begin{equation}
\lim_{L\rightarrow \infty}c_{33} = \left( V_3 + \Gamma_6 \right) \left( 1-u^2 \right).
\end{equation}
\end{subequations}

From Eq.~(\ref{c3-c33}) one can see that in the limit
(\ref{hierar}), $c_{33}$ and $c_3$ do not depend on $F_1$, $F_2$ and
$\psi$ any longer. Therefore, Eq.~(\ref{u-fix}) is reduced to an
ordinary differential equation with respect to $u$. By considering
the general constraints on a probability distribution function,
namely the non-negativity $\text{w}(u) \ge 0$ and the
normalization condition $\int{du\text{w}(u)}=1$, the solution
to Eq.~(\ref{u-fix}) is unique,
\begin{widetext}
\begin{equation}
\text{w}(u)=
\begin{cases}
\text{w}_1(u)=\frac{q_1\exp{\left( \frac{q_1}{q_2}\arctan{\frac{u}{q_2}}
\right)}}{2\sinh{\left(\frac{q_1}{q_2}\arctan{\frac{1}{q_2}}\right)}(q_2^2+u^2)},
& \Delta{V} \le 0,\\
&\\
\text{w}_2(u)=\frac{ q_1\left( \frac{q_2+u}{q_2-u}
\right)^{\frac{q_1}{2q_2}}}{ 2\sinh{\left( \frac{q_1}{2q_2}
\ln{\frac{q_2+1}{q_2-1}} \right)}(q_2^2-u^2)}, & \Delta{V} \ge 0,
\end{cases}                 \label{u-dis}
\end{equation}
\end{widetext}
where
\begin{eqnarray}       \label{q12}
q_1 &=& 2|V_1-V_2|/|\Delta{V}|, \\
q_2 &=& \sqrt{\left( V_1+V_2+4V_3\right)/|\Delta{V}|},
\end{eqnarray}
and
\begin{equation}  \label{deltV}
\Delta{V} = V_1+V_2-4V_3.
\end{equation}
Eq.~(\ref{locl-2}) and (\ref{u-dis}) are our main analytical results. The localization lengths are expressed entirely in terms of  the three Born cross sections $V_1$, $V_2$ and
$V_3$. We recall that we  made the assumptions of weak disorder, Eq.~(\ref{wdl}), and sufficiently strong coupling, Eq.~(\ref{re-str}).
\begin{figure}
\begin{center}
\includegraphics[height=5.9cm,width=8.2cm]{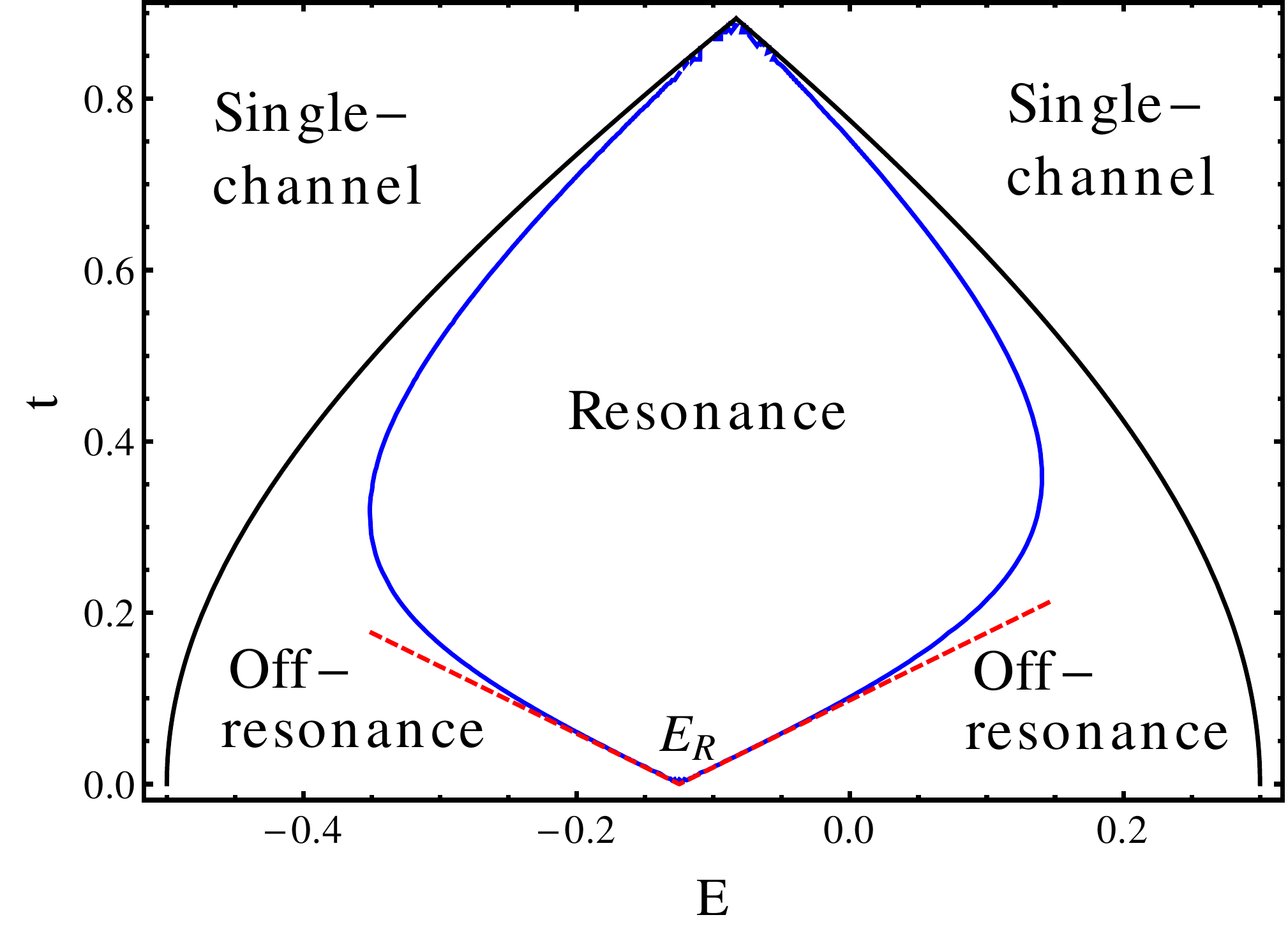}
\caption{The resonant and off-resonant regimes on the $E-t$ plane for
$t_1=1$, $t_2=0.2$, with isotropic disorder $\s_1^2=\s_2^2$ and bias $\delta{e}=-0.1$. The black
curve indicates the band edges $E=E_{1}^{-}$ and $E=E_{2}^{+}$, beyond
which only one channel exists. The blue curve marks the crossover line $\Delta{V}=0$. At
small enough $t$, $\Delta{V}=0$ can be linearized to $t \simeq
\kappa |E-E_R|$, cf. Eq.~(\ref{lin-res}), which is plotted as the red dashed line.}
\label{deltaV-0}\
\end{center}
\end{figure}

In Eq.~(\ref{u-dis}), $\text{w}_2(u)$ is simply the analytical continuation of
$\text{w}_1(u)$. To show this, we start
form $\Delta{V}>0$ side and drop the absolute value on
$\Delta{V}$. If $\Delta{V}$ crosses zero from above, namely
$\Delta{V} \rightarrow -\Delta{V}$, $q_1$ changes continuously to
$-q_1$, and $q_2$ changes to one of the two branches $\pm{i}|{q_2}|$
because of the square root. It can be easily verified that
\begin{equation}
\text{w}_1(u;-q_1,\pm{i}|q_2|)=\text{w}_2(u;q_1,|q_2|),
\end{equation}
by the formula $\arctan{z}=i/2 \ln[(1-iz)/(1+iz)]$ for a complex
number $z$. 

Given the physical meaning of the parameter $u$, it is natural to interpret the analytical continuation as describing the
\emph{crossover} between two regimes of the polarization, as controlled by the relative strength of the effective
disorders. If $\Delta V>0$ (i.e., $V_1+V_2 > 4V_3$) the
intra-channel scattering is stronger than the inter-channel
scattering, while $\Delta V<0$ means the opposite. The two regimes can be distinguished quantitatively. According to Eq.
(\ref{born-sec}), the co-efficients in the linear combination of the
effective disorder parameters, $\Var$, are determined by the bare
``mixing angle'' $\gamma$ and the rapidities, $v_\tau$. Suppose the
resonance energy $E_R$ is approached while keeping $t < t_c$ [see Eq.
(\ref{tc})]. If $E$ is in the vicinity of $E_R$, $\g \sim \pi/2$,
and $\Delta{V}<0$. Otherwise, if $E$ is far enough from $E_R$, $\g
\rightarrow 0$ or $\pi$, and $\Delta{V}>0$.
Therefore, there must be an energy interval around $E_R$, in which
the physics is similar to that at resonance, $\gamma=\pi/2$. Further
away from $E_R$ the physics is similar to the limiting cases $\gamma=0$ or
$\pi$. We call $\Delta{V}<0$ and $\Delta{V}>0$ the \emph{resonant}
and \emph{off-resonant} regimes, whose distinct behavior we will analyze below.
 %The border line between the two regimes is $\Delta{V}=0$.
%In this paper, resonance and off-resonance are crucial concepts to understand distinctly different coupling effects at different energies.
\smallskip
\subsection{Resonant and off-resonant regimes}
As shown in Fig.~\ref{deltaV-0}, for fixed $t_\nu$ and $\delta{e}$,
$\Delta V=0$ (blue curve) divides the $E-t$ plane into two regions
in the two-channel regime (below the black curve). Three important
observations are in order:

(i) At \emph{weak coupling $t$}, more precisely, for $t \ll t_c$, but still
within the condition (\ref{re-str}), the relation $\Delta V =0$ for the border of the resonance region implies the linear
relation (see the red dashed lines in Fig.~\ref{deltaV-0})
\begin{equation}
t \simeq \kappa(t_1,t_2)|E-E_R|  ,  \label{lin-res}
\end{equation}
with
\begin{equation}  \label{kappa}
\kappa(t_1,t_2)=\frac{t_1-t_2}{\sqrt{t_1^2+t_2^2}}.
\end{equation}
The slope $\kappa(t_1,t_2)$ neither depend on $\var$ nor on $\delta{e}$.

(ii) If the coupling  \emph{$t$ is strong enough}, the resonance
energy interval shrinks to zero as $t\rightarrow t_c$ (the top edge
of Fig.~\ref{deltaV-0}). This ``re-entrance'' behavior is due to the
competition between the strong coupling, which pulls $\g$ close to
$\pi/2$, and the band edge effect, which reduces the rapidity of one
of the channels.
%{\bf MM I think this is obsolete: is the preceding discussion still relevant in the light of the breakdown of coarsegraining?}
We can illustrate this behavior by considering two
limiting cases. If $t$ is weak, its effect is of first order on $\g$,
but of second order on the $v_\tau$. Therefore, the coupling wins
and the resonance energy interval follows the linear relation
(\ref{lin-res}). Alternatively, if the energy is in the vicinity of
the band edges $E=E_1^{-}$ and $E_{2}^{+}$, one of the rapidities
tends to zero. As a consequence, $V_1$ or $V_2$ is much larger than
$V_3$, which gives  a large positive $\Delta{V}$. Therefore, there
is always some region around the band edges (black curves in Fig.
\ref{deltaV-0}), which is out of resonance. As the crossover line must match the two limits $t\to 0 $ and $t\to t_c$, it is necessarily re-entrant.

(iii) In the case of a non-zero detuning energy $\delta{e}$ the resonant energy interval is slightly \emph{asymmetric} around $E=E_R$.

\subsection{Fixed point distribution $\text{w}(u=\cos\theta)$}

Let us now discuss the distribution Eq.~(\ref{u-dis}) in different
regimes and some of its consequences. For this purpose, we plot in Fig.~\ref{w-u} some representative
$\text{w}(u)$ together with the expectation
value and variance of $u$. We select various values of $E$
across the resonant and off-resonant regime. Two types of behavior can be observed in the two
regimes:
\begin{widetext}
\ \
\begin{figure}
\begin{center}
\includegraphics[height=7.4cm,width=15cm]{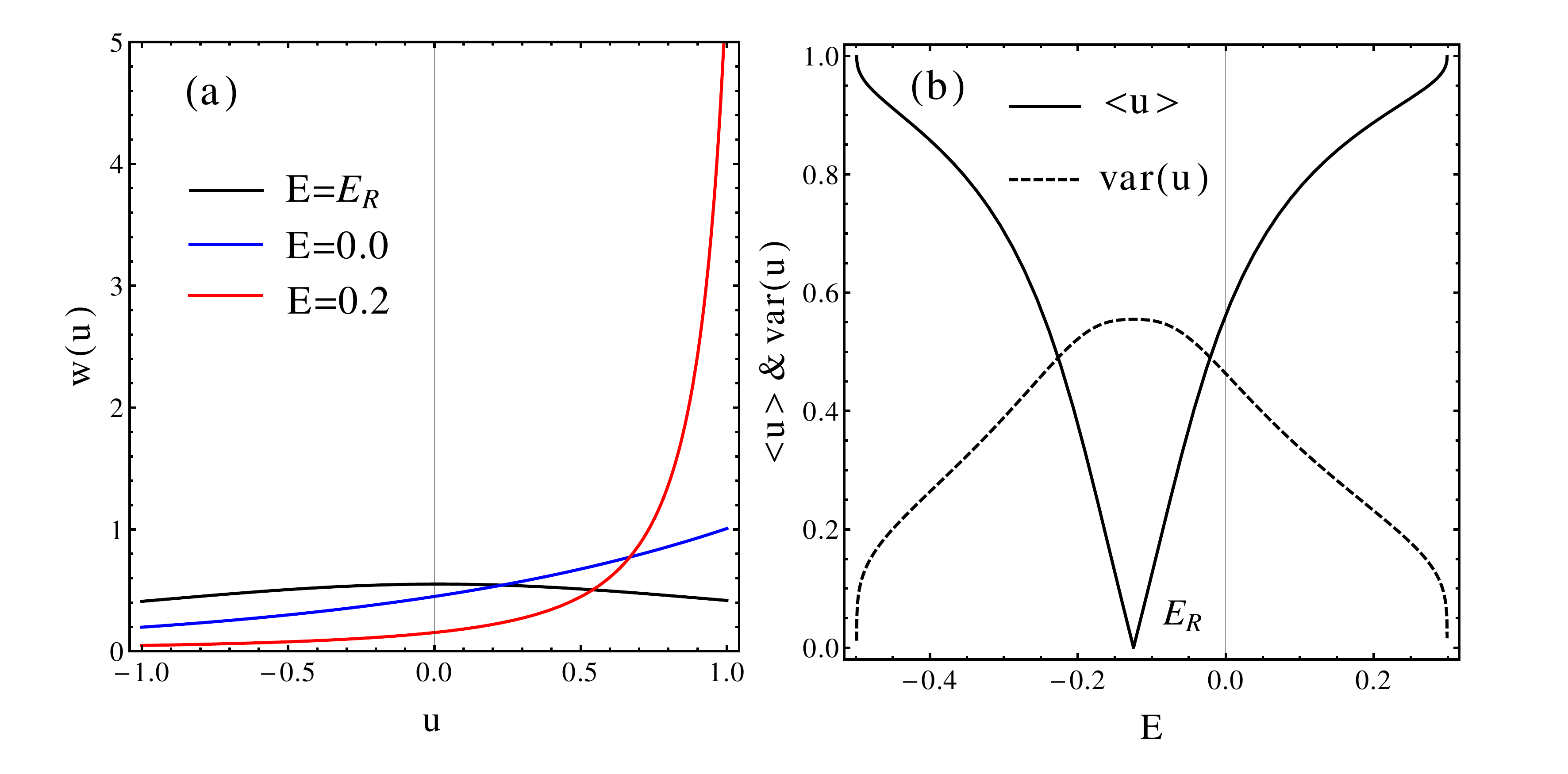}
\caption{Marginal probability distribution for the angular variable $u=\cos(\theta)$, $\text{w}(u)$, in different regimes. Here the inter-chain coupling $t=0.1$
and the other parameters are the same as in Fig.~\ref{deltaV-0}.
(a) Distributions $\text{w}(u)$ for three energies across the resonance
and off-resonance regimes, namely, $E=E_R \approx -0.13$ (i.e., resonance energy),
$E=0.2$ (off-resonance) and $E=0.0$ (crossover). On resonance the distribution is nearly uniform, while it is strongly non-uniform off-resonance (b) The expectation value and variance of $u$ as functions of $E$.} \label{w-u}\
\end{center}
\end{figure}
\ \
\end{widetext}
(i) \emph{Near the resonance}, $u=\cos\theta$ is  distributed relatively uniformly
in the interval  $(-1,1]$. Its average value is much smaller than
1, but its variance is large of order $O(1)$. However, the distribution is definitely not completely uniform. Indeed, the limit of the distribution can be obtained form Eq.~(\ref{u-dis}) in the weak coupling limit as $t\rightarrow0$ as
\begin{equation}
\text{w}(u)=\frac{3\sqrt{3}}{\pi(3+u^2)},   \label{lim-dist}
\end{equation}
which is manifestly {\it non-uniform}. A similar distribution was
obtained by Dorokhov \cite{dorok} in the case of two equivalent chains. We will discuss the difference to Eq.~(\ref{lim-dist}) later.
%{\bf MM: the difference being due to Dorokhov's neglect of forward scattering contributions. - Can we check at $E=0, t\to 0$ that our result is better than Dorokhov? This is the limit in which I believe that the continuum approximation for the discrete chain becomes accurate}

(ii) \emph{Off resonance}, the distribution function
$\text{w}(u)$ is strongly peaked at $u=1$, and its fluctuations are strongly suppressed.

At this point the difference
between the canonical DMPK equation,\cite{dorok,mpk1,mpk2} which
applies in the case $N\gg1$, and the extended DMPK equation obtained here for the case $N=2$, is clear. The isotropy  assumption, which allows one to
derive the canonical DMPK equation, states that the angular variable
distribution $\text{w}(u)$ should be uniform, i.e., independent of $u$, in contrast  to Eq.~(\ref{lim-dist}).   In order to justify the canonical DMPK equation,
we have to have a large number of equal chains. A
sufficient condition for obtaining the canonical DMPK equation is that
the probability distribution of the transfer matrices of
an ``elementary slice" is invariant under $U(N)$ rotation. This situation may be
achieved in thick wires.\cite{mpk1,mpk2} However, in few-channel
cases the localization lengths are larger, but still of the same order as
the mean free path. There is no parametric window between them that permits
the emergence of $U(N)$-invariant ensembles of transfer matrices upon coarsegraining.

The qualitative difference in the distribution function
$\text{w}(u)$ in the two regimes has important implications on the
localization lengths. To calculate the localization lengths from Eq.~(\ref{locl-2}),
we need $\langle \Gamma_6 \rangle$ and $\langle u \rangle$.
Using Eq.~(\ref{u-dis}) we obtain
\begin{equation}  \label{averg6}
\langle \Gamma_6 \rangle =
\begin{cases}
\frac{ q_1q_2|\Delta{V}| S_1(q_1,q_2)}{  2\sinh{
\left( \frac{q_1}{q_2}\arctan{\frac{1}{q_2}} \right)}}-4V_3 & \Delta{V} \le 0,\\
&\\
\frac{ q_1q_2|\Delta{V}| \tilde{S}_1(q_1,q_2)}{  2\sinh{\left( \frac{q_1}{2q_2}
\ln{\frac{q_2+1}{q_2-1}} \right)}}-4V_3 & \Delta{V} \ge 0,
\end{cases}
\end{equation}
and
\begin{equation}  \label{averu}
\langle u \rangle =
\begin{cases}
\frac{  q_1 S_2(q_1,q_2)}{  2\sinh{\left( \frac{q_1}{q_2}\arctan{\frac{1}{q_2}} \right)}} & \Delta{V} \le 0,\\
&\\
\frac{ q_1 \tilde{S}_2(q_1,q_2)}{  2\sinh{\left( \frac{q_1}{2q_2} \ln{\frac{q_2+1}{q_2-1}} \right)}} & \Delta{V} \ge 0,
\end{cases}
\end{equation}

where $S_{1(2)}$ and $\tilde{S}_{1(2)}$ are integrals defined by
\begin{equation}
\begin{split}
S_1(q_1,q_2) & =\int_{-\arctan{(1/q_2)}}^{\arctan{(1/q_2)}}{dz \sec^2{z}\,e^{\frac{q_1}{q_2}z}}, \\
S_2(q_1,q_2) & = \int_{-\arctan{(1/q_2)}}^{\arctan{(1/q_2)}}{dz \tan{z}\,e^{\frac{q_1}{q_2}z}}, \\
\tilde{S}_1(q_1,q_2) & =\int_{-1/q_2}^{1/q_2}{dz \left( \frac{1+z}{1-z} \right)^{\frac{q_1}{2q_2}}},\\
\tilde{S}_2(q_1,q_2) & = \int_{-1/q_2}^{1/q_2}{dz \frac{z}{1-z^2} \left( \frac{1+z}{1-z} \right)^
{\frac{q_1}{2q_2}}}.
\end{split}  \label{integral}
\end{equation}

\subsection{Numerical analysis}

In order to confirm our analytical results for the localization lengths in Eq.~(\ref{locl-2}) 
we calculated numerically the Lyapunov exponents of the products of transfer matrices in Eq.~(\ref{tm-l}). 
An efficient numerical method, known as the reorthogonalization method, has been developed in the study of 
dynamical systems\cite{bggs-cpv} and widely spread in the field of Anderson localization.\cite{mac-kram} The forthcoming numerical 
results in Figs.~\ref{comparison}-\ref{scale}, \ref{band-edge} and \ref{sig-chanl} are all obtained by this method.

The usefulness of the reorthogonalization method is not restricted to numerical simulations. 
It also provides the basis for the perturbative analysis 
about the Lyapunov exponents in the weak disorder limit in Sec.~\ref{wea-dis-exp}.

\section{Results for the localization lengths}

In order to reveal the effects of the transverse coupling $t$ on the localization lengths,
we define the two ratios
\begin{equation}
r_{\rho}=\xi_{\rho}/\xi_{\rho}^{(0)} , \quad \rho \in \{1,2\},
\label{ratio-def}
\end{equation}
where the $\xi_{\rho}^{(0)}$s are the
localization lengths of the decoupled legs, for which we may assume
$\xi_1^{(0)} \ge \xi_2^{(0)}$. For simplicity, we
refer to the leg $1$ and $2$ as the fast- and the slow-leg, respectively. The bare
localization lengths $\xi_\rho^{(0)}$ can easily be obtained from Eq.~(\ref{locl-2})
by taking $\gamma=0$, $\text{w}(u)=\delta(u-1)$ and
$t=0$, which yields
\begin{equation}
\xi_{\rho}^{(0)} = \frac{ 2v_{\rho}^2}{\chi^2_{\rho}}.  \label{one-chain}
\end{equation}
Eq.~(\ref{one-chain}) coincides with the well-known single-chain result.\cite{mel}

%{\bf MM: Here and below I replaced all min(max) by an index $\rho$! - please check that this is correct!}

\subsection{$E=0$ and $\delta{e}=0$: Resonant regime}
 \label{band-centre}
Consider first the case $\delta{e}=0$, in which the resonance energy vanishes
$E_R=0$. From Eqs.~(\ref{clean-angle}) and (\ref{rapidity}) it
follows that the mixing angle is $\gamma = \pi/2$ once $t \neq 0$, and the
two rapidities $v_{1}=v_{2}=v$ are equal to each other:
\begin{equation}
v^2 =4-\frac{t^2}{t_1t_2}.
\end{equation}
Consequently, the three Born cross-sections have the same value and
are equal to:
%{\bf HYX: I replaced the $\sum_\nu$ to explicit sums. But remind that $\nu$ labels ``chains'', $\tau$ labels ``channels'', $\varrho$ labels ``transmission eigenvalues'' and $\rho$ also labels the transmission eigenvalues but in descending order. Without coupling we do not need so many labels.}
 \be
 V_1=V_2=V_3=V=\frac{1}{16v^2}{\left( \Varo +\Vart \right)}.
 \ee
This gives
$q_1=0$ and $q_2=\sqrt{3}$ according to Eqs.~(\ref{q12}) and
(\ref{deltV}). Evaluating the integrals~(\ref{integral}), we obtain the two localization lengths
\begin{equation}
\xi_{\rho}=8 C_\rho v^2 /{\left( \Varo +\Vart \right)},    \label{loc-length-e-0}
\end{equation}
where
\begin{subequations}
\begin{equation}
\label{C1}
C_1=\frac{\pi}{3(\pi-\sqrt{3})} \approx 0.743,
\end{equation}
and
\begin{equation}
C_2=\frac{\pi}{\pi+3\sqrt{3}} \approx 0.377.
\end{equation}
\end{subequations}
The corresponding decoupled values ($t=0$) can easily be obtained from Eq.~(\ref{one-chain}),
\be \xi_{\rho}^{(0)}=\frac{8}{\chi^2_{\rho}}. \ee
Therefore, the ratios defined by
Eq.~(\ref{ratio-def}) read
\begin{equation}           \label{loc-ratio}
r_{\rho} = C_{\rho}v^2\chi^2_{\rho}/{\left( \Varo +\Vart \right)}.
\end{equation}

 Notice that in the resonant case, the Born cross-sections (\ref{born-sec}) are dominated by $\chi^2_{2}$, 
which gives rise to the dramatic drop of the localization length of the fast leg: The slow-leg is dominating the backscattering rate and thus the localization length.

From Eq.~(\ref{loc-ratio}) we draw several important conclusions below.
 
\subsubsection{Statistically identical chains}
\begin{figure}
\begin{center}
\includegraphics[height=5.7cm,width=8.2cm]{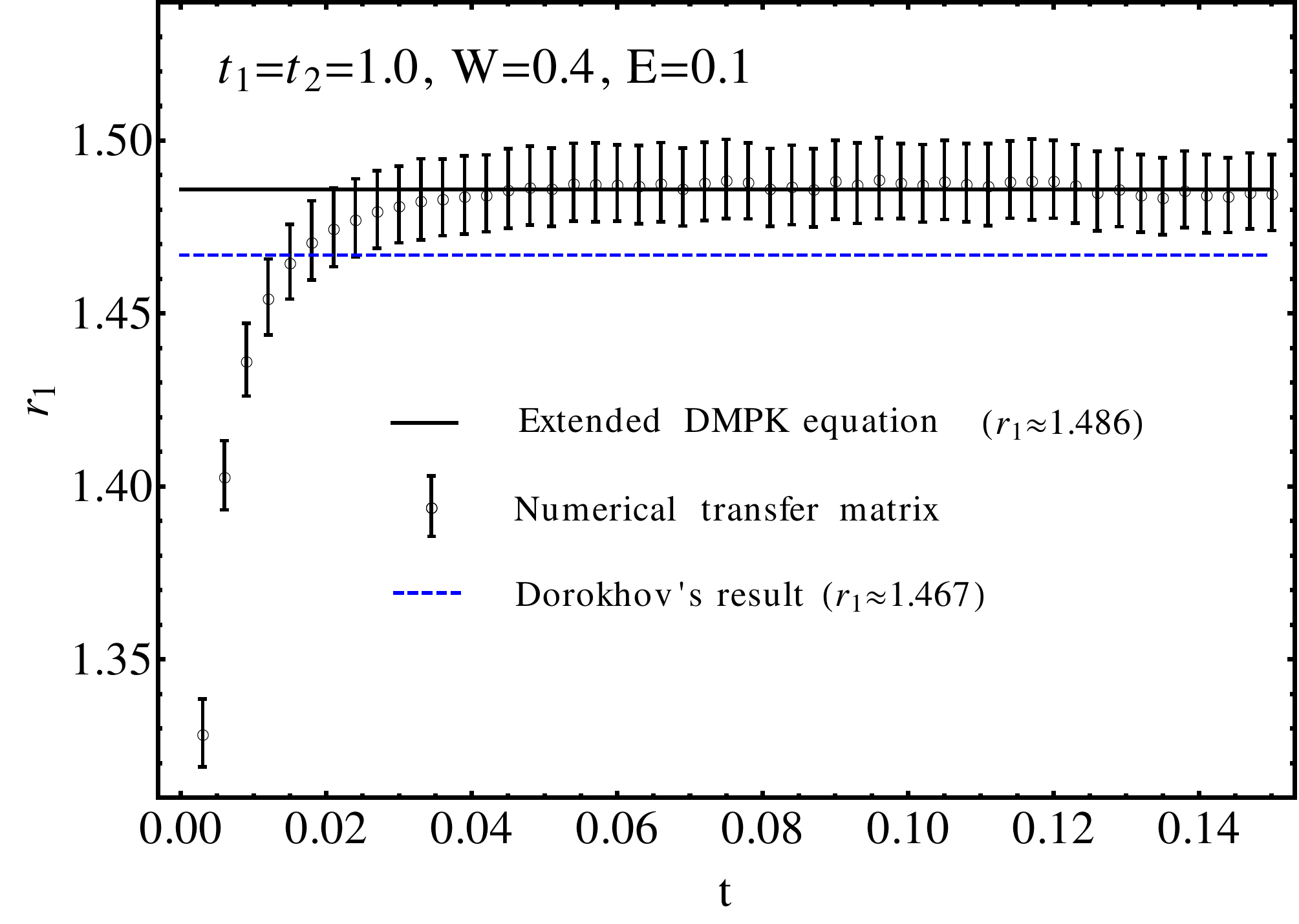}
\caption{Ratio of coupled and uncoupled localization lengths,
$r_1=\xi_1/\xi_1^{(0)}$, for statistically identical chains with $t_1=t_2=1$ and
small disorder $W$. We consider small couplings $t$, for which any energy is at resonance conditions.
 In the numerical simulation we take $E=0.1$ in order to avoid the anomaly at $E=0.0$. 
However, according to Eq.~(\ref{bceq}), $r_1$ is almost independent of 
$E$ if $t$ is weak enough. The continuum approximation becomes exact for $W^2 \ll t \ll 1$, 
as is illustrated by the convergence of the numerical data to the analytical prediction (the agreement is already good for $c W^2\lesssim t$ with $c\approx 0.25$).
For comparison we also plot Dorokhov's prediction~\cite{dorok}, which neglected forward scattering in the Fokker-Planck equation.
The result of Kasner and Weller\cite{kas-wel} ($r_1 \approx 1.776$) is in clear contradiction with these numerics.
}    \label{comparison}\
\end{center}
\end{figure}

For two coupled chains, which are statistically identical, one has $\chi^2_1=\chi^2_2$, and we
obtain
\begin{eqnarray}  \label{bceq}
\frac{\xi_1}{\xi_1^{(0)}} \equiv r_{1}=2C_{1} \approx 1.486,\nonumber\\
\frac{\xi_2}{\xi_2^{(0)}} \equiv r_{2}=2C_{2} \approx 0.754.
\end{eqnarray}
We note that $r_1$ is slightly larger than the value obtained by
Dorokhov\cite{dorok}, which is $\pi/(\pi-1) \approx 1.467$. The
reason is that we have taken into account the
\emph{forward-scattering} in the ``elementary slice'' (\ref{slice}),
which was neglected in the work by Dorokhov. Moreover the latter was restricted to $t_1=t_2$.
In Fig.~\ref{comparison} we compare our analytical prediction with Dorokhov's. 
Note that we take $E=0.1$ in the numerical simulation in order to avoid the anomaly 
at $E=0$, as mentioned in Sec.~\ref{coa-gra}.  The enhancement factor $r_1$ is essentially independent of 
the selected energy if $t$ is weak enough. This is due to the fact that any energy is at resonance conditions for $t_1=t_2$.

The effect of forward scattering, which was included in our work, is clearly visible.  
It is confirmed by the numerical simulation at resonance conditions.
However, the value $r_1 \approx 1.776$ obtained by Kasner and Weller\cite{kas-wel} deviates significantly 
from our numerical and analytical results.

% $t_1=t_2$ and small $t$, where the continuum approximation works.

\subsubsection{Parametrically different chains}
\begin{figure}
\begin{center}
\includegraphics[height=5.7cm,width=8.2cm]{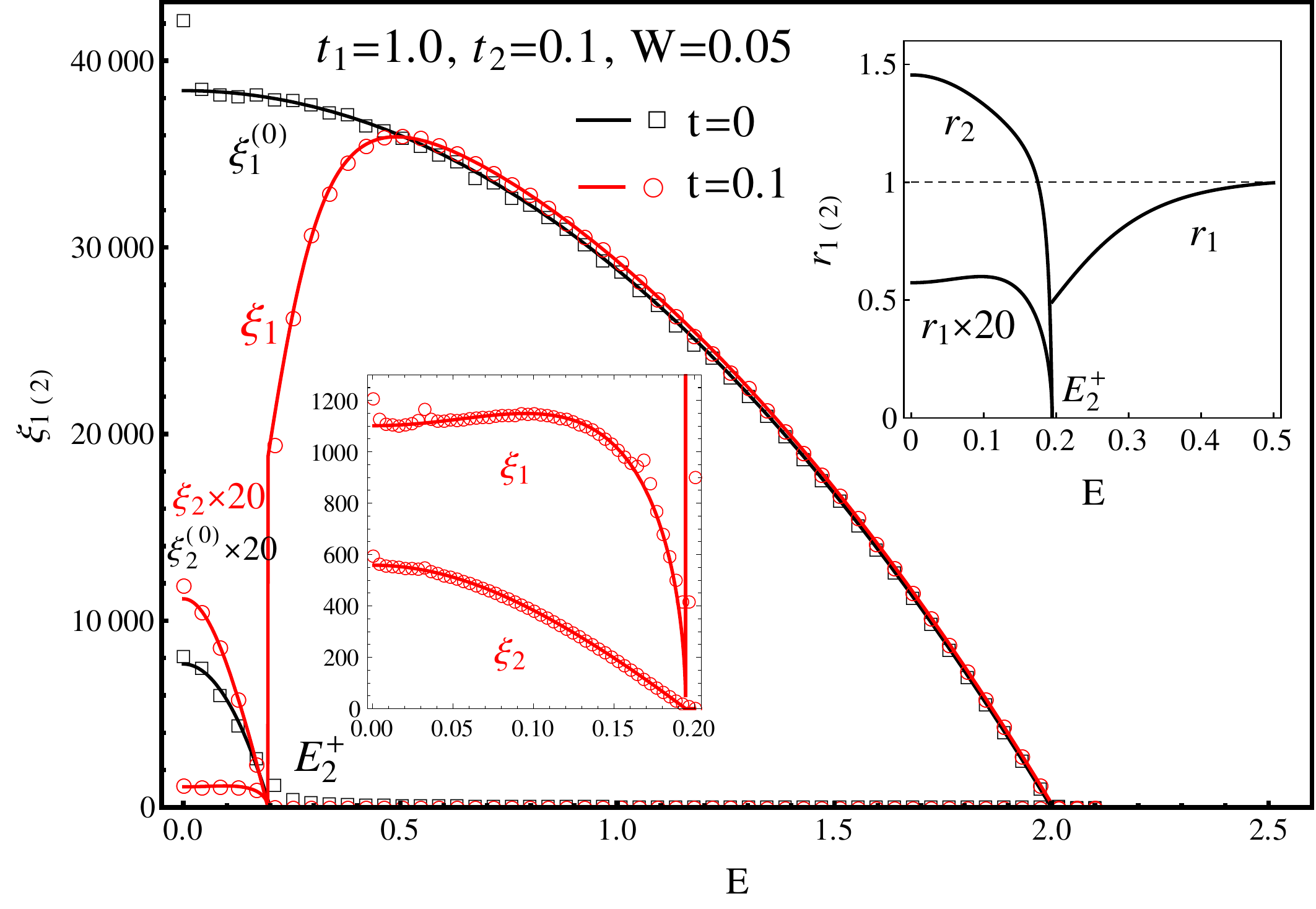}
\caption{Localization lengths for chains with different hopping strengths ($t_1=1$ and $t_2=0.1$),
but equal disorder ($W=0.05$) as a function of energy at detuning $\delta{e}=0$ and intermediate coupling $t=0.1$.
The solid curves are analytical results. Black curves correspond to uncoupled chains, red ones to the coupled chains.
The squares and circles are data of the numerical transfer matrix. $\xi_2^{(0)}$ and $\xi_2$ are amplified 20 times to
increase visibility, but $\xi_1 > \xi_2$ always holds. The lower left insert is a zoom in the two-channel region. The upper right insert shows the ratios $r_{1,2}$ of coupled to uncoupled localization lengths.
The larger localization length is very significantly suppressed due to the coupling to a slow chain.
Note the sharp recovery of the larger localization length beyond the band-edge $E_2^+$.
The analytical results coincide quantitatively with the numerical data anywhere except for specific anomalous energies:
In the uncoupled case, $E = 0$ corresponds to the commensurate wave vectors $4k_{1(2)} = 2\pi$.
In the coupled case, $E = 0$ and  $E \approx 0.03$, $0.1$ (very weak) and $0.17$ correspond to $2(k_1+k_2)=2\pi$, $3k_1+k_2=2\pi$, $4k_1=2\pi$ and $3k_2-k_1=2\pi$.
}     \label{interm-coupling}\
\end{center}
\end{figure}

\begin{figure}
\begin{center}
\includegraphics[height=5.7cm,width=8.2cm]{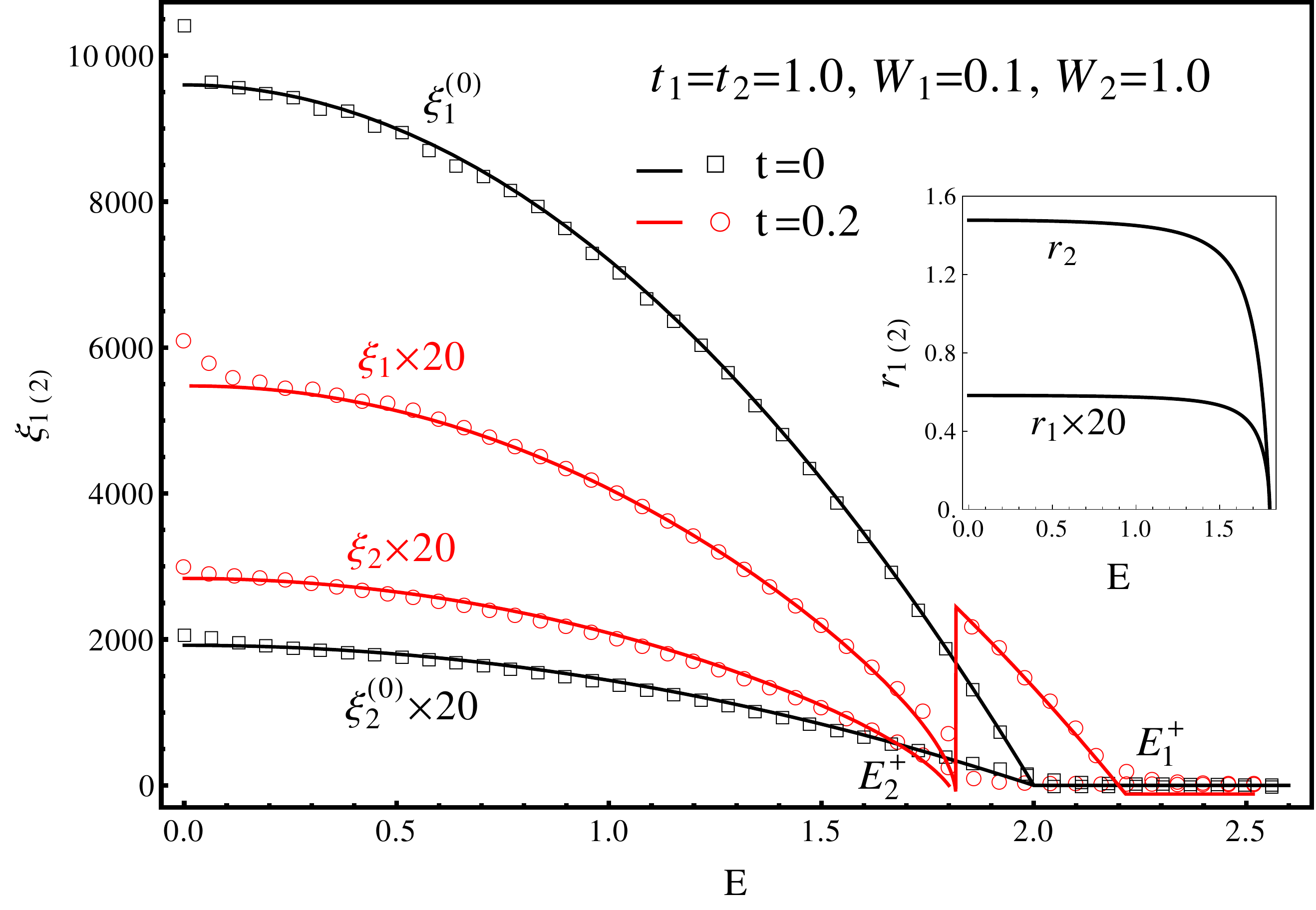}
\caption{Localization lengths for the decoupled ($t=0$, black) and coupled ($t=0.2$, red) chains with identical hopping but substantially
different disorder ($W_1=0.1$ and $W_2=1.0$) as a function of energy (detuning $\delta{e}=0$).
The solid curves are analytical results.
The squares and circles are data from the numerical transfer matrix. The values of $\xi_1$, $\xi_2^{(0)}$ and
$\xi_2$ are amplified 20 times to increase their visibility.
Without coupling $\xi_1^{(0)}/\xi_2^{(0)} \sim 10^2$. In the presence of coupling $\xi_1$ is substantially reduced,
while $\xi_2$ remains
of the same order as its decoupled value.
The insert shows the ratios $r_{1,2}$ of coupled to uncoupled localization lengths.
Since $t_1=t_2$ there is resonance at all energies, and thus dominance of the slow chain is expected. Note the sharp recovery of the larger localization length beyond the band-edge $E_2^+$. There are visible anomalies at $E= 0$
in both the uncoupled and the coupled case, which correspond to the commensurate condition $4k_{1(2)} = 2\pi$ and $2(k_1+k_2)=2\pi$.
In the coupled case further anomalies exist at the energies corresponding to $3k_1+k_2=2\pi$, $4k_1=2\pi$ and $3k_2-k_1=2\pi$.
However, they are every close to $E=0$ and too weak to be observed.}       \label{neq-disorder}\
\end{center}
\end{figure}

It is interesting to analyze what happens if the bare localization lengths of the chains are parametrically
different $\xi^{(0)}_{2} \ll \xi^{(0)}_{1}$. In the
resonant regime, for $W^{2}, |E-E_{R}|\ll t\ll t_{1},t_{2}$, we obtain:
\begin{subequations}   \label{bcnoneq}
\begin{equation}
\xi_{1} \rightarrow 4C_1 \,\xi^{(0)}_{2} \approx 2.972
\,\xi^{(0)}_{2},
\end{equation}
\begin{equation}
\xi_{2} \rightarrow 4C_2\,\xi^{(0)}_{2} \approx 1.507\,\xi^{(0)}_{2}.
\end{equation}
\end{subequations}
Eq.~(\ref{bcnoneq}) is one of the central results in this paper: In the resonant regime, the localization length of the
fast leg is \emph{dramatically} dragged down by the slow-leg. In
contrast, the localization length of the slow leg is increased by
the presence of the fast leg, but remains of the same order. As a result both
localization lengths become of the order of that for the bare {\it slow
leg}. This is illustrated for two different cases of coupled fast and slow legs
in Figs.~\ref{interm-coupling} and \ref{neq-disorder}. Fig.~\ref{interm-coupling} shows the effect in the case of legs with
equal disorder but different hopping strength, the resonance being at $E=0$. In Fig.~\ref{neq-disorder} the faster
leg has the same hopping but weaker disorder. Here the legs are resonant at \emph{every energy} below the band-edge $E_2^+$.

We note that there is \emph{no} regime where both $r_1>1$ and $r_2>1$, as this
would contradict the equality $\sum_{\rho}{r_\rho/C_\rho} = v^2$, which
follows from Eq.~(\ref{loc-ratio}). At the band center and $t \to 0$ one can achieve that both localization lengths do not decrease upon coupling the chains,
$r_1=r_2=1$. This happens when $\chi^2_{2}/\chi^2_{1} = 4C_1-1$, which assures %{\bf MM: does this assure that $r_2=1$? Doesn't $r_1=r_2=1$ simply require $t$ very small, i.e., decoupled chains?} {\bf HYX: Here $r_1=r_2=1$ is valid in the range $0\le t \lesssim W^2$. The two chains do not have to be decoupled. Actually, if two chains are decoupled, $r_1=r_2=1$ is always satisfied; if we turn on a couplingt $t \sim W^2$, only this special situation $\chi^2_{\max}/\chi^2_{\min} = 4C_1-1$ keeps $r_1=r_2=1$ (no drop/lift in Fig.~\ref{scale}. I checked this point numerically long time ago.). }
that the localization lengths do not change at coupling constants $t\lesssim W^{2}$ according to the discussion in Sec.~\ref{weak coupling}.

\subsubsection{Weak coupling limit} \label{weak coupling}

\begin{figure}
\begin{center}
\includegraphics[height=12.1cm,width=8.5cm]{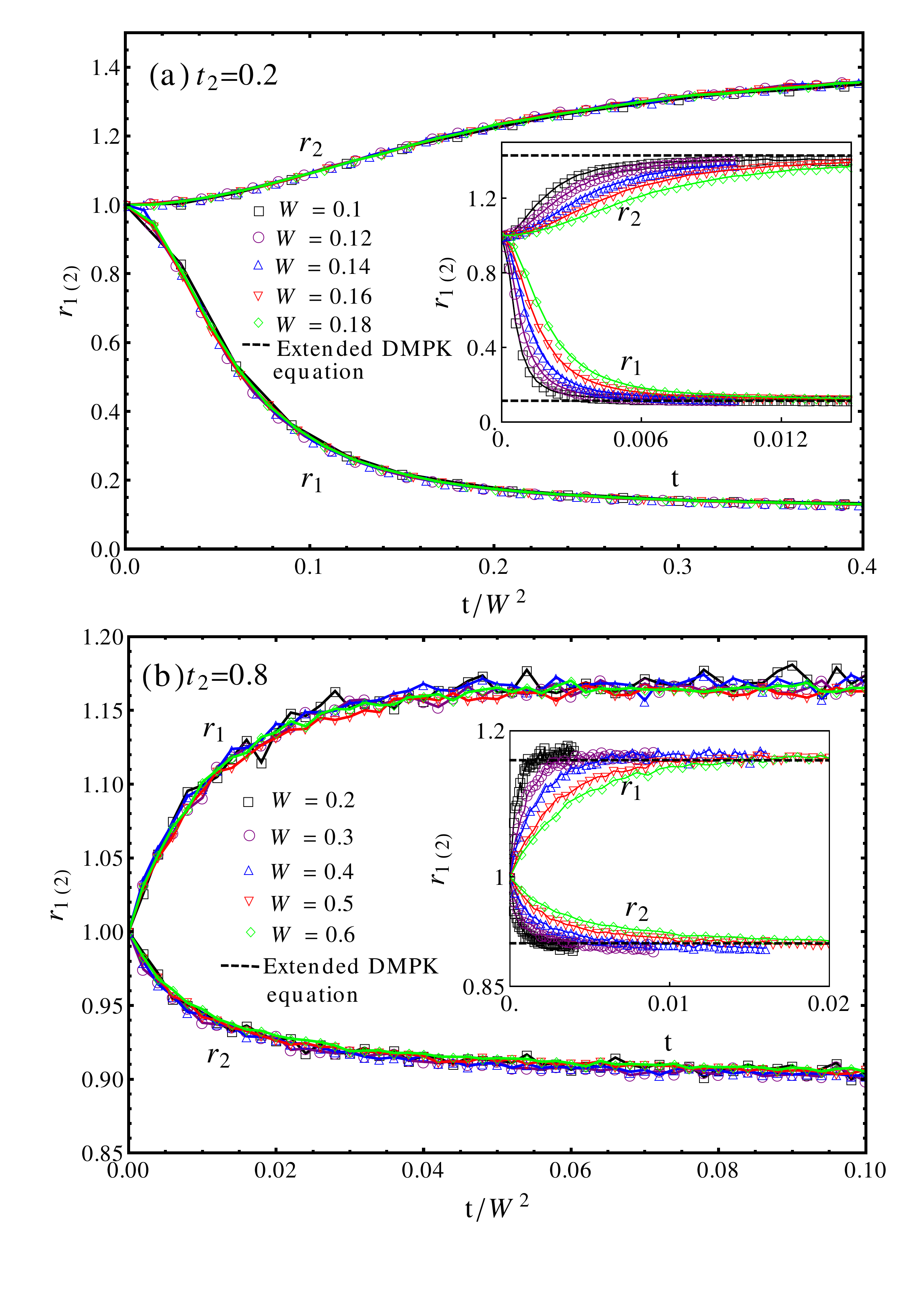}\\
\vspace{-0.7cm}
\includegraphics[height=5.6cm,width=7.7cm]{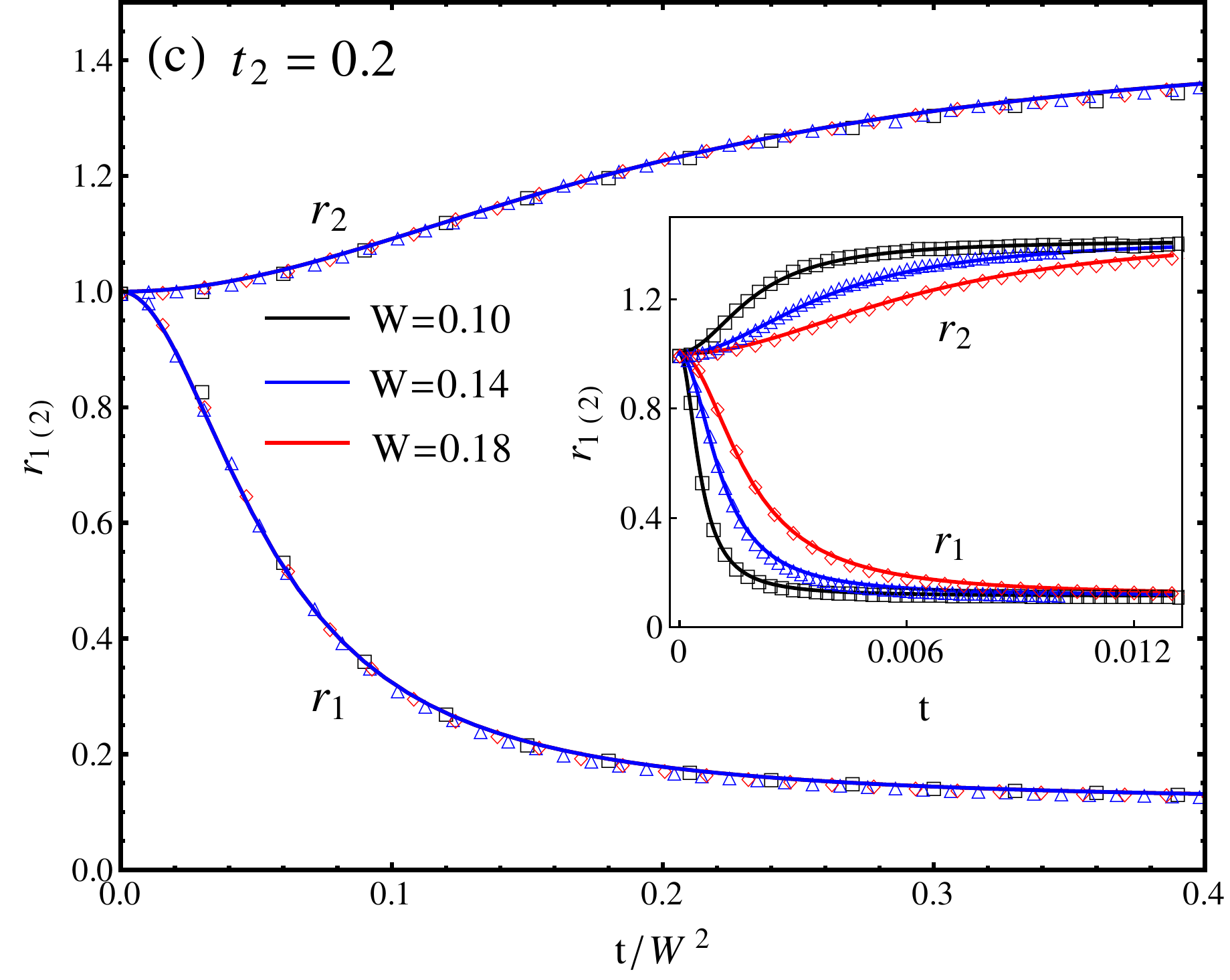}
\caption{The ratio $r_{1,2}=\xi_{1,2}/\xi^{(0)}_{1,2}$ of coupled and decoupled
localization lengths, obtained numerically as a function of the coupling constant $t$ at resonant conditions,
$E=0$ and $\delta{e}=0$. The two legs are equally disordered with a
 random potential box distributed in $[-W/2,W/2]$. (a) A slow $t_2=0.2$ and a fast
$t_{1}=1$ leg: The smaller localization length increases slightly
while the larger localization length decreases drastically, being driven
down by the slow leg, as the coupling constant $t$ increases. The
insert shows the dependence of $r_{1,2}$ on the coupling constant
$t$ at different disorder strengths; all the curves collapse to a
universal dependence on $t/W^{2}$. The dashed lines in the insert
show the analytic result given by Eq.~(\ref{loc-ratio}), which is valid under the
assumption $t\gg W^2$, Eq.~(\ref{re-str}). (b) Almost identical legs $t_{2}=0.8$,
$t_{1}=1$: In this case the localization length of the slow leg marginally decreases
 while that of the fast leg marginally increases. (c) Results obtained analytically upon replacing the mixing angle with a renormalized value, $\gamma \to \tilde{\gamma}$. 
 The parameters are the same as in (a) but with fewer realizations of disorder, 
and $\delta E/W^2 \approx 0.3$ in Eq.~(\ref{re-angle}) was optimized by fitting to the numerical data in (a). 
The scaling collapse works very well in the weak coupling limit.} \label{scale}\
\end{center}
\end{figure}

Upon simply taking the $t=0$ limit,
\begin{equation}
r_{\rho}=4 C_{\rho} \chi^2_{\rho}/{\left( \Varo +\Vart \right)},
\label{ratio-lim}
\end{equation}
one \emph{does not} recover the decoupled values $r_\rho=1$. This should indeed be expected, as we have already discussed in
Sec.~{\ref{Fokker-Planck-eq}}. The reason traces back to 
condition~(\ref{re-str}) to obtain Eq.~(\ref{fokker-planck-3}), namely that $t$ be larger than the disorder energy scale
$\delta{E}\propto W^{2}$. In order to verify
the non-commutativity of $t\to 0$ and $W\to 0$, we computed numerically  the localization lengths by
the transfer matrix approach, and obtained the values of
$r_\rho$ down to very small values of $t$, cf. Fig.~\ref{scale}). In this simulation, the re-orthogonalization
method~\cite{mac-kram} was used, and length of the ladder is $L=10^7$ with averaging over $10^3$
realizations of disorder. The hopping integral in the fast chain
$t_1=1$ was taken as the energy unit, and for simplicity, the two
legs were taken to be equally disordered. One can see that as the coupling
$t$ increases the quantities $r_{1,2}$ evolve and at $t\gg
W^{2}$ approach the limits given by Eq.~(\ref{ratio-lim}).

The insensitivity of the localization lengths to weak couplings $t\ll \delta E$ reflects
the fact that the level spacing in the chains is bigger than the coupling between the chains,
and thus wavefunctions typically do not hybridize much between the two legs.\\

Moreover,
the two families of curves for different disorder strengths seem to collapse into two universal functions
$r_{1,2}(t/W^{2})$, c.f. Fig.~\ref{scale}. This scaling shows that at weak disorder $W\ll 1$
and under resonance conditions $E=E_{R}$, the numerical results
approach the analytical ones already at a very small coupling $t\gtrsim W^{2}$.

We can rationalize the scaling by defining a regularized mixing
angle $\tilde{\gamma}$ instead of the bare
$\gamma$ defined by Eq.~(\ref{clean-angle}).  From Eqs.
(\ref{clean-angle}) and (\ref{lin-res}), we find that
\begin{equation}
\tan^2{\gamma} \propto \frac{t^2}{[\kappa(t_1,t_2)(E-E_R)]^2}.
\end{equation}
where $\kappa(t_1,t_2)$ is defined in Eq.~(\ref{kappa}). A natural way of regularizing the above result at resonance
conditions is to introduce the disorder-induced  ``width'' $\delta
E\propto W^{2}$ in the form:
\begin{equation}
\tan^2{\tilde{\gamma}} \propto \frac{t^2}
{[\kappa(t_1,t_2)(E-E_R)]^2+\delta{E}^2}  \label{re-angle},
\end{equation}
where $\delta{E}$ scales as in Eq.~(\ref{deltaE}).

Using this regularized mixing angle, the resonant
regime can be described more precisely by the condition
\begin{equation}    \label{crit-res}
t \gg \max \{\kappa(t_1,t_2)|E-E_R|,\delta{E}\},
\end{equation}
or, equivalently, $\td{\gamma} \sim \pi/2$. The observed scaling collapse in Fig.~\ref{scale}(a) suggests that in
 the weak coupling one might capture the behavior of localization lengths by replacing $\gamma$ by $\td\gamma$ in Eq. (\ref{born-sec}).
 This indeed works, as confirmed by Fig.~\ref{scale}(c) where we replot the numerical data of Fig.~\ref{scale}(a)
together with the analytical expressions, where $\tilde \gamma$ replaces $\gamma$, and the number $\delta E/W^2 \approx 0.3$
 was optimized to yield the best fit.
%$f(t_1,t_2,E,\delta{e})$ is an universal function whose value can be fitted by numerical data. }

%\begin{figure}
%\begin{center}
%\includegraphics[height=5.6cm,width=7.9cm]{analy-scale.pdf}
%\caption{The ratio $r_{1,2}$ obtained analytically with the replacement $\gamma \to \tilde{\gamma}$. 
%In this figure, the parameters are the same as in Fig.~\ref{scale}(a) but with fewer realizations of disorder, 
%and $\delta E/W^2 \approx 0.3$ in Eq.~(\ref{re-angle}) was optimized by fitting to the numerical data. 
%The scaling collapse works very well in the weak coupling limit.}       \label{analy-scale}\
%\end{center}
%\end{figure}

Note that the resonance condition can be broken either by detuning
$|E-E_{R}| \gg t$ or by increasing the disorder $\delta E\gg t$. Our analytic
approach is based on the weak-disorder expansion and is therefore valid only
in the first regime.

\subsubsection{Anomalies}
One can notice that all our numerical curves for $\xi_{1,2}$ exhibit \emph{anomalies} which are
not predicted by the analytical curves: small ``peaks'' appear
at certain energies on both $\xi_1$ and $\xi_2$. These anomalies of localization lengths are due to the commensurability discussed in Sec.~\ref{coa-gra}.
This is not captured by the extended DMPK equation (\ref{fokker-planck-3}).
However, we can identify these anomalous energies with commensurate combinations of wave vectors in Eq.~(\ref{wav-com}) (see the caption in Figs.~\ref{interm-coupling} and \ref{neq-disorder}).
The anomalies for two chains with identical hopping but different disorder have been observed numerically in Ref.~\onlinecite{ngu-kim}. 
In this case there are three anomalous energies $E =0$, $t/2$ and $t$, which correspond to commensurate combinations of wave vectors $2(k_1 + k_2)=2\pi$, $3k_1+k_2= 2\pi$ and $4k_1 = 2 \pi$.

\subsection{Solutions at $E \neq 0$: off-resonant regime}   \label{Ene0}
Without loss of generality the off-resonant regime can be
considered at $\delta e =0$ (for which the resonance is at $E_{R}=0$). 
A non-zero detuning $\delta{e}$ merely drives $E_R$ away from zero
and induces an asymmetry of the $r_\rho$ as a function of $E-E_R$.
However, the mechanism of the crossover from resonance to off-resonance is
qualitatively the same as in the case $\delta{e}=0$.

\begin{figure}[t]
\begin{center}
\includegraphics[height=12.1cm,width=8.5cm]{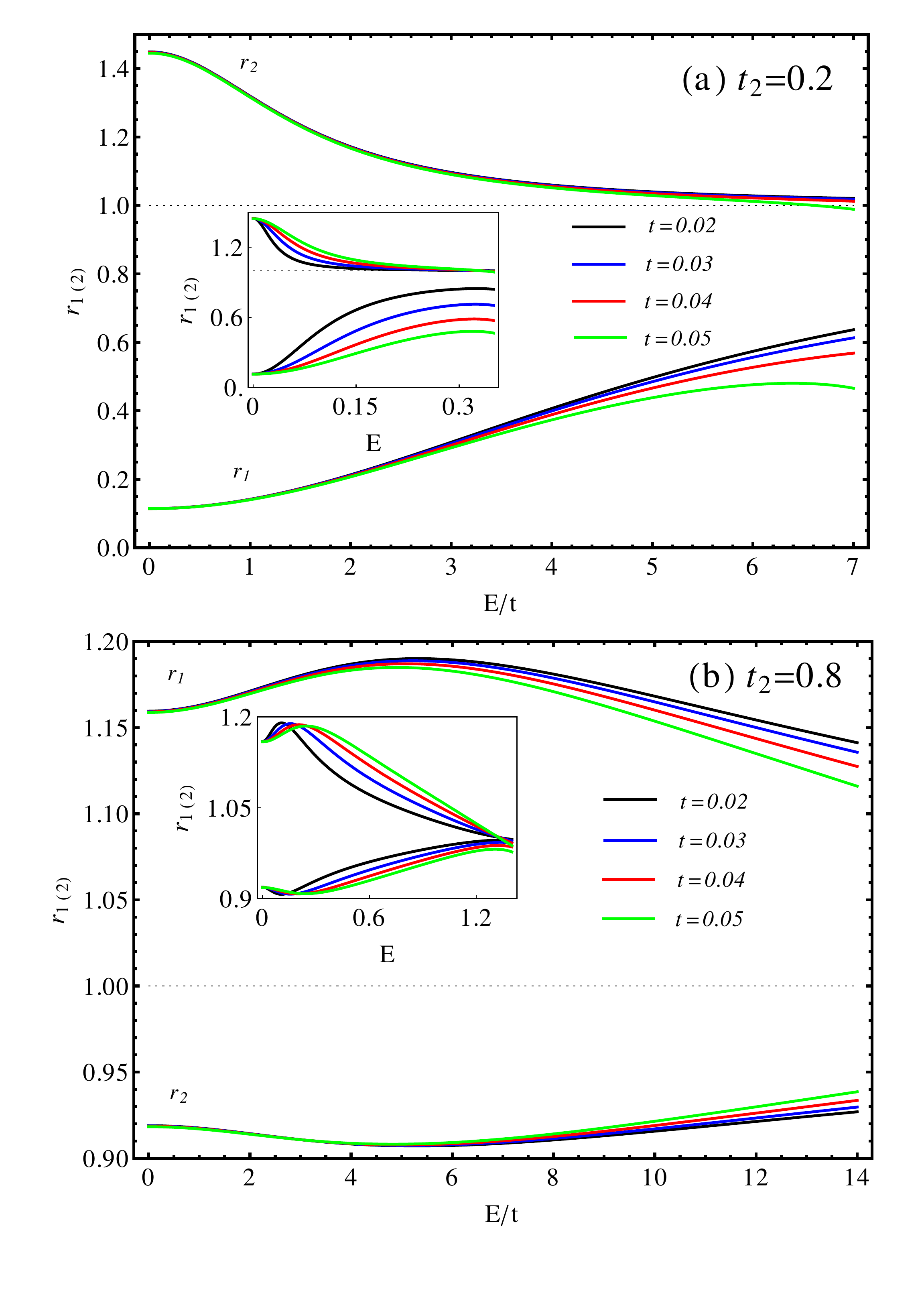}
\caption{ Analytical results for $r_{1,2}=\xi_{1,2}/\xi^{(0)}_{1,2}$
as functions of $E$ and $E/t$ obtained from the extended DMPK equation
in the weak disorder case $\delta E\ll t$ for detuning $\delta{e}=0$. The resonance energy corresponds to $E_R=0$.
(a) $t_2=0.2$, $t_{1}=1$. (b) $t_2=0.8$, $t_{1}=1$.
Close to resonance the $r_\rho$ only depend on the ratio $E/t$.} \label{r-energy}\
\end{center}
\end{figure}

Our analytical results for $r_{1,2}$ are presented in Fig.~\ref{r-energy} as functions of
the dimensionless detuning $E/t$ from resonance.

(i) \emph{Small detuning},  $|E| \ll t/\kappa(t_1,t_2)\ll 1$:
The resonance conditions are still fulfilled and the
localization lengths are close to their corresponding values at
$E=0$. The leading order expansion around $\gamma=\pi/2$ predicts that the ratios of
localization lengths, $r_{1,2}$ only depend on $E/t$, but not on $t/t_1$,
\begin{equation}    \label{on-}
\left| r_{\rho}-r_\rho(E=0) \right| \propto \left(\frac{E}{t}\right)^2,
\end{equation}
as confirmed numerically in Fig.~\ref{r-energy}.
%which comes out naturally by keeping the leading order of the expansion of $r_\rho$ at $\gamma=\pi/2$.

(ii) \emph{Very large detuning},  $|E| \gg t/\kappa(t_1,t_2)\gg 1$,
%{\bf MM: why do you need $t/\kappa(t_1,t_2)\gg 1$?}  {\bf HYX: I do not think we need this, since we only need $\tan^2\gamma \sim 0$},
$r_{1(2)}$ approaches $1$ from below (above) like
\begin{equation}
\left| r_\rho-1 \right| \propto t^2.
\end{equation}
When $t$ is small this result is obtained from the leading
order expansion of $r_\rho$ around $\gamma=0$ or $\pi$.

(iii) For chains with \emph{equal hopping}, $t_1=t_2$, resonance occurs at
\emph{any} energy and $r_\rho=2C_\rho$ is independent of $E/t$.

\subsection{Band-edge behavior}  \label{beb}
\begin{figure}[t]
\begin{center}
\includegraphics[height=5.4cm,width=7.8cm]{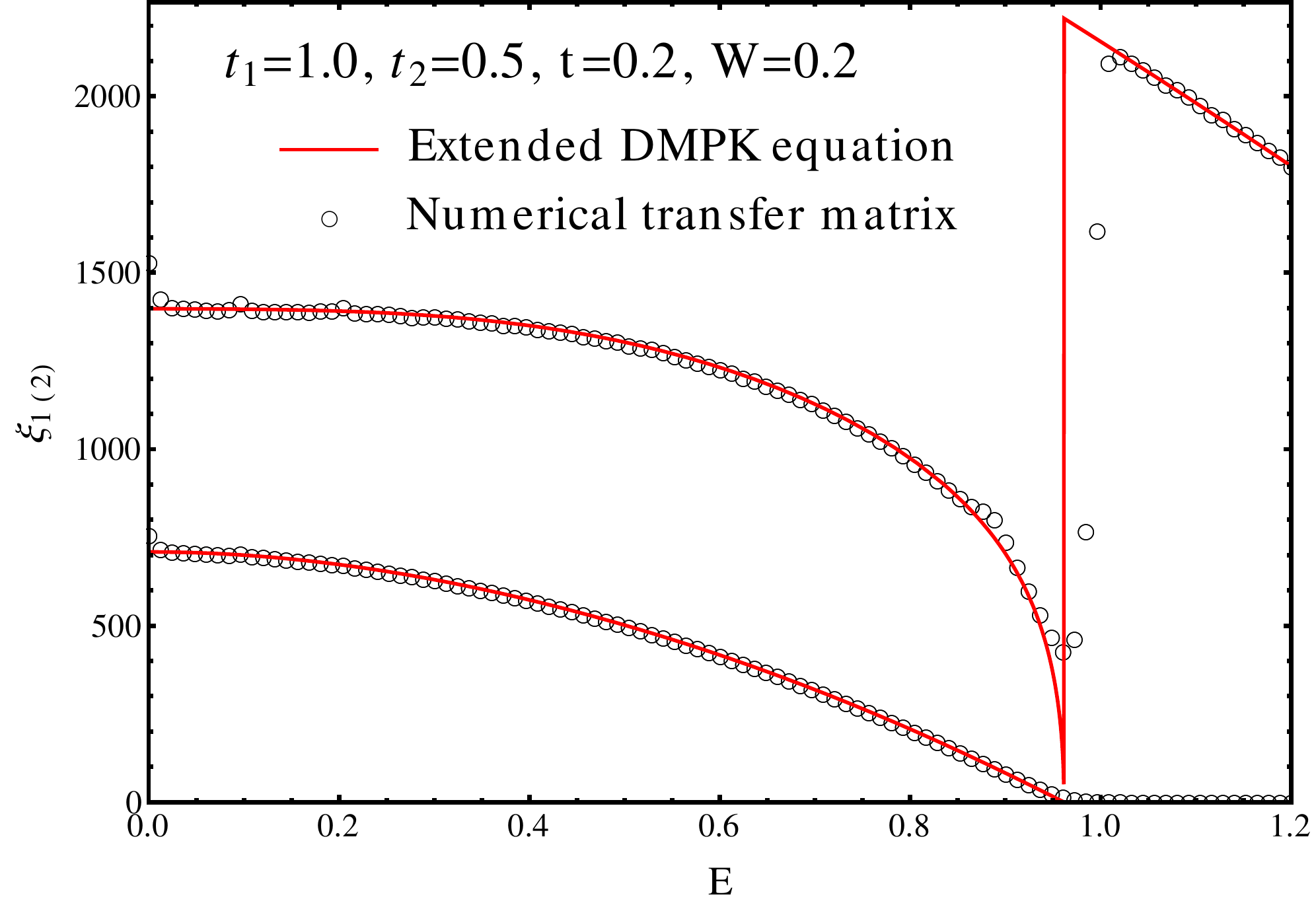}
\caption{Localization lengths as a function of energy near the edge of one of the bands. Here $W=0.2$, $t_1=1$, $t_2=0.5$, $t=0.2$
and $\delta{e}=0$. The red curve is the result of the extended DMPK equation, while the black circles are data obtained by the numerical
transfer matrix method. The quantitative agreement is significant except for four anomalous energies,
e.g. $E=0$ and $E \approx 0.1$, $0.2$ and $0.9$. The correponding commesurate combinations of wave vectors are $2(k_1+k_2)=2\pi$, $3k_1+k_2=2\pi$, $4k_1=2\pi$ and $3k_2-k_1=2\pi$. Near the termination of the lower band at $E<E_2^+$ the larger localization length is dramatically decreased, driven down
by the slow terminating channel. At $E>E_{2}^{+}$ the localization length sharply recovers. In numerical simulation the sharpness is smeared by the finite disorder.}         \label{band-edge}\
%This non-trivial behavior of the analytical solution is qualitatively confirmed by numerical simulations.
% {\bf MM: The difference is both due to the finite disorder strength and  the limits of applicability of our theory
%for $E$ within $O(t)$ from the clean band edges.} }
\end{center}
\end{figure}
Another interesting question to ask is what happens to the
localization lengths around the band-edge $E_1^{-}$ or $E_2^{+}$?
(See Fig.~\ref{disper}(a)) Especially, what is the behavior of the
localization length of the fast-leg once we turn on the coupling
$t$?
The results from the numerical transfer
matrix simulation and of the solution (\ref{locl-2})
of the extended DMPK equation, Eq.~(\ref{fokker-planck-3}), are compared in Fig.~\ref{band-edge}.

Two remarkable features can be observed in Fig.~\ref{band-edge}:

(i) Near the band edge $E=E_2^{+}$ where the system switches from
one to two propagating channels, the larger localization length
$\xi_1$ (red
curves) behaves in a singular way, as obtained from  Eq.~(\ref{locl-2}).
As the energy tends to the
band-edge $E_2^{+}$ from below, $\xi_{1}$ decreases to zero and shows a jump to a finite value for $E>E_2^+$, where only one propagating channel exists.
% and then jumps to a finite value for $E>E_2^+$, where only one channel exists. From
The numerical simulation (black circles) reproduces the same behavior, while the sharp recovering at $E=E_2^+$ is smeared by the finite disorder.
%The analytical description of the one-channel regime will be presented in Sec.~\ref{one-channel}.
%smeared out by {\bf MM: finite $t$}, finite disorder and Lifshitz tails.
This behavior is another drastic example of the dominant
effect of the slow channel.
 It can be understood from the behavior of the
Born cross-sections Eq.~(\ref{born-sec}). As we approach the
band-edge from below, the rapidities of the two channels
satisfy $v_1 \gg v_2$. As a consequence, the cross-sections obey the
hierarchy $V_2 \gg V_3 \gg V_1$. Therefore, from Eq.~(\ref{u-dis})
and (\ref{locl-2}), one can see that $\xi_1$ is dominated by the
largest cross section $V_2$ and shows qualitatively the same
behavior as $\xi_2$. We emphasize that the
mechanism of this suppression is different from that in the
resonant regime. In the latter the suppression is due to
$\gamma \sim \pi/2$, which mixes the two effective variances $\Var$
equally, while near the band-edge the suppression is due to the
vanishing rapidity, which appears in the denominators
of the cross-sections.

 (ii) Anomlies are clearly seen in the numerical data for $\xi_1$ at energies $E = 0$ and $E \approx 0.1$, $0.2$ and $0.9$.
The corresponding commensurate combinations of wave vectors are $2(k_1+k_2)=2\pi$, $3k_1+k_2=2\pi$, $4k_1=2\pi$ and $3k_2-k_1=2\pi$.

\section{one-channel regime} \label{one-channel}
So far we have discussed the localization lengths in the two-channel regime,
where the extended DMPK equation (\ref{fokker-planck-3}) applies. In the one-channel regime
(see Fig.~\ref{disper}) the second channel does not vanish but
supports \emph{evanescent} modes. In the presence of disorder a particle in propagating modes can be scattered elastically
into these evanescent modes by local impurities. Thus the evanescent channel is coupled to the propagating channel
by random potentials and may influence the transport
properties of the system. However, the effect of evanescent modes in the transport properties of
 1D disordered systems
is scarcely studied.

Bagwell\cite{bag} studied in detail the
transmission and reflection coefficients in a multi-channel wire with a single $\delta$-function impurity.
The evanescent modes renormalize the matrix elements of the impurity potential in the propagating channels. The transmission and reflection coefficients of the propagating channels
can be strongly enhanced or suppressed, nevertheless,
depending on the strength of the impurity.

The model in Ref.~\onlinecite{bag} was non-disordered but quite
relevant to disordered systems. It is reasonable to argue that in 1D
disordered systems the  effective disorder in the propagating
channels is renormalized by evanescent modes, while
the renormalization effect depends upon the strength of disorder. 

In the present two-leg Anderson model we specifically analyze the
renormalization effect of the evanescent channel in the \emph{weak
disorder} limit, which stands on an equal footing with the analysis
in the two-channel case. Actually, the special case $t_1 = t_2$ and
$\sigma_1^2 = \sigma_2^2$ has been studied analytically early on in
Ref.~\onlinecite{heinrich}. It was claimed that in the weak disorder
limit the effective disorder in the propagating channel is
significantly suppressed by the evanescent mode. As a consequence, the
localization length defined through the transmission coefficient of
the propagating channel is enhanced by a factor $\sim 2$ compared to
the value obtained if the evanescent mode is absent. However,
this conclusion was unreliable because the average of the logarithm of transmission eigenvalue was not computed correctly. In contrast,
we will prove that the evanescent channel is \emph{decoupled} from
the propagating channel to the lowest order in the effective disorder $\chi_{\nu}^2$ defined in
Eq.~(\ref{eff-var}). The coupling between the two channels
becomes relevant only at order $\chi_{\nu}^4$.

\subsection{Transfer matrix of an elementary slice}  \label{tm-sli-one}
Without loss of generality, we assume that the channel $\tau=1$ is
propagating and $\tau=2$ is evanescent (the upper branch in
Fig.~\ref{disper}). A similar analysis applies to the opposite choice
(the lower branch in Fig.~\ref{disper}). Note first of all that a
direct application of the Fokker-Planck equation approach to the transfer matrix given in Eqs.~(\ref{slice}) and (\ref{tm-l}) would be incorrect.
The reason is the following: The weak disorder expansion of the
parameters $\vec{\lambda}$, which leads to
Eq.~(\ref{fokker-planck-2}), is ill-defined in the one-channel
regime. Note that the amplitude of the evanescent basis $\psi_2(x)$
(see Eq.~(\ref{j-states-ev})) \emph{grows exponentially} $\sim
e^{\kappa_2 |x|}$.
Likewise, the elements of $\delta{\mathbf{m}}_x$ (see
Eq.~(\ref{slice})) with evanescent channel indices also grow
exponentially with factors $e^{2\kappa_2|x| }$ or $e^{4\kappa_2 |x|
}$. Therefore, $\|\delta{\mathbf{m}}_x /\ep \|$ is unbounded in the
domain of the coordinate $x$, and the formal expansion of the parameters
$\vec{\lambda}$ in disorder strength is divergent with
respect to the length $L$.\cite{eva-div}

In order to perform a weak disorder analysis, the basis of the evanescent
channel should be chosen as
\be  \label{states-ev}
\psi_{2}^{\pm}(x) = e^{\mp \kappa_2 x} /\sqrt{2 \sinh{\kappa_2}}, \quad
\kappa_2 > 0,
\ee
which replaces the current-conserving basis Eq.~(\ref{j-states-ev}),
and the basis of the propagating channel is the same as
Eq.~(\ref{j-states prop}) even though with $\tau=1$. In this newly
defined basis, the transfer matrix of elementary slice takes the form:
(see App.~\ref{app-d})
\be  \label{ele-sli-one-ch}
\mathbf{m}_x= \mathbf{m} + \delta{\mathbf{m}}_x,
\ee
with
\be \label{tran-mat-ii}
\mathbf{m} = \text{diag}\left( 1,\,\,1,\,e^{-\kappa_2},\,e^{\kappa_2}\right),
\quad \delta{\mathbf{m}}_x = \begin{pmatrix} \delta m_{1}^{1} & \delta m_{2}^{1}
\\ \delta m_{1}^{2} & \delta m_{2}^{2} \end{pmatrix},
\ee
whose blocks are
\begin{subequations} \label{delta-m-mat}
\be
\delta m_1^1= i \frac{\ep_\text{11}}{2\sin{k_1}} \begin{pmatrix} -1 & -e^{-i 2 k_1 x}
\\ e^{i 2 k_1 x} & 1\end{pmatrix},
\ee
\be
\begin{split}
\delta m_2^1= & i \frac{\ep_\text{12}}{2 \sqrt{\sin{k_1} \sinh{\kappa_2}}} \\ &
\times  \begin{pmatrix} -e^{-\kappa_2}e^{-i k_1 x} & -e^{\kappa_2}e^{-i k_1 x}
\\ e^{-\kappa_2}e^{i k_1 x} & e^{\kappa_2}e^{i k_1 x}  \end{pmatrix},
\end{split}
\ee
\be
\delta m_1^2= \frac{\ep_\text{21}}{2 \sqrt{\sin{k_1} \sinh{\kappa_2}}} \begin{pmatrix}
-e^{i k_1 x} & -e^{-i k_1 x} \\ e^{i k_1 x} & e^{-i k_1 x}  \end{pmatrix},
\ee
\be   \label{dm-2-2}
\delta m_2^2= \frac{\ep_\text{22}}{2\sinh{\kappa_2}} \begin{pmatrix} -e^{-\kappa_2} &
-e^{\kappa_2  }\\ e^{-\kappa_2} & e^{\kappa_2} \end{pmatrix},
\ee
\end{subequations}
where $\mathbf{m}$ and $\delta{\mathbf{m}}_x$ are the disorder-free and disordered part of
the elementary slice $\mathbf{m}_x$. The transfer matrix of a bulk with length $L$ is
still defined by the products in Eq.~(\ref{tm-l}). Two important points should be emphasized:

(i) Compared with Eq.~(\ref{slice}) in the two-channel case, the second and third rows
and columns of Eq.~(\ref{ele-sli-one-ch}) have been simultaneously permuted. The diagonal
blocks $\delta m_1^1$ and $\delta m_2^2$ represent the scattering in the propagating and
evanescent channel, respectively, and the off-diagonal blocks $\delta m_{2(1)}^{1(2)}$
represent the scattering between the two channels. In each block, the first and second
diagonal element describe the scattering inside right- ($+$) and left- ($-$) branch
respectively, and the off-diagonal elements describe the scattering between the two branches.
For instance, $\delta m_{1+}^{2-}$ labels the $21$-element of  $\delta m_{1}^{2}$ and stands
for a scattering event from the left evanescent channel to the right propagating channel.

(ii) The disordered part $\delta{\mathbf{m}}_x$ \emph{does not} contain exponentially
growing and/or decaying terms, and hence $\| \delta{\mathbf{m}}_x/\ep \|$ is uniformly bounded for any $x$. Instead, the disorder-free part $\mathbf{m}$, which is still diagonal
but not unity any more, contains the growing and decaying factor of the evanescent mode
per lattice spacing. The exponentially growing and decaying characteristics of evanescent
modes are represented in the products of the disorder-free part $\prod_{x=1}^{L} \mathbf{m}$.

\subsection{Weak disorder analysis of Lyapunov exponents} \label{wea-dis-exp}
In order to calculate the transmission coefficient of the
propagating channel, through which the localization length is
defined (see Sec.~\ref{loc-eva-len}), we have to know the Lyapunov
exponents of $\mathbf{M}(L)$ in Eq.~(\ref{tm-l}). We are going to
determine the Lyapunov exponents by the method introduced in
Ref.~\onlinecite{bggs-cpv}.

The Lyapunov exponents of the present model can be computed via the following recursive
relations for the four vectors $V_{i=1,\cdots,4}$:
\begin{subequations}   \label{rec}
\be \label{rec-1}
V_{1,x+1} = \mathbf{m}_x V_{1,x},
\ee
\bea \label{rec-2}
V_{i,x+1} = \mathbf{m}_x V_{i,x} -  \sum_{j=1}^{i-1}&&{\frac{ V_{j,x+1} \cdot \left( \mathbf{m}_x V_{i,x} \right)}{V_{j,x+1} \cdot V_{j,x+1}}V_{j,x+1}}, \nn \\ && \hspace{1cm} 2 \le i \le 4.
\eea
\end{subequations}
Note that the vectors are orthogonalized by Gram-Schmidt procedure after every multiplication by 
the transfer matrices (\ref{ele-sli-one-ch}). The Lyapunov exponents are extracted form the growing rate of the amplitudes of the respective vectors:
\be \label{lyap-one-ch}
\gamma_i = \lim_{L \to \infty} \frac{1}{2L} \left\langle \ln{  \frac{\left| V_{i,L} \right|^2 }{\left| V_{i,1} \right|^2}} \right\rangle%= \lim_{L \to \infty} \frac{1}{2L} \sum_{x=1}^{L} \left\langle \ln{  \frac{\left| V_{i,x+1} \right|^2 }{\left| V_{i,x} \right|^2}} \right\rangle
, \quad 1 \le i \le  4,
\ee
in which $\langle \cdot \rangle$ is the average over realizations of disorder along the strip. 
Moreover, $\{\gamma_i\}$ are in descendant order:
\be
\gamma_1 \ge \gamma_2 \ge  \gamma_3 \ge \gamma_4.
\ee
The initial vectors $V_{i,1}$ of the recursive relations (\ref{rec}) can be randomly chosen but must be \emph{linearly independent}.
In the absence of specific symmetry constraints the Lyapunov exponents are non-degenerate in the presence of disordered part of $\mathbf{m}_x$. Additionally, because of the symplecticity of $\td{\mathbf{m}}_x$ represented in Eq.~(\ref{sympl-cond}) the Lyapunov exponents are related by
\be \label{sym-lya-one}
\gamma_3 = -\gamma_2, \quad \gamma_4 = -\gamma_1,
\ee
which is proved in App.~\ref{app-d}. Therefore, only the first two recursions in Eq.~(\ref{rec}) are needed.

In the absence of disorder the four Lyapunov exponents take the values:
\be
\gamma_1|_{\ep=0}=\kappa_2, \quad \gamma_2|_{\ep=0}=\gamma_3|_{\ep=0}=0, \quad
\gamma_4|_{\ep=0}=-\kappa_2,
\ee
in which the two Lyapunov exponents corresponding to the propagating channel are \emph{degenerate}. Therefore, we make an \emph{ansatz} on the first two vectors, which separates their ``moduli'' and ``directions'',
\be \label{ansa}
V_{1,x} = v_{1,x} \begin{pmatrix} s_1(x) \\ s_2(x) \\ s_3(x) \\ 1 \end{pmatrix}, \quad V_{2,x} = v_{2,x} \begin{pmatrix} p(x) \\ q(x) \\ t_3(x)  \\ t_4(x) \end{pmatrix},
\ee
in which
\be  \label{small-s-t}
s_{1,2,3}(x), t_{3,4}(x) \sim O(\ep),
\ee
$|s_{1,2,3} (x)/\ep|$ and  $|t_{3,4}(x)/\ep|$ are bounded for all $x$, and
\be
|p(x)|^2 + |q(x)|^2 =1.
\ee
Eventually, the Lyapunov exponents are determined by the growth rate of $\{v_{i,x}\}$, which is easy to be realized from Eqs.~(\ref{lyap-one-ch}) and (\ref{ansa}). The initial vectors of Eq.~(\ref{ansa}) are chosen as the eigenvectors of the disorder-free part of the transfer matrix $\mathbf{m}$ (see Eq.~(\ref{tran-mat-ii})):
\be  \label{ini-rec}
V_{1,1} = \begin{pmatrix} 0 \\ 0 \\ 0 \\ 1 \end{pmatrix}, \quad V_{2,1} = \begin{pmatrix} p(1) \\ q(1) \\ 0  \\ 0 \end{pmatrix},
\ee
with some $p(1)$ and $q(1)$ satisfying $|p(1)|^2 + |q(1)|^2 =1$.

Note that the ansatz (\ref{ansa}) is reasonable in the sense of a perturbative analysis. 
Consider the final vectors after $L$ iterations of Eq.~(\ref{rec}) with the initial condition Eq.~(\ref{ini-rec}). 
In the absence of disorder, it is easy to obtain $V_{1,L}=e^{\kappa_2 L}V_{1,1}$ and $V_{2,L}=V_{2,1}$. On top of it weak enough 
disorder will induce perturbative effects: the direction of $V_{1,L}$ will deviate from $V_{1,1}$ \emph{perturbatively} in
 the strength of disorder. This is characterized by the smallness of $s_{1,2,3}(L)$. In other words, 
the exponential growth of $|V_{1,L}|$ is dominated by $\mathbf{m}$. Simultaneously, 
the degeneracy of the second and third exponents are lifted perturbatively. 
As a consequence, $\gamma_2 > 0$ and $v_{2,L}$ become exponentially large because of the constraint in Eq.~(\ref{sym-lya-one}). 
$p(L)$ and $q(L)$ are in general very different from their initial values $p(1)$ and $q(1)$, 
while $t_{3,4}(L)$ will be shown to remain small quantities of order $\ep$.

The orthogonality between $V_{i,x}$ in Eq.~(\ref{ansa}) gives
\be \label{orth}
t_4(x)+s_1^\ast(x) p(x) + s_2^\ast(x) q(x) + s_3^\ast(x)t_3(x) =0,
\ee
in which the first three terms $\sim O(\ep)$ and the last term $\sim O(\ep^2)$. Up to the first order in disorder strength, the recursion (\ref{rec-1}) gives
\begin{subequations}   \label{com-rec-1}
\be \label{rec-eva-ch}
 v_{1,x+1} = v_{1,x} \left[ e^{\kappa_2} + \delta{m}_{2-}^{2-} + O(\ep^2) \right],
\ee
\be
v_{1,x+1} s_1(x+1) = v_{1,x} \left[ s_1(x) + \delta{m}_{2-}^{1+} + O(\ep^2) \right],
\ee
\be
v_{1,x+1} s_2(x+1) = v_{1,x} \left[ s_2(x) + \delta{m}_{2-}^{1-} + O(\ep^2) \right],
\ee
\be
v_{1,x+1} s_3(x+1) = v_{1,x} \left[ e^{-\kappa_2} s_3(x) + \delta{m}_{2+}^{2-} + O(\ep^2) \right].
\ee
\end{subequations}
The recursion (\ref{rec-2}) gives
\begin{subequations}  \label{com-rec-2}
\be  \label{rec-prop-ch}
v_{2,x+1} \begin{pmatrix} p(x+1) \\ q(x+1) \end{pmatrix} = v_{2,x} \left[ (1 + \delta{m}_{1}^{1})\begin{pmatrix} p(x) \\ q(x) \end{pmatrix} + O(\ep^2) \right],
\ee
\be
\begin{split}
& v_{2,x+1}  t_3(x+1) \\ \, & =  v_{2,x} \left[ e^{-\kappa_2} t_3(x) + \delta{m}_{1+}^{2+} p(x)+ \delta{m}_{1-}^{2+} q(x) +O(\ep^2) \right],
\end{split}
\ee
\be
\begin{split}
& v_{2,x+1} t_4(x+1) \\ \, & = -v_{2,x} \left[ s_1^\ast(x+1) p(x)+ s_2^\ast(x+1) q(x) + O(\ep^2) \right].
\end{split}
\ee
\end{subequations}
It can be verified that the higher order terms $\sim O(\ep^2)$ \emph{do not} involve exponentially growing factors, which is guaranteed by the Gram-Schmidt re-orthogonalization procedure in the recursive relations (\ref{rec}).

We draw two important observations from Eqs.~(\ref{com-rec-1}) and (\ref{com-rec-2}):

(i) The ansatz (\ref{ansa}) is consistent with the perturbative expansion of the recursions (\ref{rec}). 
Here the consistency means that $|s_{1,2,3}(x)/\ep|$ and $|t_{3,4}(x)/\ep|$ are uniformly bounded after any number of iterations,
 and the first two Lyapunov exponents can be extracted from $v_{j,x}$.

(ii) Up to linear order in disorder strength, the recursion (\ref{rec-eva-ch}), 
which determines the first Lyapunov exponent $\gamma_1$, is decoupled from the recursion relation (\ref{rec-prop-ch}),
 which determines the second Lyapunov exponent $\gamma_2$. 
However, the coupling terms are present in higher order terms. This implies that to the leading order effect 
in disorder the evanescent and propagating channels evolve independently, the entanglement between the two channels being a higher order effect.

From Eq.~(\ref{rec-eva-ch}) one can easily calculate the first Lyapunov exponent to linear order in the effective variances $\chi_\nu^2$,
\be  \label{1st-one-ch}
\begin{split}
\gamma_1 & = \lim_{L\to \infty} { \frac{1}{L} \left\langle \ln{\prod_{x=1}^{L}{\left| e^{\kappa_2} + \delta{m}_{2-}^{2-}(x) \right|}} \right\rangle}\\
         & = \kappa_2 + \left\langle \ln{\left| 1 +  \frac{\ep_\text{22}}{2\sinh{\kappa_2}} \right|} \right\rangle\\
         & \simeq \kappa_2 - \frac{1}{8 \sinh^2{\kappa_2}}\left( \Varo\sin^4{\fr{\g}{2}}+\Vart\cos^4{\fr{\g}{2}} \right) + O(\chi_\nu^4),
\end{split}
\ee
in which $\gamma$ is the mixing angle defined in Eq.~(\ref{clean-angle}).
 The minus sign of the leading order corrections implies that the first Lyapunov exponent is \emph{reduced} in the presence of weak disorder.

Eq.~(\ref{rec-prop-ch}) is exactly the same as in a single chain Anderson model, 
for which the Lyapunov exponents are already known.\cite{mel} The second Lyapunov exponent takes the value
\be \label{2nd-one-ch}
\gamma_2 \simeq 2V_{1} + O(\chi_\nu^4),
\ee
where $V_{1}$ is the Born cross-section given in Eq.~(\ref{born-sec}).

% Alternatively, if the channel $\tau=2$ is propagating and $\tau=1$ is evanescent (the lower branch in Fig.~\ref{disper}) the same analysis leads to
%\be   \label{12-one-ch}
%\begin{split}
%\gamma_1 &\simeq \kappa_1 - \frac{1}{8 \sinh^2{\kappa_1}}\left( \Varo\cos^4{\fr{\g}{2}}+\Vart\sin^4{\fr{\g}{2}} \right) + O(\chi_\nu^4),\\
%\gamma_2 &\simeq 2V_{2} + O(\chi_\nu^4). \\
%\end{split}
%\ee
Eqs.~(\ref{1st-one-ch}) and (\ref{2nd-one-ch}) are our main results for the one-channel case, yielding the localization length and the renormalized decay rate of evanescent waves.

\subsection{Localization length and evanescent decay rate} \label{loc-eva-len}
The two Lyapunov exponents calculated above can be identified in
transport experiments. In general a two-probe experiment has the
geometry of the form ``lead--sample--lead'', in which the two leads
are semi-infinite. The current amplitudes (not the
wave amplitudes) are measured in leads. In the propagating channels both
right ($+$) or left ($-$) modes exist in both of the leads. However, the situation is rather different in the
evanescent channels: There are only growing modes ($-$) in the
left lead, and only decaying modes ($+$) in the right lead. These
modes \emph{do not} carry current at all.\cite{bag,dav} Hence the
current transmission and reflection coefficients are only defined in
propagating channels regardless of the wave amplitudes in evanescent
channels. In terms of the transfer matrix $\mathbf{M}(L)$, this
restriction on the evanescent channel implies that 
\be
\label{bound-cond}
\begin{pmatrix} a_1^{+}(L) \\ a_1^{-}(L) \\ a_2^{+}(L) \\ 0  \end{pmatrix} = \begin{pmatrix} M_\text{1}^\text{1} & M_\text{2}^\text{1} \\ M_\text{1}^\text{2} & M_\text{2}^\text{2} \end{pmatrix} \begin{pmatrix} a_1^{+}(1) \\ a_1^{-}(1) \\ 0 \\ a_2^{-}(1)   \end{pmatrix}.
\ee

From the scattering configuration (\ref{bound-cond}) one can derive 
an effective transfer matrix for the propagating channel. 
The evanescent amplitude $a_2^{-}(1)$ can be expressed in terms of the propagating amplitudes as
\be  \label{ev-amp}
a_2^{-}(1) = -\frac{1}{ M_{\text{2}-}^{\text{2}-}} \left[ M_{\text{1} +}^{\text{2}-} a_1^{+}(1)+M_{\text{1}-}^{\text{2}-} a_1^{-}(1) \right].
\ee
Substituting Eq.~(\ref{ev-amp}) into Eq.~(\ref{bound-cond}) we obtain
\be  \label{red-trans-mat}
 \begin{pmatrix} a_1^{+}(L) \\ a_1^{-}(L) \end{pmatrix} = \mathbf{X}(L) \begin{pmatrix} a_1^{+}(1) \\ a_1^{-}(1) \end{pmatrix},% \quad \mathbf{X}(L)= \begin{pmatrix} X_{+}^{+} & X_{-}^{+} \\ X_{+}^{-} & X_{-}^{-} \end{pmatrix},
\ee in which the elements of $\mathbf{X}(L)$ take the form:
\begin{subequations}  \label{eff-tm-prop}
\be  \label{ele-pp}
X_{+}^{+} = M_{\text{1} +}^{\text{1}+}+\Delta{M_{\text{1} +}^{\text{1}+}}, \quad  \Delta{M_{\text{1} +}^{\text{1}+}}=
- \frac{M_{\text{2}-}^{\text{1} +} M_{\text{1} +}^{\text{2}-}}
{M_{\text{2}-}^{\text{2}-}},
\ee
\be
X_{-}^{+} = M_{\text{1}-}^{\text{1}+} + \Delta{M_{\text{1} -}^{\text{1}+}}, \quad \Delta{M_{\text{1} -}^{\text{1}+}}=- \frac{M_{\text{2}-}^{\text{1} +} M_{\text{1}-}^{\text{2}-}}{M_{\text{2}-}^{\text{2}-}},
\ee
\be X_{+}^{-} =X_{-}^{+ \ast}, \quad X_{-}^{-} = X_{+}^{+\ast}. \ee
\end{subequations}
$\mathbf{X}(L)$ is the effective transfer matrix for the propagating channel. 
Note that its elements are modified form the values in the absence of the evanescent channel. 
One can easily verify that $\mathbf{X}(L)$ satisfies time-reversal invariance and current conservation conditions as (\ref{requ-symp}) in the single chain case:\cite{mpk1,mpk2}
\be
\mathbf{X}^\ast = \sigma_1 \mathbf{X} \sigma_1, \quad \mathbf{X}^\dagger \sigma_3 \mathbf{X} = \sigma_3.
\ee
However, $\mathbf{X}(L)$ does not evolve multiplicatively with the length $L$ any more.  The transmission coefficient is determined through $\mathbf{X}(L)$ in the usual way\cite{mpk1,mpk2}
\be   \label{tra-coe-one}
T(L) =|X_{+}^{+}|^{-2},
\ee
where $X_{+}^{+}$ is defined in Eq.~(\ref{ele-pp}).

Eqs.~(\ref{eff-tm-prop}) and (\ref{tra-coe-one}) exactly determine the transmission 
coefficient of the propagating channel. A full solution requires extensive calculations.
However, if the disorder strength is weak, as analyzed in Sec.~\ref{wea-dis-exp}, the coupling between the two channels 
%described by $M_{2}^{1}$ and $M_{1}^{2}$ can be omitted. 
is small, so that the contribution of the evanecent channel, $\Delta M_{1+}^{1+}$ is negligible. %in Eq.~(\ref{ele-pp}). 
Indeed, from Eqs.~(\ref{rec}) and (\ref{ansa}), using initial vectors $V_{1,1}=\begin{pmatrix} 0 & 0 & 0 & 1 \end{pmatrix}^T$ 
and $V_{2,1}=\begin{pmatrix} 1 & 0 & 0 & 0 \end{pmatrix}^T$, respectively, we can extract the various matrix elements of $\mathbf{M}(L)$, in particular 
\be
\frac{\Delta  M_{\text{1} +}^{\text{1}+}(L) }{  M_{\text{1} +}^{\text{1}+}(L) }= \frac{s_1(L) t_4(L)}{p(L)} \sim O(\ep^2).
\ee  
This proves that the contribution of the evanescent channel is subleading at weak disorder.
To leading order the transmission coefficient is simply given by the propagating channel as
\be
T(L) \simeq |M_{\text{1} +}^{\text{1}+}|^{-2} \simeq |v_{2,L} \,p(L)|^{-2}.
\ee
From this the localization length is obtained,
\be \label{loc-len-one-ch}
\begin{split}
1/\xi & = -\lim_{L \to \infty} \frac{1}{2L}\left\langle \ln{T(L)} \right\rangle  \simeq 2V_1 + O(\chi_\nu^4)\\
& = \fr{1}{8 \sin^2{k_1}} \left( \Varo \cos^4{\fr{\g}{2}}+\Vart \sin^4{\frac{\g}{2}} \right) + O(\chi_\nu^4)
\end{split}
\ee
Eq.~(\ref{loc-len-one-ch}) implies that to leading order in $\chi_\nu^2$ the 
localization length in the propagating channel equals the inverse of the second Lyapunov exponent obtained 
in Eq.~(\ref{2nd-one-ch}). 

Similarly to Eq.~(\ref{ratio-def}), we can introduce the localization length enhancement factor
\be  \label{enh-fac-one}
r = \xi / \xi_1^{(0)},
\ee
in which $\xi_1^{(0)}$ is the localization length of the leg 1 (with the larger hopping) in the absence of inter-chain coupling.

On the other hand, the inverse of the first Lyapunov exponent
in Eq.~(\ref{1st-one-ch}) should be associated with the evanescent decay rate which is slightly modified by disorder.

\begin{figure}[t]
\begin{center}
\includegraphics[height=5.75cm,width=8.2cm]{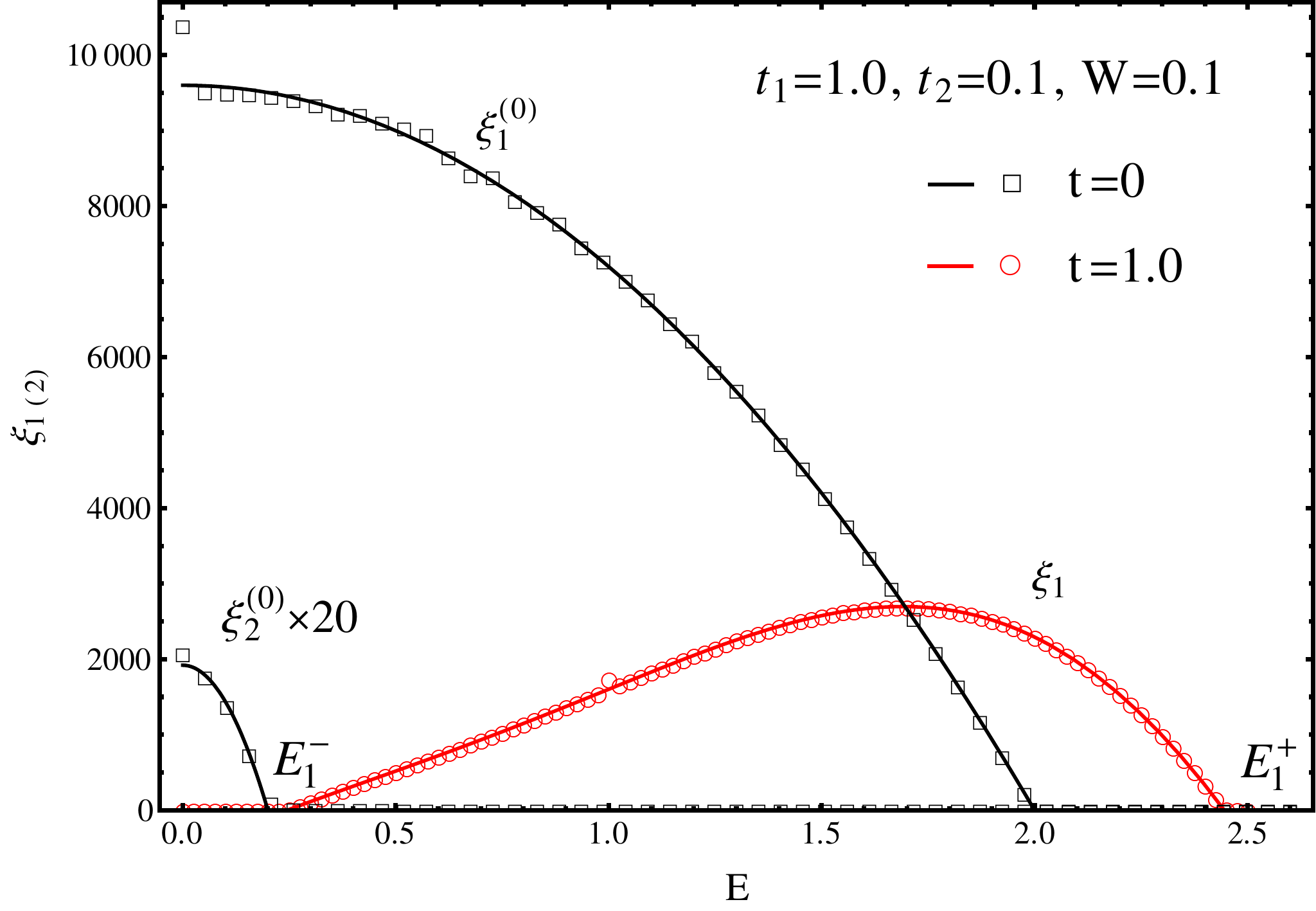}
\caption{Localization length as a function of energy in the
one-channel case. Here $t_1=1$, $t_2=0.1$, $\delta{e}=0$ and the amplitude of disorder is $W=0.1$.
The solid curves are analytical results. Black curves correspond to uncoupled chains.
The red one corresponds to the upper polariton (conduction) band (propagating channel) for strong coupling $t=1$,
which is obtained by omitting the lower polariton (valence) band (evanescent channel).
The squares and circles are data of the numerical transfer matrix.
The quantitative agreement is significant except for the anomalous energy $E \simeq 1.0$,
which correspands to the commesurate combination of wave vectors $4k_1=2\pi$. The coincidence between analytics and numerics confirms that the evanescent channel is
decoupled to the propagating channel in weak disorder limit.
} \label{sig-chanl}\
\end{center}
\end{figure}

The analytical results (\ref{loc-len-one-ch}) and/or
(\ref{enh-fac-one}) are compared with numerics in
Figs.~\ref{interm-coupling}, \ref{band-edge} and \ref{sig-chanl}.
Figs.~\ref{interm-coupling} and \ref{band-edge} correspond to
the weak coupling case $t<t_c$ (see Fig.~\ref{disper}(a)) and
Fig.~\ref{sig-chanl} to the strong coupling case $t >t_c$ (see
Fig.~\ref{disper}(b)). The remarkable agreement confirms the weak
disorder analysis developed in this section.

We specifically analyze the typical behavior of the enhancement
factor $r_1(E)$ close to the band-edge $E_2^{+}$ in the case of $t <
t_c$, where the system switch from one to two propagating channels.
From Eqs.~(\ref{clean-angle}) and (\ref{rapidity}) it is not hard to
obtain: at the band edge $E_2^+$, when coupling is weak $r_1(E_2^+)$ deviates from $1$ like
\be 1-r_1(E_2^+) \propto \left(\frac{t}{E_2^+}\right)^2. \ee
If $E$ is away from $E_2^+$,
$r_1(E)$ increases linearly, i.e., \be r_1(E)-r_1(E_2^+)
\propto t^2 (E-E_2^+), \ee 
with a fixed but weak coupling $t$. A
typical curve for $r_1(E)$ is shown in the upper right insert in
Fig.~\ref{interm-coupling}.

%%%%%%%%%%%%%%%%%%%%%%%%%%%%%%%%%%%%%%%%%%%%%%%%%%%%%
\section{Shape and polarization of the wavefunctions}  \label{wavefuntions}

In certain applications, such as exciton-polaritons, the two
linearly coupled types of excitations (represented by the two
chains) are very different in nature. This makes it in principle
possible to probe the original excitations separately from each
other. For a two-leg atomic chain one can imagine probing the
amplitude of wave function on each of the spatially  separated legs.
For polaritons the analogue would be a separate probing  of cavity
photons or excitons, e.g. by studying the 3D light emitted due to
diffraction of cavity photons at surface roughnesses or by studying
the exciton annihilation radiation or the electric current of exciton
decomposition provoked locally. Therefore it is of practical
interest to be able to manipulate the strength of localization of
one of the original excitations by coupling them to the other.

\subsection{Numerical analysis}
With this goal in mind we have carried out a numerical study of the amplitude of wave
functions on either of the chains in each of the distinct parameter regimes
discussed above.
We numerically diagonalized the
Hamiltonian (\ref{full-ham}), cf. Fig.~\ref{wavefun}, choosing $t_1=1$, $t_2=0.2$, $W=0.4$, $t=0.04$. The
length of the ladder was taken to be $L=10^3$ and periodic 
boundary conditions were used. With the above parameters the localization lengths of the
decoupled legs were of the order of $\xi_1^{(0)}\sim 10^3$ and $\xi_2^{(0)} \sim 10$, for energies close to the band-center.
In Fig.~\ref{wavefun} the black curves depict the amplitudes of eigenfunctions on the fast leg $1$, while the
red curves show the corresponding amplitudes on the slow leg $2$.

Our main findings are:

(i)  $E_2^{+}< E< E_1^{+}$: The energy is far from resonance, and
only \emph{one channel} exists. As shown in Fig.~\ref{wavefun}(a),
most of the weight is on the fast leg. The amplitude on the slow
leg is small but the spatial extension of the component $\psi_2$ is
the same as that of $\psi_1$ on the fast leg, which is almost
unaffected by the chain coupling. Thus the coupling can create a
nonzero amplitude on the chain 2, in the energy region where the
decoupled chain 2 cannot support any excitations. The spacial
extension is controlled by
the localization properties of the leg 1.\\
 (ii)  $t/\kappa(t_1,t_2)<E<
E_2^{+}$: The energy is in the  \emph{two-channel, off-resonant
regime}. The wavefunction components $\psi_1$ and $\psi_2$ are
characterized by {\it both} localization lengths $\xi_{1}$ and
$\xi_{2}$. However, the relative weights of the parts of the
wavefunction with the smaller and the larger localization lengths
fluctuate very strongly from eigenstate to eigenstate. This is shown
in Fig.~\ref{wavefun}(b) and (c), with two \emph{adjacent} energy
levels, which were properly selected. In Fig.~\ref{wavefun}(b),
$\psi_2$ consists almost entirely of a component with the smaller
localization length, while the fast leg clearly shows contributions
of both components. In Fig.~\ref{wavefun}(c), both $\psi_1$ and
$\psi_2$ consist almost entirely of a component with the larger
localization length. In brief, the former can be thought of as a
state on leg 2, which weakly admixes some more delocalized states on
leg 1, while the latter wavefunction is essentially a state of leg 1
which admixes several more strongly localized states on leg 2.

We have checked in specific cases that this interpretation is indeed
consistent (see Sec.~\ref{per-ana}): In the off-resonant regime the
wavefunctions can be obtained perturbatively in the coupling $t$,
confirming the picture of one-leg wavefunctions with small
admixtures of wavefunctions on the other leg. Off resonance, the
perturbation theory is controlled even for appreciable $t$, since
the matrix elements that couple wavefunctions of similar energy are
very small due to significant cancellations arising from the
mismatched oscillations of the wavefunctions ($k_1-k_2>
\xi_2^{(0)}$) on the two legs. Resonance occurs precisely when  at a
fixed energy $k_1-k_2$ becomes too small, so that the modes on both
legs start to mix strongly. A closer analysis of the perturbation
theory in special cases shows that the perturbative expansion is
expected to break down at the resonant crossover determined further
above.

(iii) $|E|< t /\kappa(t_1,t_2)$: If the energy is in the \emph{resonant regime}, the two
localization lengths are of the same order $\xi_{1}\sim \xi_{2}\sim
\xi^{(0)}_{2}$ and the spatial extension of both wave function components is
governed by the localization length $\xi_{2}^{(0)}$ of the decoupled slow
chain. This is illustrated in Fig.~\ref{wavefun}(d).

\begin{widetext}
\
\begin{figure}[t]
\begin{center}
\includegraphics[height=10.5cm,width=8cm]{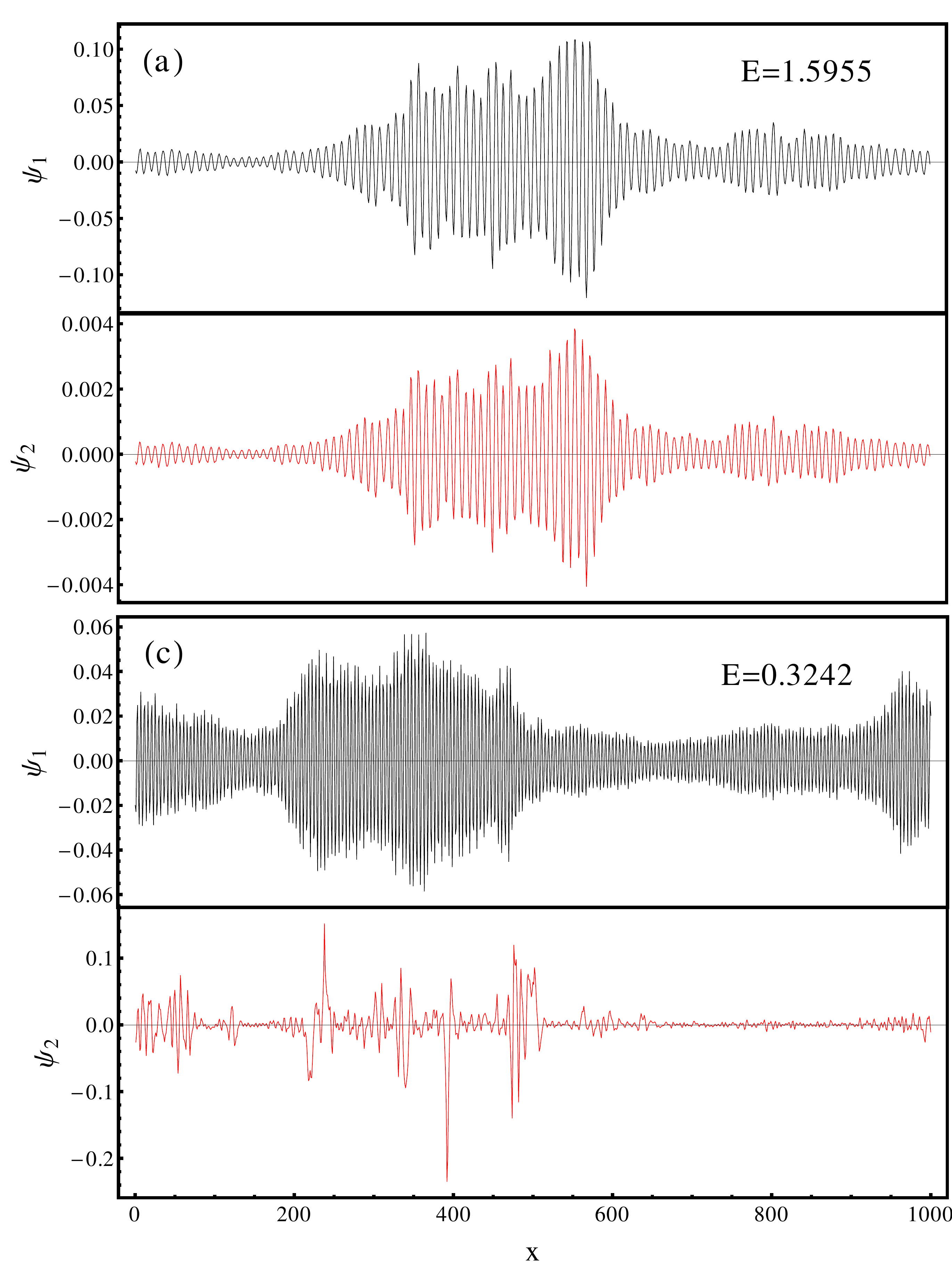}
\includegraphics[height=11cm,width=8.3cm]{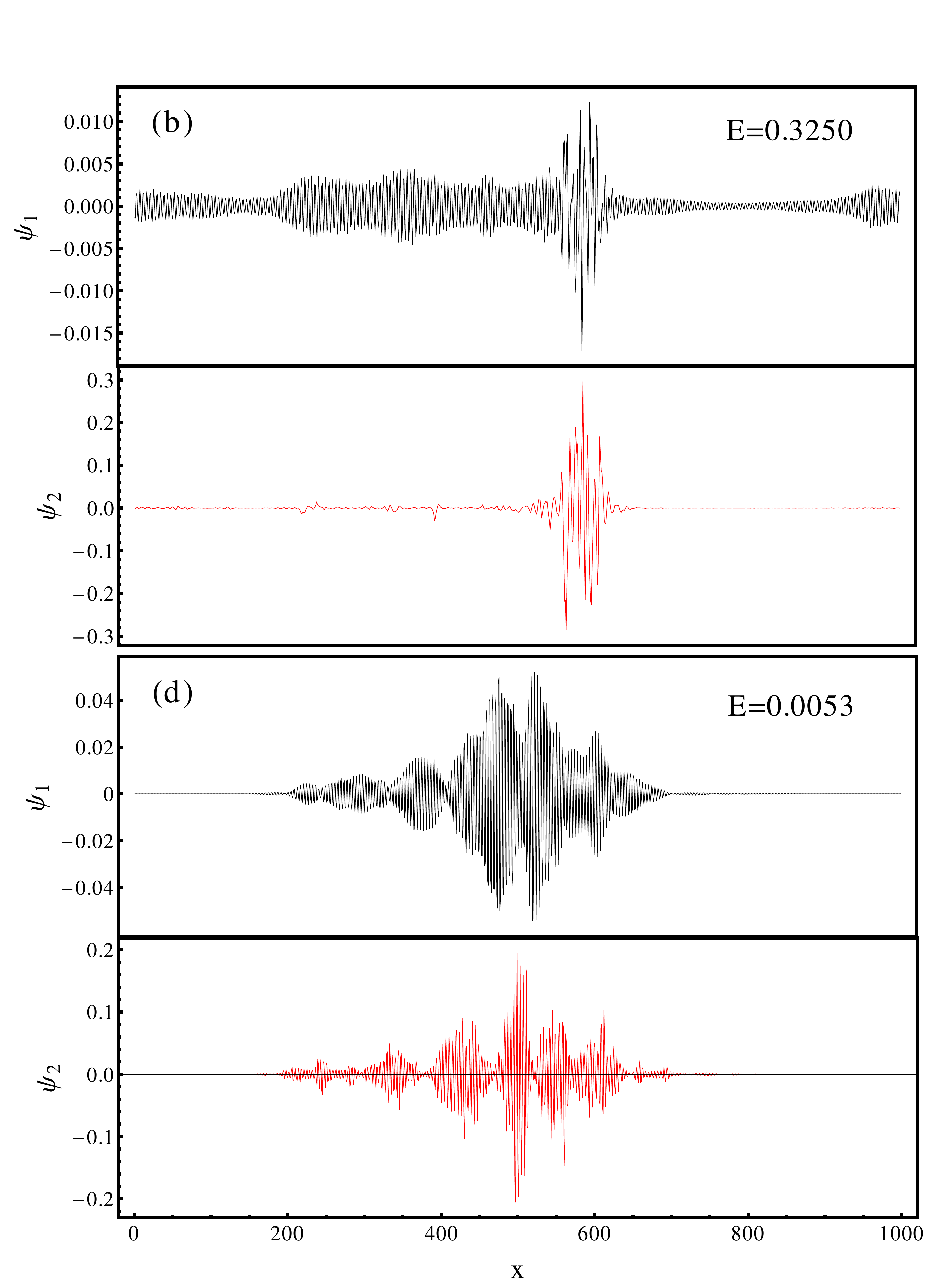}
\caption{Typical wavefunctions in different regimes. Pay attention
to the varying scales for the amplitudes in the various cases.
Parameters are $t_1=1$, $t_2=0.2$, $W=0.4$, $t=0.04$, while $E$
selects the regime. The length of the ladder is $L=10^3$ with
periodic boundary condition. The black(red) curves are the
amplitudes on the 1(2)-leg. (a) $E=1.5955$, at which only one
channel exist. (b) and (c) off-resonant regime: $E=0.3250$ and
$E=0.3242$ are a pair of adjacent levels of "opposite type". (d) $E
\simeq 0$, which is within the resonant regime.} \label{wavefun}\
\end{center}
\end{figure}
\
\end{widetext}
%Actually, the extension of amplitude on slow-leg is determined by
%the weight on fast-leg. If the amplitude on fast-leg is small, on
%slow leg a strongly localized state is weakly coupled to the others.
%This gives Fig.~\ref{wavefun}(b). Oppositely, if the amplitude on
%fast-leg is large, a localized state on slow leg is more likely
%coupled to the other localized states which is far from its
%localization radius, but the amplitude has to be substantially
%reduced. This corresponds to Fig.~\ref{wavefun}(c). This fluctuation
%might be observable. For example if we consider the conductance of
%the ladder with length $\xi_2<L<\xi_1$, we expect the ladder will be
%either metallic or insulating within a small energy variation.
\subsection{Perturbative analysis} \label{per-ana}
The properties of eigenstates at different energy regimes can
be explained by applying a perturbative analysis on the coupling
$t$. First, we define the relevant quantities of decoupled legs as
follows: the eigenstate of the $\nu$-leg with eigenenergy
$E_{\nu n}$ is $\psi_{\nu n}(x)$. The corresponding localization
length is $\xi_\nu^{(0)}$, where we assume $\xi_1^{(0)} \gg \xi_2^{(0)}$ in order to
reveal the resonance--off-resonance crossover. The mean level
spacing inside the localization volume is $\Delta_\nu$. Because in
one dimension a particle is nearly ballistic in its localization
volume, which means its wavevector is nearly conserved and its
amplitude is almost uniform, we introduce a simple ``box" approximation on
the eigenstates as the following: Inside the localization volume,
\begin{equation}   \label{box-wavf}
\psi_{\nu n}(x) \sim \frac{1}{\sqrt{\xi_\nu^{(0)}}}e^{i k_\nu x},
\end{equation}
up to a random phase, in which $\xi_\nu^{(0)}$ and $k_\tau$ are the
localization length and the wavevector at the energy $E_{\nu n}$.
Outside the localization volume $\psi_{\nu n}(x)=0$.

Now we turn on a weak enough coupling $t$ and calculate the
deviation of an energy level $E_{1n}$ on the $1$-leg. Up to second
order in $t$, the deviation is
\begin{equation}
\delta{E}_{1 n}^{(2)} = t^2 \sum_{m}{\frac{\left| \int{ dx
\psi_{1n}^{\ast}(x)\psi_{2m}(x)} \right|^2}{E_{1n}-E_{2m}}}.
\label{sec-dev}
\end{equation}
In order to estimate the value of $\delta{E}_{1 n}^{(2)}$ by the
r.h.s. of Eq.~(\ref{sec-dev}) we have to make clear three points:

(i) The summation is dominated by the terms with the smallest
denominators, whose typical value is the mean level spacing
$\Delta_2$.

(ii) The typical value of the integral on the numerator can
be estimated by the ``box'' approximation introduced above, which gives
\begin{eqnarray}
\int{dx \psi_{1n}^{\ast}\psi_{2m}} &\sim& \int_{0}^{\xi_2^{(0)}}{dx\psi_{1n}^{\ast}\psi_{2m}} \nonumber \\
                                         &\sim& \left[ \left(k_1-k_2\right)\sqrt{\xi_1^{(0)}\xi_2^{(0)}}\right]^{-1}.
\end{eqnarray}

(iii) We should consider more carefully how many dominant terms there are in the summation. We can easily realize that a state
$\psi_{1n}(x)$ on the $1$-leg can couple to about
$\xi_1^{(0)}/\xi_2^{(0)}$ states $\psi_{2m}(x)$ on the $2$-leg.
However, the value of the summation is different from a naive
deterministic evaluation because the random signs of the
denominators. If we neglect the correlation of these random signs,
according to the central limit theorem, the fluctuation of
$\delta{E}_{1 n}^{(2)}$ is
\begin{equation}
\left| \delta{E}_{1 n}^{(2)} \right| \sim
\sqrt{\frac{\xi_1^{(0)}}{\xi_2^{(0)}}} \times
\frac{t^2}{\xi_1^{(0)}\xi_2^{(0)}(k_1-k_2)^2\Delta_2} .
\label{sec-dev-est}
\end{equation}
The validity of the perturbation analysis is guaranteed if
\begin{equation}
\left| \delta{E}_{1 n}^{(2)} \right| < \Delta_1 ,
\label{mix-citr}
\end{equation}
which means there is no level-crossing in the localization volume of
the $1$-leg. To estimate the relevant quantities in Eq.
(\ref{mix-citr}), for simplicity we assume $t_1 \gg t_2$
and $\s_1^2=\s_2^2$.
If the energy $E=E_{1n}$ is close to the resonant energy
$E_R$, according to Eq.~(\ref{two-bands}) and (\ref{two-ch}), we
obtain
\begin{equation}   \label{deltak}
\left| k_1-k_2 \right| \sim \left| E-E_R \right|(t_1-t_2)/t_1t_2.
\end{equation}
The mean level spacings are
\begin{equation}  \label{mls}
\Delta_\nu \sim t_\nu/\xi_\nu^{(0)},
\end{equation}
and the localization lengths satisfy
\begin{equation}  \label{ratio-0}
\xi_1^{(0)}/\xi_2^{(0)} \sim t_1^2/t_2^2.
\end{equation}
Substituting Eqs.~(\ref{deltak}), (\ref{mls}) and (\ref{ratio-0}) to
Eq.~(\ref{mix-citr}) we obtain the condition
\begin{equation}   \label{}
t < \left| E-E_R \right|(t_1-t_2)/t_1,
\end{equation}
which is consistent with the result of Eq.~(\ref{lin-res}) with $t_1
\gg t_2$. Therefore, Eq.~(\ref{mix-citr}) is essentially equivalent to the criterion for being off-resonant ($\Delta{V} >0$) at weak
coupling $t$.

\section{Limit of vanishing hopping on the ``slow'' leg}
In the present work we are particularly interested in the case where the localization lengths of the uncoupled legs 
are parametrically different $\xi_1^{(0)} \gg \xi_2^{(0)}$. Accordingly we refer to the two legs as the ``fast'' and the ``slow'' one, respectively.
So far we have analyzed the model extensively in the limit where the disorder is weak on \emph{both} legs and thus $\xi_\nu^{(0)} \gg 1$. 

Another interesting situation is the case where the hopping strength on the slow leg vanishes $t_2=0$, or is weak enough. This is experimentally relevant for polariton systems in which the exciton hopping is weak as compared to the disorder potential. 
In this case the dimensionless disorder parameter which we introduced previously diverges $\chi_2^2 \to \infty$,  
and formally $\xi_2^{(0)} = 0$ even if the disorder strength on this second leg is arbitrarily small. For this reason the perturbative analysis in both of $\chi_\nu^2$ breaks down. Nevertheless, this limiting case can be solved exactly, too, but requires a different treatment which goes beyond the previous weak disorder analysis.

If $t_2=0$, the second leg is composed of mutually non-connected sites, which form a comb structure together with the first leg. The Schr\"odinger equation (\ref{sch-equ-1}) takes the form
\be   \label{sch-equ-2}
\begin{split}
\begin{pmatrix} -t_1 & 0 \\ 0 & 0 \end{pmatrix} & \left[ \Psi(x+1)+\Psi(x-1) \right] \\  
&= \begin{pmatrix} E - \ep_{x1} & t \\ t & E -\delta{e} - \ep_{x2} \end{pmatrix} \Psi(x),
\end{split}
\ee
where %$\Psi(x)$ has two components 
\be
\Psi(x)= \begin{pmatrix} \psi_{1}(x) \\ \psi_{2}(x) \end{pmatrix},
\ee
describes the amplitudes on the two legs, respectively. 
We obtain the effective Schr\"odinger equation for $\psi_1(x)$ by eliminating $\psi_2(x)$ in Eq.~(\ref{sch-equ-2}):
\be  \label{eff-sch-one}
-t_1 \left[ \psi_1(x+1)+\psi_1(x-1) \right] = (E- \td{\ep}_{x1} ) \psi_1(x-1),
\ee 
where
\be   \label{eff-dis-one}
\td{\ep}_{x1} = \ep_{x1} + \frac{t^2}{E-\delta{e}-\ep_{x2}}.
\ee 
Note that $\td{\ep}_{x1}$ has the meaning of an effective disorder potential on leg 1. 
Furthermore, if $|\ep_{x2}| \ll |E-\delta{e}|$ Eq.~(\ref{eff-dis-one}) can be expanded as
\be \label{eff-dis-one-c}
\td{\ep}_{x1} \simeq  \frac{t^2}{E-\delta{e}}+ \left[ \ep_{x1} + \frac{t^2}{(E-\delta{e})^2} \ep_{x2} \right]
+ O(\ep^2).
\ee
The first term on the r.h.s of Eq.~(\ref{eff-dis-one-c})  is a homogeneous potential shift. The second term is an effective disorder potential of zero mean.

Eqs.~(\ref{eff-sch-one}) and (\ref{eff-dis-one-c}) represent a single-chain problem, which can be solved exactly. 
The dispersion relation of the disorder-free part is determined by 
\be
-2t_1\cos{k} +  \frac{t^2}{E-\delta{e}} = E,
\ee
which gives the two non-overlapping bands
\be \label{ene-disp-one}
E_\tau(k) = -t_1 \cos{k} + \frac{\delta{e}}{2} -(-1)^{\tau} \sqrt{\left(t_1\cos{k} + \frac{\delta{e}}{2} \right)^2 + t^2}.
\ee
Of course, this coincides with Eq.~(\ref{dispers}) for $t_2=0$. Using the result for a single chain Anderson model\cite{mel} 
we obtain the localization length as
\be  \label{loc-len-t2-0}
1/\xi = \frac{\chi_1^2 + \td{\chi}_2^2 \tan^2{\gamma} }{8 \sin^2{k}} + O(\chi_1^4,\td{\chi}_2^4).
\ee
Here, the disorder on the leg 2 is measured by the dimensionless ratio 
\be
\td{\chi}_2 = \frac{\sigma_2^2}{t_1^2},
\ee
$\gamma$ is the mixing angle defined in Eq.~(\ref{clean-angle}) with $t_2=0$.
%, and $\td{\chi}_2$ is the effective variance of $\ep_{x2}$, which is measured in the unit of $t_1$. 
Comparing the result (\ref{loc-len-t2-0}) with Eq.~(\ref{loc-len-one-ch}), which describes the one-channel case in the weak disorder limit ($W_2\ll t_2$), one sees that the two limits do not commute. 
This is similar to the non-commutativity of the limits of weak disorder and weak inter-chain coupling in the resonant case.
Similarly as Eq.~(\ref{deltaE}) the characteristic disorder energy scale is the mean level spacing in the localization volume when $t_2=0$, which can be estimated as
\be
\delta\td{E} \sim t_1/\xi \sim \max\{\sigma_1^2,\sigma_2^2\}/t_1.
\ee
A perturbative analysis in $t_2$ is valid only if $t_2 \ll \delta\td{E}$. We expect a crossover to the regime of strong hopping on the slow leg when $t_2\sim  \delta\td{E}$.

\begin{figure}
\begin{center}
\includegraphics[height=5.75cm,width=8.2cm]{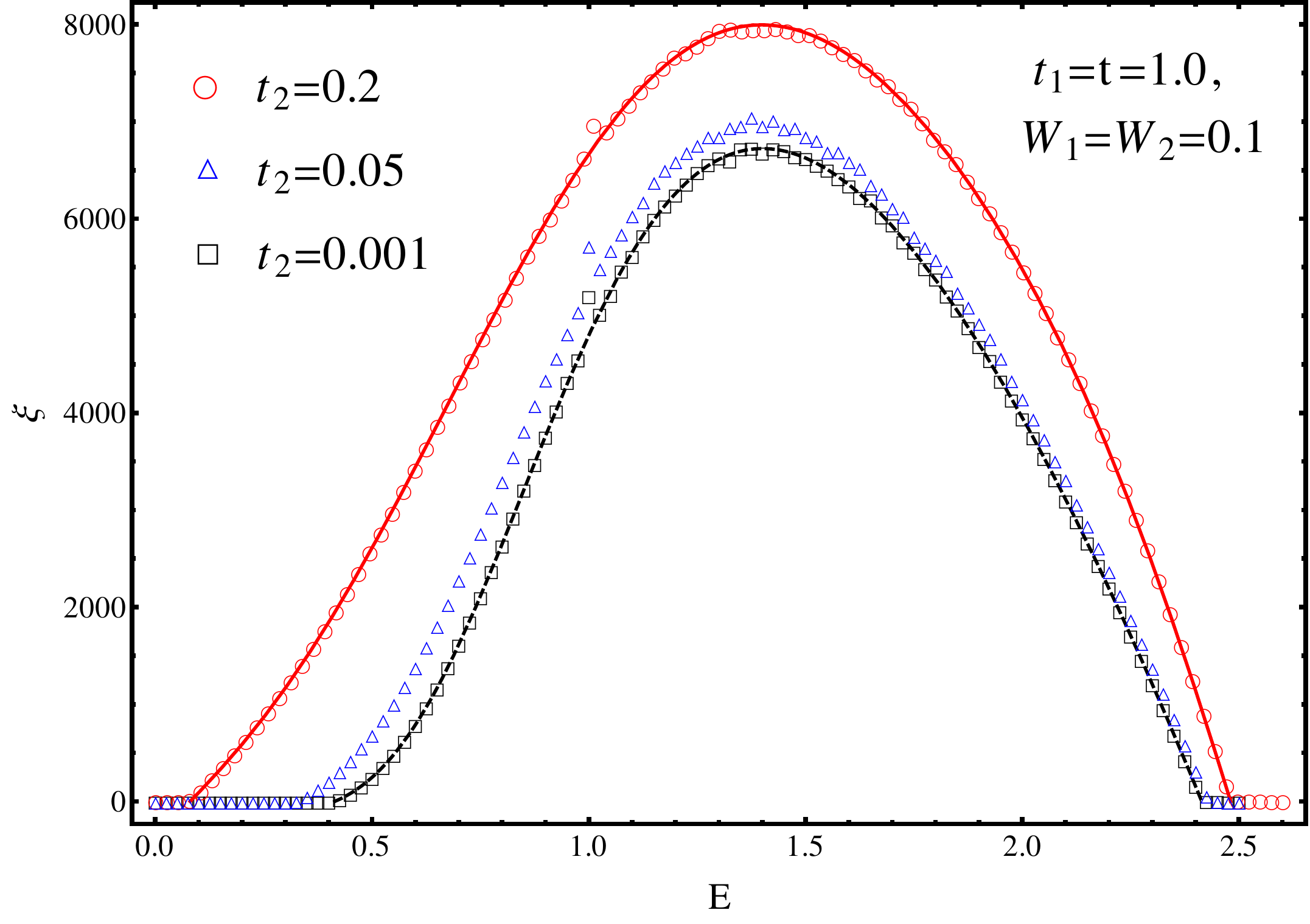}
\caption{Localization length as a function of energy for three values of $t_2$. 
$t_1=t=1$ and $W_1=W_2=0.1$ are kept fixed. $t_2 =0.2$, $0.05$ and $0.001$ capture the weak disorder limit, the intermediate regime and the limit of a nearly disconnected slow leg, respectively. 
The symbols are data of a numerical transfer matrix calculation. The solid and the dashed curves are the analytical results from 
Eqs.~(\ref{loc-len-one-ch},\ref{loc-len-t2-0}). 
The agreement with numerics in the limiting cases is very good. The localization length increases monotonically with $t_2$ for a fixed energy. 
An anomaly due to the commensurate wave vector $4k =2 \pi$ appears at $E \approx 1.0$.}  \label{ts-0}\
\end{center}
\end{figure}

The non-commutativity of the two limits is illustrated in Fig.~\ref{ts-0} where we compare the two analytical limits with numerical simulations at fixed disorder $W_1=W_2=0.1$ and hoppings $t_1=t=1$. Three values of $t_2$ are selected to cover the crossover from the weak disorder limit ($t_2 \gg W_2$) to the limit of a slow leg with disconnected sites ($t_2\to 0$). 
Eqs.~(\ref{loc-len-one-ch}) and (\ref{loc-len-t2-0}) are indeed seen to capture the two limits very well. 
Note that the localization length increases monotonically with $t_2$ for a fixed energy, as one may expect.    

\section{conclusion and possible applications}
The most important potential application of our theory is in the
realm of polaritons in quasi-one dimensional semiconductor
structures \cite{tri, man}. Here the fast chain corresponds to the
electromagnetic modes ("light") confined in a one-dimensional
structure and therefore having a parabolic dispersion with a very
small mass (large $t_1$) at small wave vectors. The slow chain
corresponds to the Wannier-Mott excitons, an electron-hole pair
coupled by Coulomb attraction. The mass of the exciton is typically
$10^{4}$ times larger than that of "light". Surface roughness of the
one-dimensional structure and impurities therein produce a disorder
potential acting on both excitons and "light"~\cite{sav}.
Experimentally, one can easily probe the intensity of the "light"
component by measuring the intensity of 3D photons that emerge due
to diffraction from the surface roughness. The amplitude of the
wavefunction of the exciton's center-of-mass  is more difficult to
access, but in principle still possible, e.g. via stimulated or
spontaneous exciton recombination and the related radiation. Another
application is related with one-dimensional structures in cold-atom
traps. By this technique one can construct and study coupled
one-dimensional chains in the same way as it was recently done for a
single chain~\cite{Aspekt, Roati}.

While all possible regimes can be achieved in a system of cold
atoms, the most relevant regime for one-dimensional cavity
polaritons is that described in Fig.~\ref{disper}(b), where for each band (the
lower and the upper polariton bands) only one channel
exists. This is because the exciton-light coupling is
 typically stronger than the narrow bandwidth $t_2\sim 1/m_{\rm exc}$ of the excitons.
The localization length in the case relevant for the upper
exciton-polaritons is shown in Fig.~\ref{sig-chanl}.

As expected, the localization length tends to zero near the bottom
and the top of the band, due to the vanishing rapidity. Note that in
the context of polaritons the upper band edge does not exist, since
the light has an unbounded continuous spectrum. In our model the top of the band
appears merely due to the discreteness of the lattice. In the
center of the band the localization length is of the order of the
localization length for the uncoupled "light" component. More
important is the distribution of amplitudes of "light" and "exciton"
wavefunctions which are similar to Fig.~\ref{wavefun}(a), with
"light" being represented by the wave function $\psi_{1}$ on the
fast leg and the "exciton" part being represented by the wave
function $\psi_{2}$ on the slow leg. One can see that the coupling
to light makes the exciton wave function spread over a distance
$\xi\sim \xi^{(0)}_{1}$ of the order of  the localization length of
light. This is much larger than the maximum exciton localization
length $\xi_{2}^{(0)}$ in the absence of coupling. The price for the
"fast transit" is that the amplitude of the exciton wave function is
small. This means that the transfer of a locally created exciton to distances of order $\xi_1^{(0)}$ is possible, but occurs with reduced probability.

In the long search for light localization  (see the paper by
A. Lagendijk, B. van Tiggelen and D. S. Wiersma in Ref.~\onlinecite{fify}) the
crucial point was to achieve a smaller localization length of light. Our results show that this
can also be achieved by coupling light to excitons near the bottom
of the upper polariton band.

We would like to emphasize that the model considered above does not
take into account an important property of polaritons, namely their
finite lifetime due to recombination of excitons, and the
out-coupling of the light from the waveguide. This limits the
coherence of polaritons and inhibits Anderson localization.  More
precisely, the effects of Anderson localization are only relevant if
the time to diffuse up to the scale of the localization length
(which by Thouless' argument is of the order of the inverse level
spacing in the localization volume)
 is smaller than the life time of the excitons. Further
crucial aspects are interactions among polaritons at finite
density, and the related possibility of interaction-induced delocalization and Bose condensation of
polaritons.~\cite{marcht,wertz}
 A complete theory of localization of hybrid particles like polaritons should take
 into account all these issues.\cite{ale}

Let us finally discuss the role of dimension for our results. We have found that under resonant conditions the 
localization lengths of two coupled chains are of the order of the localization length of the more localized, uncoupled leg. 
We may interpret this phenomenon as a manifestation of the fact that in 1d the mean free path is the relevant length scale 
that sets the localization length. Insofar it is not surprising that the backscattering rate, and thus the ``worst'' leg of 
the chains determines the localization properties of a coupled system. However, the close relation (proportionality) between 
mean free path and localization length is special for one-dimensional systems. In contrast in two dimensions the localization 
length becomes parametrically larger than the mean free path at weak disorder. In $d>2$ most eigenstates are even 
delocalized in weak disorder. Accordingly, we expect that the localization length is not so simply determined by the 
properties of the more disordered part among two coupled systems. Nevertheless, since the proliferation of weak-localization 
and backscattering leads to complete localization also in 2d (in the absence of special symmetries), we expect that 
a well propagating channel becomes more strongly localized upon resonant coupling to a more disordered channel. 
This may apply, e.g., to 2d polariton systems. However, in higher dimensions $d>2$ such a coupling might have a rather 
weak effect.  We expect that a ``fast'' channel is not affected much by a more disordered parallel channel. 
That such a trend exists indeed at high enough dimensions can be shown in the case of 
two coupled Bethe lattices,~\cite{hong-markus} which can be viewed as the limit of arbitrarily high dimensions.

We leave the investigation of problems in higher dimensions, and possible implications for interacting few-particle problems for future work.

\appendix

\section{Transfer matrix of an ``elementary slice'' in the current-conserving basis Eqs.~(\ref{j-states prop}, \ref{j-states-ev})}   \label{app-a}
In this appendix we derive Eqs.~(\ref{slice}) and (\ref{requ-symp}). In Eqs.~(\ref{sch-tran}, \ref{tran-eq}),
$\tilde{\mathbf{m}}_x$ is a \emph{symplectic} matrix which by
definition satisfies
\begin{eqnarray}
&&\tilde{\mathbf{m}}_x^{T}J\tilde{\mathbf{m}}_x=J,\label{symp}\\
&&\tilde{\mathbf{m}}_{x}=\tilde{\mathbf{m}}_{x}^{*}\label{TRS},
\end{eqnarray}
where
\begin{equation}
J=\begin{pmatrix} 0 & \mathbf{1} \\ -\mathbf{1} & 0
\end{pmatrix},
\end{equation}
and $\tilde{\mathbf{m}}_x^{T}$ is the matrix obtained from
$\tilde{\mathbf{m}}_x$ by transposition.

Define the new matrix ${\bf m}_{x}$
\begin{equation}
\label{m-def} {\bf m}_{x}={\bf U}_{x+1}^{-1}\,\tilde{{\bf
m}}_{x}\,{\bf U}_{x},
\end{equation}
where the rotation matrix is
\begin{equation}    \label{u-mat-two}
\mathbf{U}_x \equiv \begin{pmatrix} \boldsymbol{\alpha}_x &
\boldsymbol{\alpha}_x^{\ast} \\ \boldsymbol{\alpha}_{x-1} &
\boldsymbol{\alpha}_{x-1}^{\ast} \end{pmatrix},
\end{equation}
with $\boldsymbol{\alpha}_x$  defined by Eq.~(\ref{alp}). The
corresponding inverse matrix is given by:
\begin{equation}
\label{inverse} {\bf U}^{-1}_{x}=\frac{1}{\Delta}\begin{pmatrix}
\boldsymbol{\alpha}_{x-1}^{*} & -\boldsymbol{\alpha}_x^{\ast} \\
-\boldsymbol{\alpha}_{x-1} & \boldsymbol{\alpha}_{x},
\end{pmatrix}
\end{equation}
where $\Delta$ is the diagonal matrix
\begin{equation}
\label{delta}
\Delta=\boldsymbol{\alpha}_{x}\boldsymbol{\alpha}_{x-1}^{*}-
\boldsymbol{\alpha}_{x}^{*}\boldsymbol{\alpha}_{x-1}.
\end{equation}
The crucial point is that by current conservation Eq.~(\ref{curr}),
$\Delta$ is independent of coordinates and is proportional  to the
unit matrix in channel space:
\begin{equation}
\label{D} \Delta=i\,\mathbf{1}.
\end{equation}
Note also that, by construction, the rotation matrix ${\bf U}_{x}$ obeys the
disorder-free Schr\"odinger equation Eq.~(\ref{sch-tran}):
\begin{equation}
\label{U-Schr} {\bf U}_{x+1}=\tilde{{\bf
m}}_{x}|_{\tilde{\epsilon}=0}\,{\bf U}_{x}.
\end{equation}
It follows immediately from Eqs.~(\ref{U-Schr}, \ref{m-def}) that in
the absence of disorder ${\bf m}_{x}=\mathbf{1}$. In the presence of weak
disorder the matrix ${\bf m}_{x}$ acquires a small
coordinate-dependent correction proportional to
$\tilde{\boldsymbol{\epsilon}}_{x}$ which is given by Eq.~(\ref{slice}).

Next, by inverting Eq.~(\ref{m-def}) and plugging into Eq.~(\ref{TRS})
one obtains:
\begin{equation}
\label{m-TRS} \boldsymbol{\Sigma}^{(1)}_{x+1}\,{\bf
m}_{x}\,\boldsymbol{\Sigma}^{(1)*}_{x}={\bf m}_x^{*}.
\end{equation}
Using  definition of $\boldsymbol{\alpha}_{x}$ Eq.~(\ref{alp}) one
can readily show that
\begin{equation}
\label{ssig1} \boldsymbol{\Sigma}^{(1)}_{x}\equiv({\bf
U}_{x}^{*})^{-1}{\bf U}_{x}=\boldsymbol{\Sigma}_{1}
\end{equation}
is real and independent of $x$. This immediately reduces the
time-reversal symmetry condition Eq.~(\ref{m-TRS}) to the  form in
Eq.~(\ref{requ-symp}).

The same procedure applied to the symplecticity relation
Eq.~(\ref{symp}) results in the
following constraint (using $\tilde{\mathbf{m}}_x^{T}= \tilde{\mathbf{m}}_x^{\dagger}$):
\begin{equation}
\label{m-curr} {\bf
m}^{\dagger}_{x}\,\boldsymbol{\Sigma}^{(3)}_{x+1}\,{\bf
m}_{x}=\boldsymbol{\Sigma}^{(3)}_{x},
\end{equation}
where
\begin{equation}
\label{ssig3} \boldsymbol{\Sigma}^{(3)}_{x}\equiv{\bf
U}^{\dagger}_{x}\,J\,{\bf
U}_{x}=-\Delta\,\boldsymbol{\Sigma}_{3}=-i\,\boldsymbol{\Sigma}_{3}
\end{equation}
is independent of coordinate due to current conservation. Thus we
obtain the current conservation condition in Eq.~(\ref{requ-symp}).

\section{Perturbative calculation of $\delta{\vec{\l}}$ up to second order}   \la{per-cal}
Eqs.~(\ref{R-matrix}), (\ref{delta-R}) and (\ref{perturb}) fully
determine the variation of the eigen-system of the Hermitian matrix
$\mathbf{R}$. It is given by $\delta{\vec{\lambda}}$, which characterize the
``perturbation'' $\dR$. We can therefore use standard
perturbation theory to expand $\delta{\vec{\lambda}}$ into powers
of disorder on the additional slice. In this Appendix we calculate
$\delta{\vec{\l}}$ up to the second order, which is necessary to
derive the Fokker-Planck equation (\ref{fokker-planck-2}).

We introduce some quantities which are convenient to present the
results. Analogously to $\boldsymbol{\alpha}_x$, defined by Eq.~(\ref{alp}), we define
\begin{equation}
\boldsymbol{\beta}_x=\boldsymbol{\alpha}_x\mathbf{u},   \label{polbas}
\end{equation}
where $\mathbf{u}$ is the unitary matrix in Eq.~(\ref{u-matrix}).
Since $\boldsymbol{\alpha}_x$ describes the propagation in the plane-wave basis on the individual chains, and
$\mathbf{u}$ is the ``polarization'' matrix, we can consider
$\boldsymbol{\beta}_x$ as describing clean propagation in the ``polarized'' plane-wave basis.
Furthermore, analogously to the blocks in Eq.~(\ref{slice}), we can define two
quantities  on the ``polarized" basis, related with the forward- and back-scattering of the
right-moving particle off the slice:
\begin{equation}   \label{polfb}
\begin{split}
\boldsymbol{\Lambda}_x & =i\boldsymbol{\beta}^{\dagger}_x
\td{\boldsymbol{\epsilon}}_x \boldsymbol{\beta}_x, \\
\boldsymbol{\Sigma}_x & =i\boldsymbol{\beta}^{\dagger}_x
\td{\boldsymbol{\epsilon}}_x \boldsymbol{\beta}_x^{\ast},
\end{split}
\end{equation}
which are $2 \times 2$ matrices. It is easy to realize that
$\boldsymbol{\Lambda}_x$ is \emph{anti-Hermitian} and
$\boldsymbol{\Sigma}_x$ is \emph{symmetric}. The
corresponding left-moving quantities are complex conjugates of them.
The perturbative series of $\delta{\vec{\lambda}}$ are functions of
elements of $\boldsymbol{\Lambda}_x$ and $\boldsymbol{\Sigma}_x$.
For simplicity of further notations,
 we define
\begin{equation}   \label{def-ft}
\td{\mathbf{F}} = \sqrt{\mathbf{F}^2-1},
\end{equation}
and
\begin{equation} \label{del-f}
\Delta{F}=F_{1}-F_{2}.
\end{equation}
In order to facilitate the perturbative calculation, we adopt a parametrization of
$\mathbf{R}+\delta \mathbf{R}$ as in (\ref{R-matrix}), but with $\mathbf{F} \to \mathbf{F}+\delta \mathbf{F}$, $\tilde{\mathbf{F}} \to \tilde{\mathbf{F}}+\delta \tilde{\mathbf{F}}$ and $\mathbf{u} \to \mathbf{u}+\delta \mathbf{u}$,
in which
\be
\delta\td{\mathbf{F}} = \sqrt{(\mathbf{F}+\delta{\mathbf{F}})^2-1}-\sqrt{\mathbf{F}^2-1}.
\ee
Substituting this into Eq.~(\ref{perturb}) and
(\ref{delta-R}) we obtain two coupled equations for $\delta \mathbf{F}$ and the $2 \times 2$ matrix $\bf S$ which captures the incremental change of the polarization basis,
\begin{equation}
\mathbf{S} = 1+\mathbf{u}^{\dagger}\delta{\mathbf{u}},
\end{equation}
as
\begin{subequations}
\begin{equation}        \label{pe1}
\mathbf{S}(\mathbf{F}+\delta\mathbf{F})\mathbf{S}^{\dagger}=
\mathbf{F}+\mathbf{F}^{(1)}+\mathbf{F}^{(2)},
\end{equation}
\begin{equation}        \label{pe2}
\mathbf{S} (\td{\mathbf{F}}+\delta\td{\mathbf{F}})\mathbf{S}^{T}=
\td{\mathbf{F}}+\td{\mathbf{F}}^{(1)}+\td{\mathbf{F}}^{(2)},
\end{equation}
\end{subequations}

%, which is diagonal, defines the
We have introduced perturbation terms on the r.h.s. of the two equations as
\begin{subequations}     \label{per-term}
\begin{equation}
\mathbf{F}^{(1)} = -\mathbf{F} \mathbf{\Lambda}_{L} + \mathbf{\Lambda}_{L}
\mathbf{F} + \mathbf{\Sigma}_L \td{\mathbf{F}} + \td{\mathbf{F}}
\mathbf{\Sigma}_L^{\ast},
\end{equation}
\begin{equation}
\mathbf{F}^{(2)} = -\mathbf{\Lambda}_{L}\mathbf{F}\mathbf{\Lambda}_{L} +
\mathbf{\Sigma}_L \mathbf{F} \mathbf{\Sigma}_L^{\ast} + \mathbf{\Lambda}_L
\td{\mathbf{F}} \mathbf{\Sigma}_L^{\ast}- \mathbf{\Sigma}_L \td{\mathbf{F}}
\mathbf{\Lambda}_L,\end{equation}
\begin{equation}
\td{\mathbf{F}}^{(1)} = -\td{\mathbf{F}} \mathbf{\Lambda}_{L}^{\ast} +
\mathbf{\Lambda}_{L} \td{\mathbf{F}} + \mathbf{\Sigma}_L \mathbf{F} +
\mathbf{F}\mathbf{\Sigma}_L,
\end{equation}
\begin{equation}
\td{\mathbf{F}}^{(2)} = -\mathbf{\Lambda}_{L} \td{\mathbf{F}}
\mathbf{\Lambda}_{L}^{\ast} + \mathbf{\Sigma}_L \td{\mathbf{F}}
\mathbf{\Sigma}_L + \mathbf{\Lambda}_L \mathbf{F} \mathbf{\Sigma}_L -
\mathbf{\Sigma}_L \mathbf{F} \mathbf{\Lambda}_L^{\ast},
\end{equation}
\end{subequations}\\
where $\mathbf{F}^{(1)}$ and $\td{\mathbf{F}}^{(1)}$ are linear in
disorder, while $\mathbf{F}^{(2)}$ and $\td{\mathbf{F}}^{(2)}$ are
quadratic. Additionally, $\mathbf{F}^{(1)}$ and $\mathbf{F}^{(2)}$
are \emph{Hermitian}, but $\td{\mathbf{F}}^{(1)}$ and
$\td{\mathbf{F}}^{(2)}$ are \emph{symmetric}.

We expand $\delta{\mathbf{F}}$ and $\delta\mathbf{u}$ in
disorder strength. From the latter we calculate the corresponding variations of
angular variables.
%The calculation is based on the \emph{canonical perturbation theory}, which is introduced in quantum mechanics.
Without going into the details of the calculation, we present the
results up to the second order in disorder.

To first order the corrections
\be \delta{\vec{\lambda}}^{(1)} =
(\delta{F}_1^{(1)}, \delta{F}_2^{(1)}, \delta{\theta}^{(1)}, \delta{\psi}^{(1)}, \delta{\phi}^{(1)}, \delta{\varphi}^{(1)})
\ee
are given by
\begin{subequations}  \label{1st}
\be
\delta{F}_\varrho^{(1)} = \mathbf{F}^{(1)}_{\varrho,\varrho} ,
\quad  \varrho \in \{1,2\},
\ee
\be
\delta{\theta}^{(1)} =\frac{2}{\Delta{F}} \re{\left(
\mathbf{F}^{(1)}_{2,1} e^{i \psi} \right)},
\ee
\be
\delta{\psi}^{(1)}  = \frac{1}{2} \left(\frac{\im\td{\mathbf{F}}^{(1)}_{2,2}}
{\td{F}_2} - \frac{\im\td{\mathbf{F}}^{(1)}_{1,1}}{\td{F}_1}\right)-\delta{\varphi}^{(1)} \cos{\theta},
\ee

\be
\delta{\phi}^{(1)} = -\frac{1}{2} \left(\frac{\im\td{\mathbf{F}}^{(1)}_{2,2}}
{\td{F}_2} + \frac{\im\td{\mathbf{F}}^{(1)}_{1,1}}{\td{F}_1}\right),
\ee
\be
\delta{\varphi}^{(1)} = \frac{2}{\Delta{F}}
\im{\left(\mathbf{F}^{(1)}_{2,1}e^{i \psi}\right)} \csc{\theta},
\ee
\end{subequations}
where the subscripts denote the matrix elements of the ``perturbations''  $\mathbf{F}^{(1)}$ in Eq.~(\ref{1st}). %$\re(z)$ and $\im(z)$ are real and imaginary part of the complex argument $z$.
 We recall that the  ``perturbations'' in Eq.~(\ref{1st}) are $L$-dependent.

The second order corrections
\be \delta{\vec{\lambda}}^{(2)} =
(\delta{F}_1^{(2)}, \delta{F}_2^{(2)}, \delta{\theta}^{(2)}, \delta{\psi}^{(2)}, \delta{\phi}^{(2)}, \delta{\varphi}^{(2)})
\ee
are more
complicated. However, we recall that our aim is to calculate the
correlators
$\overline{\delta{\lambda_i^{(1)}}\delta{\lambda_j^{(1)}}}$ and
$\overline{\delta{\lambda_i^{(2)}}}$ in Eq.~(\ref{fokker-planck-2}).
To avoid repeating calculation, we should express the
$\delta{\vec{\lambda}}^{(2)}$'s in terms of the first order corrections
Eq.~(\ref{1st}) as far as possible. We obtain
\begin{widetext}
\begin{subequations}                    \label{2nd}
\be
\delta{F}_\varrho^{(2)}=\mathbf{F}^{(2)}_{\varrho,\varrho} -(-1)
^{\varrho} \frac{|\mathbf{F}^{(1)}_{2,1}|^2} {\Delta{F}}  ,\quad
\varrho \in \{1,2\},
\ee
\be
\delta{\theta}^{(2)}= \frac{2}{\Delta{F}} \re{\left
( \mathbf{F}^{(2)}_{2,1} e^{i \psi} \right)} +  a^{(1)}
\delta{\theta}^{(1)}+\frac{1}{4}\sin{2\theta}
\left( \delta{\varphi}^{(1)} \right)^{2},
\ee
\be
\delta{\psi}^{(2)} = a^{(2)}_{-} + b^{(1)}_{-}
\delta{\phi}^{(1)}+b_{+}^{(1)} d^{(1)} +c_{+} e^{(2)} -a^{(1)}_{-} c^{(1)}_{+} -
\frac{1}{2}\sin{\theta}\delta{\theta}^{(1)}
\delta{\varphi}^{(1)}-\cos{\theta}\delta{\varphi}^{(2)},
\ee
\be
\delta{\phi}^{(2)} = -a_{+}^{(2)}-b_{+}^{(1)} \hph^{(1)}-b_{-}^{(1)} d^{(1)} +c_{-} e^{(2)},
\ee
\be
\delta{\varphi}^{(2)} = \frac{2}{\Delta{F}}
\im{\left(\mathbf{F}^{(2)}_{2,1} e^{i \psi}\right)}
\csc{\theta}+ a^{(1)} \delta{\varphi}^{(1)} -  \cot{\theta}\hth^{(1)}\hvp^{(1)},
\ee
\end{subequations}
\end{widetext}
in which
\begin{subequations}
\be
a^{(1)}= \frac{1}{\Delta{F}}\left( \delta{F}_2^{(1)}
-\delta{F}_1^{(1)} \right),
\ee
\begin{equation}  \label{a2}
a^{(2)}_{\pm}=\frac{1}{2} \left( \frac{\im\td{\mathbf{F}}^
{(2)}_{2,2}}{\td{F}_2} \pm \frac{\im\td{\mathbf{F}}^{(2)}_{1,1}}{\td{F}_1} \right),
\end{equation}
\begin{equation}
b_{\pm}^{(1)}= \frac{1}{2} \left( \frac{F_1}{\td{F}_1^2}
 \delta{F}_1^{(1)} \pm \frac{F_2}{\td{F}_2^2} \delta{F}_2^{(1)} \right),
\end{equation}
\be
c_{\pm}=\fr{1}{2}\left( \frac{\td{F}_1}{\td{F}_2} \pm \fr{\td{F}_2}{\td{F}_1} \right),
\ee
\be
d^{(1)}=\hvp^{(1)}\cos{\theta}+ \delta\psi^{(1)},
\ee
\bea
e^{(2)} & = & \frac{1}{4}\left[ \left(\hvp^{(1)}\right)^2 \sin^2{\theta} -\left( \hth^{(1)} \right)^2 \right]\sin{2\psi} \nn \\
        && +\frac{1}{2} \hvp^{(1)} \hth^{(1)} \sin{\theta} \cos{2\psi}.
\eea
\end{subequations}
In practice, we first calculate all the correlators
$\overline{\delta{\lambda_i^{(1)}}\delta{\lambda_j^{(1)}}}$ by Eq.
(\ref{1st}). At the same time we obtain the correlators relevant
for the products of first order terms on the r.h.s. of Eq.~(\ref{2nd}). Finally,
after evaluating the disorder average of $a^{(2)}_{\pm}$ in Eq.
(\ref{a2}), $\overline{\delta{\lambda_i^{(2)}}}$s can be obtained.\\

\section{co-efficients of Eq.~(\ref{fokker-planck-3})}     \label{coe}
The co-efficients of Eq.~(\ref{fokker-planck-3}) are
\begin{widetext}
\begin{subequations}
\begin{equation}
c_1=-2\frac{\td{F}_1^2}{\Delta{F}}\Gamma_6,
\end{equation}
\begin{equation}
c_2=2\frac{\td{F}_2^2}{\Delta{F}}\Gamma_6,
\end{equation}
\begin{eqnarray}
c_3  = && \frac{1}{\Delta{F}^2} \left[ \left(F_1^2-F_2^2-2\right)
\left(\Gamma_5+\Gamma_4\right)+2F_1F_2\left(\Gamma_5-\Gamma_4\right) \right. \nonumber \\
       && \left. -\left( \tilde{F}_{1}^2+ \tilde{F}_{2}^2 \right)
       \left( 1-u^2 \right)\partial_u\Gamma_6\right] -4 \frac{1}
       {\Delta{F}^2} \tilde{F}_{1}\tilde{F}_{2}u\Gamma_6\cos{2 \psi},
\end{eqnarray}
\begin{eqnarray}
c_4  = && \frac{1}{\Delta{F}}\left[ \left( \tilde{F}_{1}\tilde{F}_{2} +
\frac{F_{2}}{\tilde{F}_{2}}\tilde{F}_{1} - \frac{F_{1}}{\tilde{F}_{1}}\td{F}_{2}\right)
\Gamma_3 + 2\frac{\td{F}_{1}\td{F}_{2}}{\Delta{F}} \frac{u}{\sqrt{1-u^2}}
\left(\Gamma_5-\Gamma_4\right) \right. \nonumber \\
       &&  \left. + \frac{F_{1}}{\tilde{F}_{1}}\td{F}_{2}
       \partial_u\Gamma_4+\frac{F_{2}}{\tilde{F}_{2}}\td{F}_{1}
       \partial_u \Gamma_5 -2\frac{\tilde{F}_{1}\tilde{F}_{2}}
       {\Delta{F}} \left( \Gamma_6+2u \partial_u\Gamma_6 \right) \right]\sin{2 \psi},
\end{eqnarray}
\begin{equation}
c_{11} =\tilde{F}_{1}^2 \Gamma_{1},
\end{equation}
\begin{equation}
c_{12}=2 \tilde{F}_1 \tilde{F}_2 \Gamma_3 \cos{2\psi},
\end{equation}
\begin{equation}
c_{13} = \frac{2\tilde{F}_{1}}{\DF} \left( \tilde{F}_{1}+\tilde{F}_{2}
\cos{2\psi} \right)\Gamma_{4},
\end{equation}
\begin{equation}
c_{14} = - \tilde{F}_{1} \left( \frac{F_{2}}{\tilde{F}_{2}}\Gamma_3 +
2\frac{\tilde{F}_2}{\DF} \frac{u}{\sqrt{1-u^2}} \Gamma_4 \right) \sin{2\psi},
\end{equation}
\begin{equation}
c_{22} =\tilde{F}_{2}^2 \Gamma_{2},
\end{equation}
\begin{equation}
c_{23} =  2\frac{\tilde{F}_{2}}{\DF} \left( \tilde{F}_{2} + \tilde{F}_{1}
\cos{ 2 \psi } \right) \Gamma_{5},
\end{equation}
\begin{equation}
c_{24} = - \tilde{F}_{2}\left(\frac{ F_{1}}{\tilde{F}_{1}}\Gamma_3 +
2\frac{\tilde{F}_1}{\DF} \frac{u}{\sqrt{1-u^2}} \Gamma_{5} \right) \sin{2 \psi},
\end{equation}
\begin{equation}
c_{33} = \left[ V_3+\frac{1}{\DF^2}\left(\tilde{F}_1^2+\tilde{F}_2^2+
2\tilde{F}_1\tilde{F}_2\cos{2 \psi}\right) \Gamma_6 \right] \left( 1-u^2 \right),
\end{equation}
\begin{equation}
c_{34} = -\frac{1}{\DF} \left( \frac{F_1 \tilde{F}_2}{\tilde{F}_1}
\Gamma_4+ \frac{\tilde{F}_1 F_2}{\tilde{F}_2} \Gamma_5 -
4\frac{\tilde{F}_1\tilde{F}_2}{\DF} u \Gamma_6 \right) \sin{2\psi},
\end{equation}
\begin{equation}
\begin{split}
c_{44} = & \frac{1}{2}\left[ 1+ \frac{1}{2}
\left( \frac{F_1}{\tilde{F}_1} \right)^2 \right]
\Gamma_1+ \frac{1}{2} \left[ 1+ \frac{1}{2}
\left( \frac{F_2}{\tilde{F}_2} \right)^2 \right]
\Gamma_2-\Gamma_6\\
& +\frac{u}{\sqrt{1-u^2}} \left[ \left( 2+ \frac{F_1}{\DF}\right)
\Gamma_4+\left(2+ \frac{F_2}{\DF}\right)\Gamma_5 \right] +\frac{u^2}{1-u^2}
\left[ \Gamma_3+\left( 1+\frac{\tilde{F}_1^2+\tilde{F}_2^2}{\DF^2} \right)
\Gamma_6 \right]\\
     &- \left[ \frac{1}{2}\frac{F_{1}F_{2}}{\td{F}_1\td{F}_2}
     \Gamma_3 + \frac{u}{\sqrt{1-u^2}} \frac{1}{\Delta{F}}
     \left(\frac{F_1\tilde{F}_2}{\tilde{F}_1}\Gamma_1+
     \frac{\tilde{F}_1 F_2}{\tilde{F}_2}\Gamma_2\right) +
     2 \frac{u^2}{1-u^2} \frac{\td{F}_1\td{F}_2}{\Delta{F}^2} \right] \cos{2\psi},
\end{split}
\end{equation}
\end{subequations}
\end{widetext}
where the $\td{F}_\varrho$'s are the two diagonal elements of the matrix
(\ref{def-ft}) and $\Delta{F}$ is defined in Eq.~(\ref{del-f}). The
new quantities introduced in Eq.~(\ref{coe}) are defined below.
Notice first that
%\begin{equation}
%\td{\psi}=2\psi.
%\end{equation}
The $\Gamma_n$  are functions of $u$ defined by
\begin{widetext}
\begin{subequations}   \label{gammas}
\begin{equation}
\Gamma_{1(2)}(u)=V_1+V_2+4V_3-(+)2\left(V_2-V_1\right)u+\left(V_1+V_2- 4 V_3\right)u^2,
\end{equation}
\begin{equation}
\Gamma_{3}(u)=\left(V_1+V_2- 4 V_3\right)\left(1-u^2\right),
\end{equation}
\begin{equation}
\Gamma_{4(5)}(u)=\left[ V_1-V_2 +(-)\left(V_1+V_2-4V_3\right)u \right] \left(1-u^2\right),
\end{equation}
\begin{equation}  \label{gamma6}
\Gamma_{6}(u)=V_1+V_2-\left(V_1+V_2-4V_3\right)u^2,
\end{equation}
\end{subequations}
\end{widetext}
where  $V_1$, $V_2$ and $V_3$ are the three Born cross-sections defined in
Eq.~(\ref{born-sec}). In order to solve Eq.~(\ref{u-fix}) in the limit $L\gg
1$, we need the values of $c_3$ and $c_{33}$ in the limit $F_{\max} \gg F_{\min} \gg 1$,
\begin{subequations}
\begin{equation*}
\lim_{L\rightarrow \infty}c_{3} = \left( \left|V_1-V_2\right|-
\partial_u \Gamma_6 \right) (1-u^2),
\end{equation*}
\begin{equation*}
\lim_{L\rightarrow \infty}c_{33} = \left( V_3 + \Gamma_6 \right) \left( 1-u^2 \right),
\end{equation*}
\end{subequations}
which is Eq.~(\ref{c3-c33}).

\section{Transfer matrix of an ``elementary slice'' in the basis Eqs.~(\ref{j-states prop}, \ref{states-ev}) } \label{app-d}
The derivation of Eq.~(\ref{ele-sli-one-ch}) goes in parallel with the derivation of Eq.~(\ref{slice}) in App.~\ref{app-a}. However, the rotation $\mathbf{U}_{x}$ is constructed in such a way that
\be  \label{u-mat-one}
\mathbf{U}_{x} = \begin{pmatrix} \psi_{1}^{+}(x) & \psi_{1}^{-}(x) & 0 & 0 \\ 0 & 0 & \psi_{2}^{+}(1) & \psi_{2}^{-}(1) \\
                                   \psi_{1}^{+}(x-1) & \psi_{1}^{-}(x-1) & 0 & 0  \\ 0 & 0 & \psi_{2}^{+}(0) & \psi_{2}^{-}(0)
 \end{pmatrix},
\ee
where $\psi_{1,2}^{\pm}(x)$ are defined by Eqs.~(\ref{j-states prop}) and (\ref{states-ev}). 
Compared with Eq.~(\ref{u-mat-two}) for the two-channel case, the second and third columns of Eq.~(\ref{u-mat-one}) have been permuted, and the columns corresponding to the evanescent channel are coordinate-independent. The inverse of $\mathbf{U}_{x}$ is
\be  \label{inv-u-mat}
\mathbf{U}_{x}^{-1} = \begin{pmatrix} -i \psi_{1}^{-}(x-1) & 0 & i\psi_{1}^{-}(x) & 0 \\ i\psi_{1}^{+}(x-1) & 0 &  -i\psi_{1}^{+}(x) & 0 \\
                                  0 & -\psi_{2}^{-}(0) & 0 & \psi_{2}^{-}(1)  \\ 0 &  \psi_{2}^{+}(0) & 0 & -\psi_{2}^{+}(1)
 \end{pmatrix}.
\ee
The transfer matrix of an elementary slice (\ref{ele-sli-one-ch}) is
\be
\begin{split}
\mathbf{m}_x = & \mathbf{U}_{x+1}^{-1} \td{\mathbf{m}}_x \mathbf{U}_{x}
             = \mathbf{m} + \delta{\mathbf{m}}_x,
 \end{split}
\ee
where $\mathbf{m}$ and $\delta{\mathbf{m}}_x$ take the form given in Eq.~(\ref{tran-mat-ii}).
Following the same procedure as in App.~\ref{app-a}, one can also obtain the symmetry constraints on the matrix $\mathbf{m}_x$, 
which is imposed by the reality and symplecticity of the matrix $\td{\mathbf{m}}_x$. Without going into details we present the following results: The reality relation (\ref{TRS}) gives
\be
\mathbf{m}_x^\ast = \boldsymbol{\Lambda}_1 \mathbf{m}_x \boldsymbol{\Lambda}_1, \quad \boldsymbol{\Lambda}_1= \begin{pmatrix} \sigma_1 & 0 \\ 0 & \mathbf{1} \end{pmatrix};
\ee
The symplecticity relation (\ref{symp}) gives
\be  \label{symp-one}
\mathbf{m}_x^\dagger \boldsymbol{\Lambda}_3 \mathbf{m}_x = \boldsymbol{\Lambda}_3, \quad \boldsymbol{\Lambda}_1= \begin{pmatrix} \sigma_3 & 0 \\ 0 & \sigma_2 \end{pmatrix}.
\ee
Finally, it is not hard to show that the Lyapunov exponents of the products (\ref{tm-l}) satisfy the symmetry property stated in Eq.~(\ref{sym-lya-one}).

\begin{acknowledgments}
We are grateful to B. L. Altshuler, D. Basko, F. Marchetti, M. Richard, Y. Rubo and V.~I.~Yudson for
stimulating discussions.
\end{acknowledgments}


\smallskip
\begin{thebibliography}{99}
\bibitem{and}P. W. Anderson, Phys. Rev. \textbf{109}, 1492 (1958).

\bibitem{fify}\emph{50 years of Anderson localization}, ed. by E. Abrahams, World Scientific, 2010.

\bibitem{abra-and-Licc-rama} E.~Abrahams, P.~W.~Anderson, D.~C.~Licciardello, and
T.~V.~Ramakrishnan, Phys. Rev. Lett. \textbf{42}, 673 (1979).

\bibitem{berez}V. L. Berezinskii, Zh. Exp. Tero. Fiz. \textbf{65}, 1251 (1973)
[Sov. Phys. JETP \textbf{38}, 620 (1974)].

\bibitem{mel}V. I. Mel'nikov, Fiz. Tverd. Tela (Leningrad) \textbf{23}, 782 (1981)
[Sov. phys. Solid State \textbf{23}, 444 (1981)].

\bibitem{efet}K. B. Efetov, \emph{Supersymmetry in disorder and chaos},
Cambridge University Press, 1999.

\bibitem{ger-vas}M.~E.~Gertsenshtein and V.~B.~Vasil'ev, Theor. Probab. Appl., \textbf{IV}, 391 (1959).

\bibitem{dorok}O. N. Dorokhov, Solid State Commun. \textbf{44}, 915 (1982).

\bibitem{kas-wel} M. Kasner and W. Weller, Phys. Stat. Solidi (b) \textbf{148}, 635 (1988).

\bibitem{tar}A.~V.~Tartakovski, Phys. Rev. B \textbf{52}, 2704 (1995).

%\bibitem{lifz}T. M. Lifshitz, Adv. Phys. \textbf{13}, 483 (1964).

\bibitem{mpk1}P. A. Mello, P. Pereyra and N. Kummar, Ann. Phys. \textbf{181}, 290 (1988).

\bibitem{mpk2}P. A. Mello and N. Kumar, \emph{Quantum Transport in Mesoscopic Systems:
Complexity and Statistical Fluctuations} (Oxford University Press, 2004).

\bibitem{bee}C. W. J. Beenakker  and B. Rejaei, Phys. Rev. Lett. \textbf{71}, 3689 (1993);
Phys. Rev. B \textbf{49}, 7499 (1994).

%\bibitem{bee2}For example, Eq.~(154) in C. W. J. Beenakker, Rev. Mod. Phys. \textbf{69}, 731 (1997).

\bibitem{tri}A. Trichet et al. Phys. Rev. B \textbf{83}, 041302(R) (2011).

\bibitem{man}F. Manni, et al. Phys. Rev. Lett. \textbf{106}, 176401 (2011).

\bibitem{sav}V. Savona, et al. Phys. Rev. Lett. \textbf{78}, 4470 (1997).

\bibitem{kos}V. A. Kosobukin, Fiz. Tverd. Tela \textbf{45}(6), 1091 (2003) [Phys. Solid State \textbf{45}(6), 1145 (2003)].

\bibitem{whi}D. M. Whittaker, et al. Phys. Rev. Lett. \textbf{77}, 4792 (1996).

\bibitem{wei}W. Zhang, et al. Phys. Rev. B \textbf{81}, 214202 (2010).

%\bibitem{guo-xia}A.~M.~Guo and S.~J.~Xiang, Phys. Rev. B \textbf{83}, 245108 (2011).

\bibitem{tho}A. U. Thomann, V. B. Geshkenbein and G. Blatter, Phys. Rev. B \textbf{79}, 184515 (2009).

\bibitem{hamsh}M. Hamermesh, \emph{Group Theory and This Applications to Physical Problem},
(Addison-Wesley, 1962).

\bibitem{krav-yud}V. E. Kravtsov and V. I. Yudson, Phys. Rev. B \textbf{82}, 195120 (2010);
Annals of Physics, \textbf{326}, 1672 (2011).

\bibitem{ngu-kim}B.~P.~Nguyen and K.~Kim, J.~Phys.: Cond.~Matter \textbf{24}, 135303 (2012).

\bibitem{imp-trassm-mat}The transmission matrix $\mathbf{t}$ we calculate here describes
transmission from left to right of the sample. Its transpose describes the reverse transmission, as assured by time reversal symmetry.
 Its element $t_{\tau \tau^\prime}$ denotes the out-going amplitude
on the r.h.s of the sample in the $\tau$-channel when there is a unit current incident from
the l.h.s in the $\tau^\prime$-channel.
\bibitem{shi-gen} Z.~Shi and A.~Z.~Genack, Phys.~Rev.~Lett. \textbf{108}, 043901 (2012).
\bibitem{fgp} S. Fishman, D. R. Grempel and R. E. Prange, Phys. Rev. Lett. \textbf{49}, 509 (1982).

\bibitem{mac-kram} A. MacKinnon and B. Kramer, Z. Phys. B \textbf{53}, 1 (1983);
A. MacKinnon, in \emph{Anderson Localization and Its Ramifications: Disorder,
Phase Coherence and Electron Correlations} [Lect. Notes Phys. Vol. 630], edited
by T. Brandes and S. Kettemann (Springer, 2003), p. 21.

\bibitem{bag} P.~F.~Bagwell, Phys.~Rev.~B {\bf 41}, 10354 (1990).
\bibitem{heinrich} J. Heinrichs, Phys.~Rev.~B {\bf 68}, 155403 (2003).
\bibitem{eva-div} Repeating the perturbative calculation in App.~\ref{per-cal}, we can obtain three cross-sections similar to Eq.~(\ref{born-sec}), in which $V_2$ and $V_3$ acquire exponential growing factors $e^{4\kappa_\nu |x|}$ and $e^{2\kappa_\nu |x|}$. Thus the obstacle of divergence can be rephrased in this way: The intensities of intra-evanescent-channel and inter-channel scattering (either forward or backward) are amplified exponentially with respect to the coordinate of the elementary slice, which invalidates the weak disorder expansion of $\vec{\lambda}$.

\bibitem{bggs-cpv} G.~Benettin, L.~Galgani, A.~Giorgilli, J.~M.~Strelcyn,  Meccanica \textbf{15}, 9,21 (1980); A.~Crisanti, G.~Paladin, A.~Vulpiani, \emph{Products of Random Matrices in Statistical Physics} [Springer Series in Solid-State Sciences 104], (Springer-Verlag, 1993).
\bibitem{dav} J.~H.~Davies, \emph{The Physics of Low-Dimensional Semiconductors: An Introduction}, (Cambridge University Press, 1998).
\bibitem{hh-swk}H. Huang and S. W. Koch, \emph{Quantum Theory of the Optical
and Electronic Properties of Semiconductors}, (World Scientific, 2004).

\bibitem{Aspekt} J. Billy, V. Josse, Z. Zuo, A. Bernard, B. Hambrecht,
P. Lugan, D. Clement, L. Sanchez-Palencia, P. Bouyer, and A. Aspect,
Nature (London), {\bf 453}, 891 (2008).

\bibitem{Roati} G. Roati, C. D'Errico, L. Falani, M. Fattori, C. Fort,
M. Zaccanti, G. Modugno, M. Modugno, and M. Inguscio, Nature (London),
{\bf 453}, 895 (2008).

\bibitem{marcht} F. M. Marchetti, J. Keeling, M. H. Szyma\'{n}ska and P. B. Littlewood, Phys. Rev. B {\bf 76}, 115326 (2007).

\bibitem{wertz}E.~Wertz, L.~Ferrier, D.~D.~Solnyshkov, R.~Johne, D.~Sanvitto, A.~Lema\^{i}tre, I.~Sagnes, 
R.~Grousson, A.~V.~Kavokin, P.~Senellart, G.~Malpuech and J.~Bloch, Nature Phys. \textbf{6}, 860 (2010).

\bibitem{ftn} Note that we are using the semiconductor terminology of valence and conduction band rather loosely to
denote the lower and the upper energy band. They will typically not be separated by a gap,
but overlap in some energy range, see Fig.~\ref{disper}. The latter "two-channel regime" is at the focus of our attention in this paper.

\bibitem{ale} I.~L.~Aleiner, B.~L.~Altshuler and Y.~G.~Rubo, Phys.~Rev.~B {\bf 85}, 121301(R) (2012).
\bibitem{hong-markus} H.~Y.~Xie and M. M\"uller (unpublished).
\end{thebibliography}
\end{document}